\documentclass[aps,pra,reprint,twocolum,amsmath,superscriptaddress,amssymb,showpacs,nofootinbib]{revtex4-1}
\usepackage{amsmath,braket,amssymb}
\usepackage[final]{graphicx}
\usepackage{subfigure}
\usepackage[english]{babel}
\usepackage[utf8]{inputenc}
\usepackage{mathtools}
\usepackage{empheq}
\usepackage{tensor}
\usepackage{cancel}

\usepackage{simplewick}

\usepackage[usenames,dvipsnames]{color}

\usepackage{enumitem}
\usepackage{slashed}
\usepackage{graphicx}
\usepackage{xcolor}

\colorlet{Green}{black!30!green}

\usepackage{tikz}
\usetikzlibrary{calc,fadings,decorations.pathreplacing,shapes,shapes.multipart,arrows,shapes.misc,intersections,positioning,patterns}
\tikzset{arrow data/.style 2 args={%
		decoration={%
			markings,
			mark=at position #1 with \arrow{#2}},
		postaction=decorate}
}
\usepackage{bm}
\usepackage[colorlinks=true,citecolor=blue,linkcolor=blue,urlcolor=blue]{hyperref}
\usepackage{dsfont}
\usepackage{changepage}
\usepackage{array}
\usepackage{soul}
\usepackage[capitalise]{cleveref}
\crefname{section}{Sec.}{Secs.}
\Crefname{section}{Sec.}{Secs.}
\usepackage{xifthen}
\usepackage{xargs}
\usepackage{dynkin-diagrams}

\usepackage{bm}
\renewcommand{\vec}[1]{\boldsymbol{\mathbf{#1}}}

\graphicspath{{./}{./figures/}}

\newcommand{\bit}{\begin{itemize}}
\newcommand{\eit}{\end{itemize}}

\newcommand{\f}{\frac}
\renewcommand{\>}{\right\rangle}
\newcommand{\<}{\left\langle}
\newcommand{\ba}{\begin{align}}
\newcommand{\ea}{\end{align}}
\newcommand{\be}{\begin{equation}}
\newcommand{\ee}{\end{equation}}
\newcommand{\bi}{\begin{itemize}}
\newcommand{\ei}{\end{itemize}}
\newcommand{\lf}{\left(}
\newcommand{\ri}{\right)}
\newcommand{\dd}{\mathrm{d}}

\newcommand{\Tr}{\operatorname{Tr}}
\newcommand{\tr}{\operatorname{tr}}

\DeclareMathAlphabet{\mymathbb}{U}{BOONDOX-ds}{m}{n}

\newcommand{\rvline}{\hspace*{-\arraycolsep}\vline\hspace*{-\arraycolsep}}

\renewcommand{\log}{\ln}

\newcommand{\red}{\color{red}}
\newcommand{\blue}{\color{blue}}

\newcommand{\Hs}{\mathcal{H}}

\usepackage{braket}
\usepackage[normalem]{ulem}

\begin{document}
\date{\today}

\newcommand{\bbra}[1]{\<\< #1 \right|\right.}
\newcommand{\kket}[1]{\left.\left| #1 \>\>}
\newcommand{\bbrakket}[1]{\< \Braket{#1} \>}
\newcommand{\pll}{\parallel}
\newcommand{\nn}{\nonumber}
\newcommand{\transp}{\text{transp.}}
\newcommand{\nor}{z_{J,H}}

\newcommand{\hL}{\hat{L}}
\newcommand{\hR}{\hat{R}}
\newcommand{\hQ}{\hat{Q}}

\title{Nonlinear sigma models for monitored dynamics of free fermions}

\begin{abstract}
We derive field theory descriptions for measurement-induced phase transitions in free fermion systems. We focus on a multi-flavor Majorana chain, undergoing Hamiltonian evolution with continuous monitoring of  local fermion parity operators.  Using the replica trick, we map the dynamics to the imaginary time evolution of an effective spin chain, and use the number of flavors as a large parameter for a controlled derivation of the effective field theory. This is a nonlinear sigma model for an orthogonal $N\times N$ matrix, in the replica limit $N\to 1$. (On a boundary of the phase diagram, another sigma model with higher symmetry applies.) Together with known results for the renormalization-group beta function, this derivation establishes the existence of stable phases --- nontrivially entangled and disentangled respectively --- in the physically-relevant replica limit $N\to 1$. In the nontrivial phase, an asymptotically exact calculation shows that the bipartite entanglement entropy for a system of size $L$ scales as $(\log L)^2$, in contrast to findings in previously-studied models. Varying the relative strength of Hamiltonian evolution and monitoring, as well as a dimerization parameter, the model's phase diagram { contains} transitions out of the nontrivial phase, which we map to vortex-unbinding transitions in the sigma model, { and also contains} separate  critical points on the measurement-only axis. 
We highlight the close analogies as well as the differences with the replica approach to Anderson transitions in disordered systems. 
\end{abstract}

\author{Michele Fava}
\affiliation{Philippe Meyer Institute, Physics Department, \'{E}cole Normale Sup\'{e}rieure (ENS), Universit\'{e} PSL, 24 rue Lhomond, F-75231 Paris, France}

\author{Lorenzo Piroli}
\affiliation{Philippe Meyer Institute, Physics Department, \'{E}cole Normale Sup\'{e}rieure (ENS), Universit\'{e} PSL, 24 rue Lhomond, F-75231 Paris, France}

\author{Tobias Swann}
\affiliation{Rudolf Peierls Centre for Theoretical Physics, Clarendon Laboratory, Parks Road,
Oxford OX1 3PU, UK}

\author{Denis Bernard}
\affiliation{Laboratoire de Physique de l’\'Ecole Normale Sup\'erieure, CNRS, ENS \& Universit\'e PSL, Sorbonne Universit\'e, Universit\'e Paris Cit\'e, 75005 Paris, France}

\author{Adam Nahum}
\affiliation{Laboratoire de Physique de l’\'Ecole Normale Sup\'erieure, CNRS, ENS \& Universit\'e PSL, Sorbonne Universit\'e, Universit\'e Paris Cit\'e, 75005 Paris, France}
\maketitle

\section{Introduction}

This paper develops field theory descriptions
 for systems of free fermions that are continuously monitored.
We may imagine a chain of fermions evolving under a quadratic hopping Hamiltonian
(perhaps time-dependent), 
and an experimentalist who makes repeated 
measurements of local fermion bilinears \cite{carollo2017fluctuating,Bernard_2018,cao2019entanglement} at all positions throughout the  chain.
The  evolving state of the fermions then depends on the specific random outcomes of these measurements, 
but we can ask about the statistical ensemble of  evolving states 
(quantum trajectories).
To answer the question ``what are the universal properties of this ensemble?'' 
we need appropriate long-wavelength descriptions.

The effect on local expectation values can often be understood in terms of an effective Markovian dynamics \cite{carollo2017fluctuating,Bernard_2018}.
But the entanglement structure of an evolving quantum state can also change qualitatively when the rate of measurement is increased.
Monitored interacting systems show a volume-law entangled phase at low measurement rate and an area-law entangled phase at high measurement rate~\cite{skinner2019measurement,li2018quantum}.
For free fermions the volume law phase is  generically destroyed even by very weak measurement \cite{cao2019entanglement}, 
but there can be critical points, and even critical phases, 
with (logarithmically) super-area-law entanglement \cite{nahum2020entanglement,
alberton2021entanglement,jian2022criticality,alberton2021entanglement,turkeshi2021measurementinduced,buchold2021effective,sang2021measurement,kells2021topological,fidkowski2021dynamical,turkeshi2022enhanced,turkeshi2022entanglement,piccitto2022entanglement,muller2022measurement,turkeshi2022entanglementand,lucas2022generalized,coppola2022growth,gal2022volume,merritt2022entanglement,chen2020emergent,kells2021topological,paviglianiti2023multipartite}.
Such phases and  transitions have been found in a quantum circuit for a Majorana chain  with projective measurements \cite{nahum2020entanglement}, 
and in a more generic continuous-time process 
in Ref~\cite{alberton2021entanglement}, and in simulations of many other models \cite{jian2022criticality,turkeshi2021measurementinduced,buchold2021effective,sang2021measurement,kells2021topological,turkeshi2022enhanced,turkeshi2022entanglement,muller2022measurement,turkeshi2022entanglementand,lucas2022generalized,coppola2022growth,gal2022volume,merritt2022entanglement,chen2020emergent,bao2021symmetry}.

It is clear that there is a landscape of phases and critical points to explore here
(and these quadratic models are much more accessible numerically than generic interacting systems \cite{cao2019entanglement,bravyi2004lagrangian}).
There are also intriguing connections with other  
low-dimensional critical phenomena. 
{ 
Here we will construct effective field theories that allow universal results. As we will show, these effective field theories} are  close cousins of familiar models from magnetism and disordered systems,  but with differences reflecting
 the state-dependent randomness of quantum measurements.

So far, however, free fermion measurement transitions (FFMTs) in generic models have been established only numerically.
Majorana circuits with  ``swap'' operations and projective measurements \cite{nahum2020entanglement,sang2021entanglement} are solvable by a classical mapping, 
but that mapping does not extend to general circuits or Hamiltonians.
(Many qualitative features of this model are nevertheless shared with generic models \cite{merritt2022entanglement}, as we will discuss later.) 
Models in which the randomness of quantum measurements is eliminated by complete ``postselection'' of outcomes can also be solvable \cite{turkeshi2021measurementinduced, turkeshi2022entanglementand,gal2022volume,granet2022volume}, but are in a different regime.

 A more general approach is to derive an effective  model by using the replica trick to average over the randomness, in analogy to monitored interacting circuits \cite{jian2020measurement,bao2020theory}
(cf. also unitary circuits \cite{zhou2019emergent} and tensor networks \cite{vasseur2019entanglement} and related mappings 
\cite{nahum2018operator,hayden2016holographic,nahum2021measurement}).
This approach works with an effective dynamics for  $N$ identical \textit{replicas} of the system. 
Since the dynamics is anyway random due to the random measurement outcomes, a natural simplification is to take the Hamiltonian also random in space and time, so that the averaging is over both types of randomness.

The replica trick introduces a symmetry: 
in the interacting case this involves discrete permutations of replicas, 
but for free fermions it is possible to make continuous rotations between replicas 
\cite{nahum2021measurement,bao2021symmetry,jian2022criticality,buchold2021effective}.
As we will discuss, this is in close analogy to other disordered free fermion systems \cite{evers2008anderson,jian2022criticality},  but a distinctive feature of the measurement problems is that the required replica limit is ${N\to 1}$, rather than ${N\to 0}$ as in more familiar disordered systems. 
The ``additional'' replica of the system is used to express the Born-rule probability of a quantum trajectory, which must be included in averages.

Ref.~\cite{bao2021symmetry} examined a simple quantum circuit for a 1D Majorana chain involving weak measurements of fermion parity $i\gamma_i\gamma_{i+1}$ for adjacent sites, and by averaging over randomness obtained an effective model that was analyzed for the case ${N=2}$. 
(See also Ref.~\cite{buchold2021effective}, where a different kind of effective model was derived.) However, obtaining an effective theory for the physically relevant case ${N\to 1}$ requires the replica limit to be addressed, and this has proved challenging.  

Building on the models of Refs.~\cite{bao2021symmetry,nahum2020entanglement}
(cf. also \cite{sahu2022entanglement,zhang2022universal,merritt2022entanglement}),
 our starting point will be a generalized Majorana chain with continuous-time evolution and with an arbitrary number $N_F$ of ``flavors'' at each site (not to be confused with the number of  replicas).
In this context it will be possible to obtain a continuum description in a controlled way, using $1/N_F$ as the control parameter. 
On symmetry grounds, we expect the resulting field theories to extend to all $N_F$ (including the case ${N_F=1}$, which is dual to a monitored Ising chain).
The actions we obtain have the schematic form:
\be\label{eq:introaction}
\mathcal{S}  = \f{1}{2g} \int \dd x \dd t \, \Tr \, (\partial_\mu Q)^T (\partial_\mu Q) + (\ldots.),
\ee
where the matrix $Q$ lives on an appropriate manifold.
In the most generic case that we discuss,
 $Q_{ab}(x,t)$ is an ${N\times N}$ orthogonal matrix, where ${N\to 1}$ is the number of replicas. 
(The ``$\ldots$'' in Eq.~\ref{eq:introaction} stands for a topological term that appears for the distinct theory that applies on a boundary of the model's phase diagram.)

\begin{figure}
    \centering
    \includegraphics[width=0.85\linewidth]{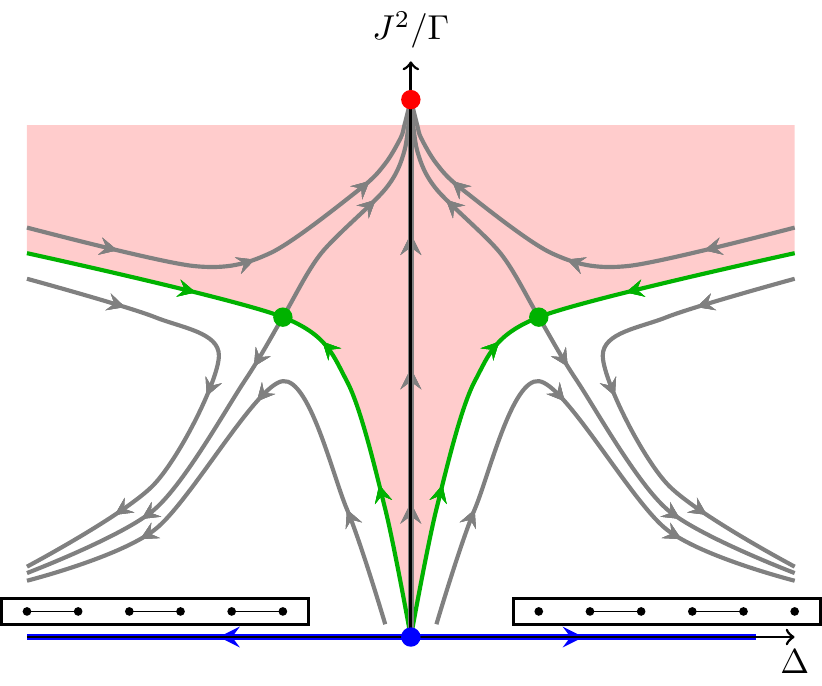}
    \caption{Structure of the phase diagram for the model with $N_F=1$, 
    superimposed  with a schematic illustration of the RG flows. Here, $J^2$ is the strength of the stochastic nearest-neighbor hopping term. Weak measurements are performed on odd/even bonds with strength $\Gamma(1\pm\Delta)$ respectively.
    The pink region is a stable phase with nontrivial entanglement scaling (Eq.~\ref{eq:introSn}).
    It is separated by continuous phase transitions (green lines) from  two area-law phases (white regions),
     where the steady states are dressed versions of the 
fully dimerized states depicted at the bottom left and right of the diagram.
The nontrivial phase is governed by marginal flow to the ${g=0}$ fixed point of the ${{\rm SO}(N)}$ NL$\sigma$M (Eq.~\ref{eq:introaction}) and the green lines represent disordering transitions of this model.
The  ${J=0}$ axis (blue) 
has a higher  symmetry and has an unstable critical point at $\Delta=0$ (blue dot).}
    \label{fig:phase-diag-with-flows}
\end{figure}

These NL$\sigma$Ms yield universal results for  classes of monitored free fermion systems, as we will discuss. 
The monitored Majorana chain
(and other models with the same symmetry)
 has a stable phase in which the $n$th R\'enyi  entanglement entropy across a cut, in a system of size $L$, has the universal form
\be\label{eq:introSn}
S_n \sim \f{n+1}{96 n} (\ln L)^2
\ee
{ Note that, unusually,  this model allows an exact result for the von Neumann entropy (${n=1}$).} 
We will also check this scaling numerically.
The model shows phase transitions out of the stable phase
whose exponents are not known exactly, but which can be understood  qualitatively using the RG.
The model also has disentangled phases, which are disordered phases for the sigma model field, driven by proliferation of topological defects (though 
unlike the transitions in the $N=2$ model \cite{bao2021symmetry} these are not   Kosterlitz-Thouless transitions).
In Fig.~\ref{fig:phase-diag-with-flows} we indicate schematically the topology of the phase diagram in the simplest case ${N_F=1}$. 

An interesting feature of these effective field theories is the close analogy to Anderson localization,
where the same  hierarchies of NL$\sigma$Ms 
describe the eigenfunctions of disordered 2D Hamiltonians
\cite{senthil2000quasiparticle,
read2000paired,bocquet2000disordered,evers2008anderson,fendley2001critical,efetov1999supersymmetry}.
In particular the  two NL$\sigma$Ms that we discuss correspond at ${N=0}$ to the Anderson localization classes DIII and D,
describing localization of Majorana fermions
\cite{senthil2000quasiparticle,
read2000paired,bocquet2000disordered,chalker2001thermal,read2000absence,gruzberg2001random,bocquet2003network,gruzberg2005localization,mildenberger2007density,cho1997criticality,merz2002two}.

In a notable step, Ref \cite{jian2022criticality} made a  symmetry classification of Gaussian random tensor networks 
(these can also be interpreted as evolution operators, 
though they do not correspond to monitored dynamics with Born's rule). The authors showed that the most generic Gaussian networks have symmetry matching that of 
network models \cite{chalker1988percolation} in symmetry class   DIII \cite{PhysRevB.55.1142}, 
and proposed that the DIII NL$\sigma$M at $N\to 0$ 
should apply to such networks on grounds of symmetry. 
This is consistent with the symmetry we find in our explicit derivation of a field theory for a generic continuous time model.
The universal behavior we find is however different, most importantly because of the change in $N$ arising from Born's rule 
(this agrees with the claim in \cite{nahum2021measurement} and answers a question raised in \cite{jian2022criticality}).
Analyzing the field theory also shows that the  scaling of entanglement in  ``metallic'' phases differs from the scale-invariant form assumed previously in the tensor network context \cite{jian2022criticality}.

The different value of $N$ in the NL$\sigma$M leads to different amplitudes and exponents and in some cases to a different topology for phase diagram. 
However, key ideas from localization carry over to the ${N\to 1}$ theories. 
(The localization field theory involves two spatial coordinates, rather than a space and a time coordinate, so the analogy is with ``Anderson localization in spacetime''.)
Very loosely, entanglement in the measurement problem
plays a similar role to  conductance in the localization problem, 
because  in the measurement problem the inverse of the running 
sigma model coupling, $g(L)^{-1}$ (cf. Eq.~\ref{eq:introaction})
is a measure of the strength of 
entanglement at a certain lengthscale $L$,
whereas in the localization context $g(L)^{-1}$ is proportional to a conductivity.

A well-known  phenomenon in localization is that 
the coupling can flow the ``wrong way'' in the replica limit, 
with ${g(L)^{-1}\propto \ln L}$; so that the flow approaches the ``ordered'' fixed point at $g=0$,\footnote{In the localization context this flow gives a ``supermetal'' since at large scales the conductance is larger than the Ohm's law prediction \cite{hikami1980spin,evers2008anderson}.} contrary to conventional 2D models with continuous symmetry.
In the present context, this flow leads to the bipartite entanglement scaling as in Eq.~\ref{eq:introSn}.
As a result of the flow of $1/g$ 
(the increase in the ``strength of entanglement'' with scale)
 this entanglement is larger than the ${\ln L}$ that holds in scale-invariant theories~\cite{Calabrese_2009}. 
 The more standard scale-invariant behavior holds at  critical points (as opposed to the stable phase) that we find in the phase diagram of the measurement model.

It is interesting that the structure of the phase diagram in Fig.~\ref{fig:phase-diag-with-flows} for the fairly generic chain studied here
is similar to that in the much simpler quantum circuit that maps to a classical loop model \cite{nahum2020entanglement}. 
 This robustness of that phase diagram was previously found numerically  in Ref.~\cite{merritt2022entanglement}, which explored a different (but also relatively generic) model.
 We explain this similarity by close structural similarities between the sigma models describing the loop model and those describing more generic Majorana chains. However, the two kinds of problems are in different universality classes.

The recipe outlined in this paper, for deriving continuum nonlinear sigma model descriptions, can be applied to models with other symmetries or with other dimensionalities, and 
opens the way to a more general classification of monitored free fermion systems. (It could also be used to address other phenomena such as boundary decoherence \cite{PhysRevLett.123.110601,PhysRevX.9.021007}.)
A panoply of critical behaviors are shown by NL$\sigma$Ms for Anderson localization \cite{fendley2001critical}:
it will be very interesting to explore the corresponding landscape 
at the ``${N\to 1}$'' level that describes monitored dynamics.

This paper is organized as follows.  Sec.~\ref{sec:model} introduces our model and its hybrid dynamics, and derives the replica Hamiltonian. In Secs.~\ref{sec:mapping_stat_mech} and~\ref{sec:mapping-to-NLSM} we carry out the mapping to an effective spin chain and derive the effective ${\rm SO}(N)$ NL$\sigma$M  description in the limit of large flavor index. In Sec.~\ref{sec:firstlookatphasediagram} we discuss the qualitative features of the phase diagram, while in Sec.\ref{sec:entanglement}
we use this theory to compute the basic universal properties of the stable critical phase, which we verify in Sec.~\ref{sec:numerics}.
In Sec.~\ref{sec:dimerization-and-vortices} we discuss the transition out of this phase via proliferation of $\mathbb{Z}_2$ instantons.
In Sec.~\ref{sec:measurement-only} we discuss the different NL$\sigma$M, on the manifold ${\rm SO}(2N)/{\rm U}(N)$, that we propose for the measurement-only version of the model, and the RG flows between models with different symmetries. Finally, we conclude in
Sec.~\ref{sec:conclusions} with an outlook. The most technical parts of our work are consigned to several appendices.

\tableofcontents

\section{Model and  replica approach}
\label{sec:model}
\subsection{Definition of the model}
\label{sec:model_def}

We consider a chain of Majorana fermions subject to continuous monitoring --- in other words to repeated weak measurement in the limit where the measurements are very frequent and very weak.
We expect  the universal results also to apply to a larger family of monitored quantum circuit models~\cite{bao2021symmetry, merritt2022entanglement}, but the continuous time formulation simplifies the subsequent mappings.

Let us start with the Hamiltonian that generates  the unitary part of the dynamics:
 \be\label{eq:Hamiltonian}
H(t) =  \sum_{j=1}^L \sum_{\mu, \nu=1}^{N_F} J_j^{\mu\nu}(t) \, i \gamma_{j,\mu} \gamma_{j+1,\nu}.
\ee
Here $\gamma_{j,\mu}$ are standard Majorana operators satisfying 
\begin{equation}
\{\gamma_{j,\mu},\gamma_{k,\nu}\}=2\delta_{j,k}\delta_{\mu,\nu}.
\end{equation}
The fermions carry a site index $j=1,\ldots, L$, and we have also given them a flavor index ${\mu = 1,\ldots, N_F}$.
Flavors are a tool to aid the derivation of the  NL$\sigma$M, which will be controlled at large $N_F$. However we also expect this  continuum theory to apply for $N_F=1$, when there is just a single Majorana at each site. This case is also dual to a monitored Ising chain.

We consider the limit in which the couplings $J_j^{\mu\nu}(t)$ are white noise, i.e.  are random variables with vanishing mean and with variance
\begin{equation}\label{eq:variance}
\mathbb{E}_G [J_j^{\mu_1\nu_1}(t)J_k^{\mu_2\nu_2}(t^\prime)]=\frac{J^2}{N_F}
\delta_{j,k}
\delta^{\mu_1,\mu_2}\delta^{\nu_1,\nu_2}\delta(t-t^\prime)\,.
\end{equation}
$\mathbb{E}_G [\cdot]$ denotes the Gaussian average.
Note that $J^2$ has the units of a rate.

We now include monitoring of 
all the fermion-parity operators ${i\gamma_{j,\mu} \gamma_{j+1,\nu}}$ for adjacent sites. 
Physically, continuous-time monitoring can be thought of as the limit ${\Delta t\to 0}$ of a discrete-time process, with the strength of the measurements simultaneously approaching zero.
A nice feature is that the resulting dynamics in the limit ${\Delta t\to 0}$ is independent of essentially all the details of the discrete-time measurement protocol, and is characterized solely by a measurement \textit{rate} for each measured operator ${i\gamma_{j,\mu}\gamma_{j+1,\nu}}$.
We will take this rate to be independent of the flavor indices,
but we allow for a staggered dependence on the spatial position~\cite{nahum2020entanglement}:
\be
\Gamma_j= \left[ 1+ (-1)^j \Delta \right] \Gamma.
\ee
We include the dimerization $\Delta$ in order to be able to drive a phase transition into a disentangled phase. 
In the regime where ${\Gamma\gg J^2}$, 
and where $\Delta$ is close to either $+1$ or $-1$,
the dynamics consists almost entirely of measurements of a single sublattice of bonds. When ${N_F=1}$, this manifestly leads at long times to area-law states 
of the forms shown in Fig~\ref{fig:phase-diag-with-flows}. 
In Fig~\ref{fig:phase-diag-with-flows} we anticipate the schematic phase diagram for $N_F=1$ that will result from our analysis (we will discuss the $N_F$-dependence in later sections). When $\Delta=0$ the model has translation symmetry by one site: we will find that when $N_F$ is odd this symmetry guarantees that the model is nontrivially entangled, consistent with Fig~\ref{fig:phase-diag-sketch}.

The evolution of the density matrix that results from the combined dynamics can be written down in many ways.
In this section it will be convenient for us to define it directly in continuous  time, using a loose notation%
\footnote{{A standard fact about white-noise (such as  $J^{\mu\nu}_j(t)$) is that it is not a well-defined function of $t$, due to divergences in the limit of vanishing correlation time. 
Therefore to give a precise meaning to the notion of a ``realization'' of $J^{\mu\nu}_j(t)$ we should discretize time.  But ultimately all we will need are averages such as Eq.~\ref{eq:variance}, which  are well-defined in the ${\Delta t\to 0}$ limit.  The quantity $M^{\mu\nu}_j(t)$  below (the measurement record) has a similar status: ultimately we will only need its  expectation values. Appendix~\ref{app:sec:discrete-time} gives Eqs.~\ref{eq:defnrhounnormalized},~\ref{eq:timeorderedexp} a precise meaning by discretizing time.}} analogous to Eq.~\ref{eq:Hamiltonian},
in terms of a non-Hermitian extension of the Hamiltonian.
This formulation is equivalent to the (perhaps more familiar) stochastic Schr\"odinger equation formalism~\cite{caves1987quantum,diosi1998non,gisin1992quantum} (we will use the latter for simulations in Sec.~\ref{sec:numerics}).
In Appendix~\ref{app:sec:discrete-time} we give a more careful definition of the evolution as the $\Delta t\rightarrow 0$ limit of a discrete time process, and explain how the non-Hermitian evolution  below arises.

The required  non-Hermitian extension of the Majorana Hamiltonian is:
\be
\label{eq:HamiltonianNH}
H_{J,M}(t) =  \sum_{j=1}^L \sum_{\mu, \nu=1}^{N_F} \left[
J_j^{\mu\nu}(t) 
+  i M_j^{\mu\nu}(t)   \right]
\, i \gamma_{j,\mu} \gamma_{j+1,\nu}.
\ee
Schematically,
the mapping to non-Hermitian Hamiltonian evolution arises because {
a discrete weak measurement of ${i\gamma_{j,\mu} \gamma_{j+1,\nu}}$ involves conjugating the density matrix with a Kraus operator that is proportional to 
${\exp(-\dd M_j^{\mu\nu}  \gamma_{j,\mu} \gamma_{j+1,\nu})}$. Here  }
$\dd M_j^{\mu\nu}$
is proportional to the measurement outcome multiplied by an infinitesimal measurement strength: see Appendix~\ref{app:sec:discrete-time}.
In Eq.~\ref{eq:HamiltonianNH}, $M_j^{\mu\nu}(t)$ 
is the ``continuum limit'' 
of the list of  measurement outcomes 
(indexed by $t$)
of the associated observable 
${i \gamma_{j,\mu} \gamma_{j+1,\nu}}$.
We will define  
$M_j^{\mu\nu}(t)$ in the continuum limit through its statistics.
While $J_j^{\mu\nu}(t)$ is simple white noise,  the statistics of 
$M_j^{\mu\nu}(t)$ are nontrivial as a result of Born's rule.

First, however, consider the evolution of the state
$\rho_{J,M}(t)$ conditioned on a given realization of the couplings $J$ and given measurement outcomes $M$ (i.e. on a given quantum trajectory), and starting in some initial state $\rho(0)$.
It is convenient first to define the evolution for an \textit{un-normalized}  version of the density matrix, which we denote by~${\check \rho(t)}$:
\be\label{eq:defnrhounnormalized}
\check \rho_{J,M}(t)  = K_{J,M}(t) \rho(0) K_{J,M}(t)^\dag.
\ee
This has the same structure as for unitary evolution, but the  time-evolution operator $K_{J,M}(t)$ is nonunitary:
\be\label{eq:timeorderedexp}
 K_{J,M}(t) \equiv 
 \mathcal{T}
 \exp \lf   - i
 \int^t_0 \dd t' H_{J,M}(t') 
\ri.
\ee
($\mathcal{T}$ is time-ordering.)
Note that $\check \rho_{J,M}(t)$
depends on 
$J(t')$ and $M(t')$ over the full time interval $[0,t]$, i.e. on the complete trajectory.
The physical density matrix is obtained by normalizing $\check \rho(t)$,
i.e. as ${\rho_{J,M} (t) = {\check \rho_{J,M} (t)}/{\tr \check \rho_{J,M}(t)}}$.

To complete the definition of the continuous time dynamics we must specify how to average over ${M_j^{\mu\nu}(t)}$.
To be concrete, let us consider the physical  average  ${\mathbb{E} [\cdot]}$ of some quantity that depends on the state
$\rho_{J,M}(t)$,
and therefore on $J(t')$ and $M(t')$ for ${t'\in [0,t]}$.
We may define these expectation values in two steps.

First, we define Gaussian averages, denoted $\mathbb{E}_\mathrm{G}[\cdot]$, in which both $J$ and $M$ are treated as white noise:
the variance of $J$ is given in Eq.~\ref{eq:variance}, and the variance 
of $M$ is set by the measurement rate:
\be\label{eq:varianceM}
\mathbb{E}_\mathrm{G} [M_j^{\mu_1\nu_1}(t^\prime) 
M_k^{\mu_2\nu_2}(t^{\prime\prime})]=
\frac{\Gamma_j}{N_F}
\delta_{j,k}
\delta^{\mu_1,\mu_2}\delta^{\nu_1,\nu_2}\delta(t^\prime-t^{\prime\prime})\,.
\ee
The correct measure for $M$ is related to this white-noise measure by a factor that comes from Born's rule. 
In outline, the probability for a given measurement record 
${\{M_k^{\mu\nu}(t')\}}$
is proportional 
to ${\tr \check \rho_{J,M}(t)}$. 
(See Eq.~\ref{eq:defnrhounnormalized} for the definition of $\check \rho$.)
This modifies expectation values by the same factor of ${\tr \check \rho_{J,M}(t)}$:
\be\label{eq:expvalwithBorn1}
\mathbb{E} [\cdot]
=
\frac{\mathbb{E}_\mathrm{G} 
\left[
(\cdot) \times \Tr \check \rho_{J,M}(t)
\right]}{
\mathbb{E}_\mathrm{G} 
\left[
 \Tr \check \rho_{J,M}(t)
\right]
}
\ee
The denominator here is a trivial constant ensuring that the measure on trajectories is normalized. For notational simplicity we can set this denominator to 1 simply by absorbing an additive constant into $H$, so that: 
\be\label{eq:expvalwithBorn}
\mathbb{E} [\cdot]
={\mathbb{E}_\mathrm{G} 
\left[
(\cdot) \times \Tr \check \rho_{J,M}(t)
\right]}.
\ee
For details of the mappings above, see Appendix~\ref{app:sec:discrete-time}, where we make the formulas more precise by starting with a discrete-time measurement process defined in terms of Kraus operators.

Eq.~\ref{eq:expvalwithBorn} is convenient because it expresses averages of physical quantities in terms of simple Gaussian averages. In particular, the basic object in the replica approach is the following Gaussian average of \textit{$N$ copies} of the un-normalized density matrix:
\be\label{eq:Nreplicastate}
{\rho}^{(N)}(t) \equiv \mathbb{E}_\mathrm{G} \left[ 
\check \rho_{J,M}(t)^{\otimes N} 
\right].
\ee
Formally, this can be viewed as a density matrix for $N$ copies of the Majorana chain.
We will use two properties of this object.
First, it has a simple time evolution,
specified by an effective Hamiltonian that we will derive shortly. 
Second, by taking the replica limit ${N\rightarrow 1}$, 
all the physical averages of interest can be expressed as traces of  ${\rho}^{(N)}(t)$~\cite{jian2020measurement}.

\subsection{Review of replica formalism}
\label{sec:replicareviewmaintext}

In this section, we recall how the replica approach works for an illustrative class of observables. We simplify the discussion with respect to standard expositions by focusing on generating functions for the entropy, instead of averages involving logarithms.

The full probability distribution of the $n$th R\'enyi entropy is encoded in the generating function
\be
\mathbb{E} \left[ 
e^{ -k(n-1)S_n(t)}
\right]
=
 \mathbb{E}   \left[
  \left( \Tr \rho(t)^n\right)^k
 \right].
\ee
In particular, the behavior near ${k=0}$    gives the mean of $S_n$.
Here it will be sufficient to treat $k$ and $n$ as positive integers, since we can analytically continue to real values at the very end. We have suppressed the subscripts on ${\rho(t) = \rho_{J,M}(t)}$ to avoid clutter.

Using Eq.~\ref{eq:expvalwithBorn}, the expectation value above becomes
\be
\mathbb{E}_\mathrm{G} \left[ 
\lf \Tr \check \rho(t)^n \ri^k
\lf \Tr \check \rho(t) \ri^{1-nk}
\right],
\ee
where the factors of ${\Tr \check \rho(t)}$ arise both from the normalization of $\rho(t)$ and from the nontrivial factor in Eq.~\ref{eq:expvalwithBorn}.
We now write this as 
\be\label{eq:limtracetrace}
\lim_{N\rightarrow 1}
\mathbb{E}_\mathrm{G} \left[ 
\lf \Tr \check \rho(t)^n \ri^k
\lf \Tr \check \rho(t) \ri^{N-nk}
\right].
\ee
We then study the limit by analytically continuing from integer values of $N$ with $N\geq nk$.\footnote{Since the dynamics under consideration is Gaussian, 
an alternative to the replica formulation is to use supersymmetry, as is familiar  in the context of Anderson localization \cite{evers2008anderson}. For our purposes here the replica approach is simpler.}
The utility of this is that the density matrices then appear only to positive powers, so that the above average can be written in terms of $\check \rho(t)^{\otimes N}$:
\be\label{eq:genfnreplica}
\mathbb{E} \left[ 
e^{ -k(n-1)S_n(t)}
\right]
=
\lim_{N\rightarrow 1}
 \mathbb{E}_\mathrm{G}  \Tr \left[
  \check \rho(t)^{\otimes N}
  \lf 
    \mathcal{C}_{n}^{\otimes k} \otimes \mathbb{I} 
  \ri
 \right].
\ee
The trace is now taken in the Hilbert space of $N$ copies of the system. 
$\mathcal{C}_n$ is an operator that cyclically permutes $n$ out of the $N$ copies of the system: see Appendix~\ref{app:sec:boundary-stabilizers} for an explanation of this notation. 
While the notation on the right hand side of~\eqref{eq:genfnreplica} may appear formal, it is just a way of expressing the pattern of index contractions needed to give the traces in Eq.~\ref{eq:limtracetrace}.

Finally, we can push the expectation value in~\eqref{eq:genfnreplica} inside the trace, so that the generating function for the entropies is written in terms of a trace of
the object
$\rho^{(N)}(t)$ defined in Eq.~\ref{eq:Nreplicastate}.

\subsection{Effective replica Hamiltonian}
\label{sec:effectivereplicahamiltonian}

In order to write the evolution of $\rho^{(N)}(t)$ in a concise way, 
it will be convenient to implement a standard operator-to-state mapping.
Instead of thinking of $\rho^{(N)}(t)$ as a density matrix for $N$ copies of the Majorana chain, we will think of it as a wavefunction (a ket) for $2N$ copies of the Majorana chain:
\be
\rho^{(N)} \longrightarrow \ket{\rho^{(N)}}.
\ee
This is a standard transformation for bosonic Hilbert spaces, and can be adapted to fermionic ones by mapping them to a bosonic one first. Here we follow the convention of Ref.~\cite{sunderhauf2019quantum}, which we summarize in Appendix~\ref{app:sec:replica-majorana}. 

Under this operator-state mapping,  traces of the replicated density matrix (which we will need to compute entanglement entropies)  are mapped to state overlaps: schematically,
\be
\Tr \big[ \rho^{(N)} \mathcal{C} \big] \longrightarrow 
\big\langle \mathcal{C} \big| \rho^{(N)} \big\rangle,
\ee
where $\mathcal{C}$ is any operator on the  $N$-copy Hilbert space and $\ket{\mathcal{C}}$ a corresponding state in the  $2N$-copy space.
From Eq.~\ref{eq:genfnreplica}, we then see that the generating function for $S_n(t)$ becomes such a transition amplitude, ${\langle \mathcal{C}_{k,n} | \rho^{(N)} \rangle}$, for a state
 $\ket{\mathcal{C}_{k,n}}$ that we 
define   precisely~later.

After this mapping, the evolution of the state $\ket{\rho^{(N)}}$ 
(for any fixed natural number $N$)
has the form of conventional imaginary time evolution with a Hamiltonian~$\Hs$:
\be\label{eq:imaginarytimeevo}
\ket{\rho^{(N)}(t)}
=
\exp \lf 
- N_F t \Hs 
\ri \ket{\rho^{(N)}(0)}.
\ee
More precisely, there is a separate Hamiltonian $\Hs$ for each value of $N$,
acting on the Hilbert space of $2N$ copies of the Majorana chain.
Since the random quantities $J$ and $M$ have been averaged out, $\Hs$ is non-random, and independent of time (so depends only on $N$).

The evolution~\eqref{eq:imaginarytimeevo} of the replicated state  follows from Eq.~\ref{eq:defnrhounnormalized}, and is of the form\footnote{Our convention for complex conjugation 
is specified in Appendix~\ref{app:sec:replica-majorana}, where we explain these mappings in more detail.}
\be\label{eq:replicaevolutionbeforeaverage}
\exp \lf 
- N_F t \Hs 
\ri
=
\mathbb{E}_G \left[
K_{J,M}^{\otimes N} \otimes  
 (K^*_{J,M})^{\otimes N}
\right].
\ee
The term inside the average involves   
the exponential of $2N$ copies of the original Hamiltonian~\eqref{eq:HamiltonianNH}:
\be
\label{eq:H-JM-many-copies}
H_{J,M}^{(N)} = 
\sum_{\sigma=\pm}\sum_{a=1}^N \sum_{j,\mu,\nu} \left[
J_j^{\mu\nu}(t) 
+  i \sigma  M_j^{\mu\nu}(t)   \right]
\, i \gamma_{j,\mu}^{(\sigma a)} \gamma_{j+1,\nu}^{(\sigma a)}.
\ee
Here, for each physical Majorana operator $\gamma_{j,\mu}$,
 we now have $2N$ ``replicated'' Majoranas.
 We have labelled these replicas by an index $\sigma=\pm$ which  distinguishes the first $N$ copies in Eq.~\ref{eq:replicaevolutionbeforeaverage} from the last $N$ copies,
together with an index ${a=1,\ldots,N}$.

The Gaussian average~\eqref{eq:replicaevolutionbeforeaverage} yields:
\be
\label{eq:Heff-first-expr}
\Hs  = \f{1}{2}
\sum_{j,\mu,\nu} 
\left[
J^2  (H_\text{U})^{\mu \nu}_{j} 
-
 \Gamma_j   \, (H_{\text{non-U}})^{\mu \nu}_{j}
\right].
\ee
We have separated terms coming from the unitary and non-unitary parts of the dynamics:
\ba\notag
(H_\text{U})^{\mu \nu}_{j} &= \frac{1}{N_F^2}  \Big(  \sum_{\sigma,a}  
i \gamma^{(\sigma a)}_{j \mu}  \gamma^{(\sigma a)}_{j+1 \nu}
\Big)^2
\\ 
(H_{\text{non-U}})^{\mu \nu}_{j} & =  \frac{1}{N_F^2} \Big(  \sum_{\sigma,a}  \sigma
i \gamma^{(\sigma a)}_{j \mu}  \gamma^{(\sigma a)}_{j+1 \nu}
\Big)^2
\end{align}
This interacting fermion Hamiltonian is much more conveniently written as a spin model. We describe this next.

\section{Mapping to a spin chain}
\label{sec:mapping_stat_mech}

This Hamiltonian takes a much simpler form when reinterpreted in terms of generators of $\mathrm{SO}(2N)$ rotations.
Above we have used normalization conventions so that $N_F$ appears explicitly in~\eqref{eq:imaginarytimeevo}: this is convenient because $1/N_F$ will play the role of $\hbar$ in the semi-classical treatment. 

There is a natural action of $\mathrm{so}(2N)$ at each site of the chain~\cite{bao2021symmetry}.
For a moment, let us write the replica multi-index $(\sigma a)$ as a single index ${\alpha=1,\ldots, 2N}$. 
Then the $\mathrm{so}(2N)$ rotations are generated by the quantum operators 
\begin{align}
	\label{eq:spin-S-def}
	S_j^{\alpha\beta} &= \frac{i}{2 N_F}  \sum_\mu \left[ \gamma_{j,\mu}^{\alpha }, \, \gamma_{j,\mu}^{\beta} \right].
\end{align}
The generator $S_j^{\alpha\beta}$
corresponds to an infinitesimal rotation between Majoranas $\gamma^{\alpha}$ and $\gamma^{\beta}$ from distinct replicas.
They have commutation relations
\be\label{eq:so2Ncommutation}
[S^{\alpha\beta}, S^{\alpha' \beta'}] = \frac{i}{N_F} 
        \left[
              \delta_{\beta, \alpha'}  S^{\alpha\beta'}
            - \delta_{\alpha,\alpha'}  S^{\beta\beta'}
            - (\alpha'\leftrightarrow \beta')
        \right].
\ee
Apart from the factor of $1/N_F$ arising from our normalization convention, these are the standard commutation relations of rotation generators in $2N$ dimensions.

The set of generators $S_j^{\alpha\beta}$ at a site $j$ make up an antisymmetric $2N\times 2N$ matrix of quantum operators, which we denote by $S_j$
(since it will play a role analogous to the spin operator $\vec S_j$ in a Heisenberg antiferromagnet). Returning to the notation $(\sigma a)$  that distinguishes forward and backward replicas, we order the replicas as 
 $(+1), \dots, (+N), (-1),\dots, (-N)$.
Then we denote the blocks of $S_j$ as:
\begin{equation}\label{eq:blocks}
	S_j = 
	\begin{pmatrix}
		\, L_j & & Q_j\\
		-Q_j^T & & R_j
	\end{pmatrix},
\end{equation}
so that $L=-L^T,\, R=-R^T,\, Q$ are $N\times N$ matrices whose entries are operators.

In terms of these operators, the Hamiltonian is
(up to an additive constant that we neglect)
\begin{align}
	\label{eq:H_N-spin}
	\Hs = - \sum_j \tr \left[
 J_\parallel (L^T_j L_{j+1} + R^T_j R_{j+1} )
 +
 2 J_\perp Q_j^T Q_{j+1}
 \right].
\end{align}
We use ``$\tr$'' to denotes traces for  ${2N\times 2N}$ (and later ${N\times N}$) matrices, reserving ``$\Tr$'' for the many-body Hilbert space. The couplings in $\Hs$ are
\begin{equation}
	J_{\parallel} = \frac{J^2 -\Gamma}{2}, \qquad J_\perp = \frac{J^2 +\Gamma}{2}.
\end{equation}
For now we have specialized to the undimerized case, $\Gamma_j=\Gamma$; we will return to the effect of dimerization later.

For generic $J$, $\Gamma$ the symmetry of this  Hamiltonian is only a subgroup of $\mathrm{SO}(2N)$~\cite{bao2020theory},  because the $Q$ generators and the $L$, $R$ generators appear with different coefficients in~\eqref{eq:H_N-spin}. 
This is analogous to the fact that the Hamiltonian of the XXZ spin chain, 
which is written in terms of the $\mathrm{so}(3)$ generators $S^x$, $S^y$ and $S^z$, is invariant only under a subgroup of $\mathrm{SO}(3)$.
The symmetry of $\Hs$ is enlarged when either $J^2=0$, or $\Gamma=0$, as we will discuss in Sec.~\ref{sec:measurement-only}.

To complete the definition of the spin chain, we must specify which representation of $\mathrm{SO}(2N)$ the generators in $\Hs$ act on: we do this in the next section.
In the XXZ chain analogy, this would correspond to specifying the magnitude of the spin at each site.

In Sec.~\ref{sec:mapping-to-NLSM} we will show that the low-energy dynamics of the spin chain  are described by the $\mathrm{SO}(N)$ non-linear sigma model (NL$\sigma$M).
The limit of large $N_F$ will allow a quantitatively controlled derivation 
(but, on grounds of symmetry, this  continuum theory may be extended to small values of $N_F$).
We see from Eq.~\ref{eq:so2Ncommutation} that
the commutators of the $S$ operators  vanish when ${\text{``$\hbar$''} := 1/N_F\rightarrow 0}$, so that at large $N_F$ there is a regime where semiclassics is accurate.
This is analogous to large spin in an $\mathrm{su}(2)$ spin chain.

\subsection{Symmetries and local conserved quantities}

The effective Hamiltonian $\Hs$ above possesses a large set of symmetries. We can split them into global symmetries, 
which will be present in the NL$\sigma$M, 
and an extensive number of local integrals of motion,
which will be completely fixed by the temporal boundary conditions for the evolution. 
For concreteness, we take the physical initial state to be the maximally mixed state, though, as we will discuss, the specific choice is not crucial.

$\Hs$ is invariant under an ${[\mathrm{O}(N)\times \mathrm{O}(N)]\rtimes \mathbb{Z}_2}$ global replica symmetry.
The two orthogonal groups correspond to rotations among Majorana operators within the same sector (i.e. within the $\sigma=+$ or $\sigma=-$ sector),
and the corresponding symmetry generators are $\sum_j L_j$ and $\sum_j R_j$ respectively (replica indices omitted).
The $\mathbb{Z}_2$ operation is an exchange of forward and backward replicas: $\gamma_{j,\mu}^{+ a} \leftrightarrow \gamma_{j,\mu}^{- a}$.

In addition,
each site possesses local integrals of motion that label choices of symmetry representation.
The on-site Hilbert space splits into different representations of $\mathrm{so}(2N)$, and the choice of representation is conserved in time.
This conservation is due to the fact that $\Hs$ is written entirely in terms of local $\mathrm{so}(2N)$ generators $S_j^{\alpha\beta}$. In fact, since $S_j^{\alpha\beta}$ is itself a sum of generators acting on different flavors (see Eq.~\ref{eq:spin-S-def}),
the choice of representation is separately conserved for each site $j$ and for each flavor~$\mu$. 

Fortunately, boundary conditions greatly simplify the Hilbert space, by isolating a unique choice of representation at each site. 
The states that are in other representations can be discarded, since they have no overlap with the initial state that we have chosen.
In fact, the same choice of representation is also fixed by the boundary states that we impose at the \textit{final} time in order to compute entanglement entropies
(Sec.~\ref{sec:entanglement}), which is why the specific choice made for the initial state is not crucial.

It suffices to consider a single site $j$.
First, for a given flavor index $\mu$, the matrices $\frac{1}{2}[\gamma^\alpha_{j,\mu},\gamma^\beta_{j,\mu}]$ { form} a representation of $\mathrm{so}(2N)$, isomorphic to the spin representation or its complex { conjugate}, with highest weight $\omega_s$ or $\omega_{\bar s}$ respectively,  depending on the value of the fermion parity number $\mathcal{R}_{j,\mu}=\pm 1$. In our case the initial state fixes ${\mathcal{R}_{j,\mu} =+1}$ for every site and  flavor index, so that we are dealing with spin representations, as we explain in Appendix~\ref{app:sec:boundary-state-irreps}.

Next, the operators \eqref{eq:spin-S-def} act on the tensor product of those representations, so that the $N_F$ flavors combine into various possible irreducible representations of $\mathrm{so}(2N)$. Among those, the relevant irreducible representation is that with maximal highest weight, namely with weight $N_F\,\omega_s$, cf. Appendix~\ref{app:sec:boundary-state-irreps}. This is analogous to combining $N_F$ spin-1/2s  into a  state with maximum possible spin $N_F/2$. This choice of representation is an invariant of the dynamics.

This representation is the one that maximizes the value of the quadratic Casimir on each site
\begin{align}
    \label{eq:casimir}
	C_j &
 = \mathrm{tr} \left[ L_j^T L_j + R_j^T R_j + 2 Q^T_j Q_j\right].
\end{align}
By a direct computation
this is  (see Appendix~\ref{app:sec:boundary-state-irreps})
\be\label{eq:casimirvalue}
C_j = 2N \lf 1 + \frac{2(N-1)}{N_F} \ri,
\ee
with $C_j\simeq 2N$ at large $N_F$.

\section{Mapping to nonlinear $\sigma$-model}
\label{sec:mapping-to-NLSM}

Next we show that the low-energy dynamics of the spin chain  \eqref{eq:H_N-spin} 
is  captured by a NL$\sigma$M 
for a field $Q(x,t)$ that lives on the compact manifold $\mathrm{SO}(N)$. 
More precisely, this applies for the 
generic case with  nonzero $J^2$ and $\Gamma$: the case ${J=0}$ has  a higher symmetry and will be discussed separately in Sec.~\ref{sec:measurement-only}.

An intuitive way to obtain the continuum theory   is 
via the  equations of motion that arise in the semiclassical (large $N_F$ or small $\hbar$) limit \cite{affleck1985quantum}.
These equations of motion allow us to deduce the Lagrangian,
which may then be used to ``requantize'' the theory by writing the path integral. 
This is the route we will follow in this section.
(An alternative route would be to start with a coherent states path integral on the lattice, cf.   Appendix~\ref{app:coherentstates}.)
We expect that our derivation of the continuum field theory, including the values of nonuniversal constants, is quantitatively controlled at large $N_F$.

Let us outline the steps in a little more detail. First, in Sec.~\ref{subsec:path-integral-deg} we identify
the ``ultraviolet'' 
degrees of freedom
as antisymmetric matrices in $\mathrm{SO}(2N)$.
These are the fields 
(either in a path integral or in the semiclassical equations of motion)
corresponding to the spin $S^{\alpha\beta}$ that appears in $\Hs$.

Next, by analyzing the classical ground-state manifold (i.e. at large $N_F$), 
we see that only the field 
 ``$Q$'' in the block decomposition~\eqref{eq:blocks} of the  matrix $S$ is a massless degree of freedom. 
This suggests that we should
eliminate the modes $L$ and $R$ to obtain an effective Lagrangian $\mathcal{L}(Q)$. 
This can be done at the level of the equations of motion.

We obtain the semiclassical equations of motion from $\Hs$ in the usual way in Sec.~\ref{subsec:equations-of-motion}
(the Heisenberg equation of motion for the operator $S$ becomes a classical differential equation at large $N_F$).
The large $N_F$ limit also allows us to take a controlled spatial continuum limit, and then to eliminate $L$, $R$. This gives a simple equation of motion for $Q$, from which we can identify the Lagrangian.

\subsection{Degrees of freedom and classical ground states}
\label{subsec:path-integral-deg}

In the limit ${N_F\rightarrow \infty}$ the matrix of quantum operators  $ S^{\alpha\beta}$ on a site becomes an antisymmetric matrix of classical phase space coordinates.
In this limit we also have the constraint 
\be\label{eq:orthogonalityofS}
S^T S = \openone,
\ee
or in terms of the  ${N\times N}$ block decomposition~\eqref{eq:blocks},
\ba
\label{eq:constraints-on-S}
L Q + Q R & = 0, 
&
Q^T Q - R^2 &  
=
Q Q^T - L^2   = \openone.
\end{align}
Eq.~\ref{eq:orthogonalityofS} follows from applying the  quantum operator $( S^T S)^{\alpha\beta}$
to an arbitrary state in the representation determined in the previous section,  and taking the large $N_F$ limit. We provide further details in Appendix~\ref{app:sec:boundary-state-irreps}.

Eq.~\ref{eq:orthogonalityofS}  shows that the appropriate  semiclassical degree of freedom  $S$ lives on the space of   antisymmetric orthogonal matrices. 
More precisely, $S$ lives on the part of this space which is continuously connected to the point ${Q=\openone}$, ${L=R=0}$, as can be seen by noting that the Pfaffian of $S$ --- fixed by the $\mathrm{so}(2N)$ representation --- is ${\operatorname{Pf} S} = {(-1)^{N(N-1)/2}}$  for $N_F\to\infty$.
Such matrices form a compact symmetric space isomorphic to ${\mathrm{SO}(2N)/\mathrm{U}(N)}$, cf. Appendix~\ref{app:coherentstates}.

We note that it is possible to formulate a coherent-state path integral for the spin chain
as a functional integral over the local degrees of freedom $S$ living in the same symmetric space, as we discuss in  Appendix~\ref{app:coherentstates}.%
\footnote{The procedure is standard \cite{stone2001note}: we define  coherent states $\ket{S}$ that are labelled by the expectation value of  $S$,
and construct the path integral using resolutions of the identity of the schematic form $\int \dd \mu(S) \ket{S}\bra{S}$.}
This is reassuring as it confirms that 
the field theory degrees of freedom that we identify in the semiclassical limit also make sense at finite $N_F$.

Next, consider the classical ground states that are obtained by minimizing our Hamiltonian $\Hs$~\eqref{eq:H_N-spin}
with the relevant on-site representation.
If the rates $J^2$ and $\Gamma$ for the unitary and nonunitary parts of the dynamics are both nonzero, then ${J_\perp > |J_\parallel|}$ in Eq.~\ref{eq:H_N-spin}. 
In this case the energy is minimized by taking\footnote{Recall that the Casimir ${C=\tr (L^TL + R^T R + 2 Q^TQ)}$ is fixed to ${C=2N}$ (Eq.~\ref{eq:casimirvalue}), so taking $L=R=0$ maximizes ${\tr Q^T Q}$.} $L,R=0$,
 and $Q_j$ independent of $j$.
By the constraints described above, the ``order parameter'' $Q$ is then a (proper) rotation matrix:
\be\label{eq:QinSO(N)}
Q\in \mathrm{SO}(N).
\ee
In passing, we note that
there are more general ground states, 
with the same energy in the semiclassical limit,
but which do not satisfy the constraint on the on-site representation imposed by the boundary conditions.\footnote{These states have ${L=R=0}$ and 
${Q_j = M_L 
\operatorname{diag} (s_1, \ldots, s_N)
M_R^T}$,
with $M_L, M_R\in \mathrm{SO}(N)$ and ${\sum s_a^2 = N}$.
These states have the same value of the quadratic Casimir, but in general have different values of the higher Casimirs. Our case~\eqref{eq:QinSO(N)} corresponds to taking ${s_a=1}$ for all $a$.}

\subsection{Equations of motion and continuum limit}
\label{subsec:equations-of-motion}

The Heisenberg evolution of an operator $O$~is\footnote{To avoid imaginary units in the equations of motion, we derive them for real time evolution, and Wick rotate back to imaginary time (as required for Eq.~\ref{eq:imaginarytimeevo}) after obtaining the Lagrangian.}%
\be\label{eq:heisenberggeneral}
    \frac{\dd O}{\dd t} = i N_F [\Hs, O]. 
\ee
The factor of $N_F$ plays the role of an effective ``$1/\hbar$'', and is consistent with the normalization of the Hamiltonian in Eq.~\ref{eq:imaginarytimeevo}. 
In the $N_F\rightarrow \infty$ limit, Eq.~\ref{eq:heisenberggeneral}
 yields 
classical equations of motion for the variables $L_j^{ab}$, $R_j^{ab}$, and $Q_j^{ab}$.

Using the commutation relations in Eq.~\ref{eq:so2Ncommutation} yields

\ba 
    \frac{\dd Q_j}{\dd t} =&-2 J_\pll
    \left[ (L_{j-1}+L_{j+1}) Q_j - Q_j (R_{j-1}+R_{j+1})\right] +\nn\\ \notag
    &+ 2J_\perp \left[ L_j (Q_{j-1} + Q_{j+1})  - (Q_{j-1}+ Q_{j+1}) R_j \right],\\
    \frac{\dd L_j}{\dd t} =& -2 J_\pll \left[ (L_{j-1} + L_{j+1}) L_j - \transp \right]\nn\\
    &+ 2 J_\perp \left[(Q_{j-1}+Q_{j+1}) Q_j^T - \transp \right],
\end{align}
where ``$\transp$'' denotes the transpose of the previous terms
(ensuring $L_j$ and $R_j$ remain antisymmetric matrices).
The equation for ${\dd R_j/\dd t}$ is obtained from that for ${\dd L_j/\dd t}$ by the exchanges ${L\rightarrow R}$, ${Q\rightarrow -Q^T}$, as required by the $\mathbb{Z}_2$ symmetry exchanging forward and backward replicas.

These equations of motion simplify in the continuum limit, i.e. after keeping only lowest orders in a derivative expansion. 
Taking this  limit amounts to isolating low-momentum fluctuations of the fields. This is quantitatively accurate at large $N_F$.\footnote{ In more detail:
The justification for dropping higher derivative terms
can be understood in field theory language (anticipating slightly).
Dropping these terms amounts to modifying the Lagrangian for modes with momentum comparable to $1/a$.
We could worry that this will affect the IR Lagrangian, since integrating out high-momentum modes can renormalize the Lagrangian for the low-momentum modes.
However, these renormalizations
are of order ${\text{``$\hbar$''}\sim 1/N_F}$
relative to the bare Lagrangian.
This guarantees that there is a momentum scale $\Lambda(N_F)$, which is parametrically small for large $N_F$, above which the renormalizations are negligible.
This small momentum scale justifies the derivative expansion. 
At still smaller scales, the nontrivial RG flow \textit{does} become important: see  Eq.~\ref{eq:renormalized-stiffness}.}

If we denote the  lattice spacing by $a$, then the continuum limit is formally  equivalent to an expansion in $a$, 
so that a lattice operator ${O_{j\pm1}}$ is expanded as
${O(x_j) \pm a \partial_x O(x_j) + \frac{a^2}{2} \partial_x^2 O(x_j)}$. Retaining the leading terms in $a$ in the equations of motion, 
and making the rescaling  $L\mapsto L/a$ and $R\mapsto R/a$, we find:
\ba
\label{eq:eqs-of-motion-continuum}
	\partial_t Q &= 2 a \Gamma \,
	\left[  L(x) Q(x) - Q(x) R(x) \right],\\
 \label{eq:eqs-of-motion-continuumL}
	\partial_t L &= 2 a J_\perp \left[(\partial_x^2 Q) Q^T - \transp \right],\\
 \label{eq:eqs-of-motion-continuumR}
	\partial_t R &= 2 a J_\perp  \left[(\partial_x^2 Q^T) Q - \transp \right].
\end{align}
We used the fact that $2(J_\perp-J_\pll)=\Gamma$ is the measurement  rate. The rescaling of $L$ and $R$ is possible because their value is zero in the ground-state manifold.
In the following, we return to units where $a=1$.

In passing we note that the continuum equations of motion could have been equivalently obtained, following Ref.~\cite{affleck1985quantum}, by first taking the continuum limit at the level of the Hamiltonian
\be
\label{eq:continuum-H}
    \Hs =  \int \dd x\,
    \Tr
    \big(
    J_\perp (\partial_x Q^T) (\partial_x Q)  + \f{\Gamma}{2} ({L}^T {L} + {R}^T {R})
    \big),
\ee
up to a constant, and noting that  only the commutation relations which  involve  ${L}$ or ${R}$ remain nontrivial in the continuum limit, while $[Q(x),Q(x')]\to0$.

The equations above should be supplemented with the  kinematic condition arising from the $S^T S=\mathds{1}$ constraint, which in terms of the rescaled variables gives
\ba
\label{eq:constraintctm}
    Q^T Q &= \mathds{1},
    &
    {L} Q + Q {R} = 0.
\end{align}
Using these constraints, we can eliminate ${L}$ and ${R}$ from the equation of motion for ${\partial_t^2 Q}$
that is obtained by differentiating \eqref{eq:eqs-of-motion-continuum}.
The terms 
 with ${\dd {L}/ \dd t}$ and ${\dd {R}/ \dd t}$ can be eliminated using the relevant equations of motion \eqref{eq:eqs-of-motion-continuumL},~\eqref{eq:eqs-of-motion-continuumR}, 
and the terms $L$ and $R$ can be eliminating using the identities
\begin{align}
    {L} &= ({4\Gamma})^{-1} (\partial_t Q) Q^T,
    &
    {R} &= -({4\Gamma})^{-1} Q^T (\partial_t Q),
\end{align}
which follow from \eqref{eq:eqs-of-motion-continuum} together with the constraint \eqref{eq:constraintctm}.
Finally,
\be
    \partial_t^2 Q -  Q (\partial_t^2 Q)^T  Q  = v^2 \left( \partial_x^2 Q - Q (\partial_x^2 Q)^T Q \right)
\ee
with $v = 4\sqrt{\Gamma J_\perp}$.
Equivalently, we can rewrite this equation in the more suggestive form
\be
\label{eq:Q-eq-of-motion}
    Q^T \left(v^{-1} \partial_t^2 Q - v \partial_x^2 Q \right) - \transp = 0.
\ee
As we will see in the next section this is the equation of motion of the NL$\sigma$M.

\subsection{$\mathrm{SO}(N)$ NL$\sigma$M Lagrangian}
\label{subsec:lagrangian}

The equation of motion above arises from a nonlinear sigma model action for $Q$, which, after rotating back to imaginary time, is
\be
\label{eq:action-real-time}
    \mathcal{S}[Q] = \frac{1}{2 g_B} \int \dd x\, \dd t\, \Tr \left[ \f{1}{v} \partial_t Q^T \partial_t Q + v \, \partial_x Q^T \partial_x Q \right]
\ee 
subject to the constraint ${Q^T Q = \mathds{1}}$ at each point. 
The equations of motion can be recovered by noting that the allowed variations of $Q$ are of the form ${Q\mapsto Q(1+\delta q)}$, with ${\delta q}$ antisymmetric, and  requiring 
${\delta \mathcal{S}/{\delta q}=0}$.

The coupling constant $g_B$ does not appear in the equations of motion, 
but it is fixed by noting that the  coefficient of the 
${\Tr(\partial_x Q^T \partial_x Q)}$ term in the action is  inherited directly from the Hamiltonian in Eq.~\ref{eq:continuum-H} (see e.g. Eq.~\ref{eq:coherentstatespathintegral}). Together with the result of the previous section for the velocity, this gives:
\ba
\label{eq:gandv}
    g_B & = 
    \f{2}{N_F} \sqrt{\f{\Gamma}{ J_\perp}},
    &
    v & = 4\sqrt{\Gamma J_\perp}.
\end{align}
This is the bare value of the coupling $g$: below we will consider its RG flow.

This almost completes
our derivation of the effective Langrangian for smooth $Q$ configurations.
The final point is that 
 ${\pi_2(\mathrm{SO}(N))=0}$, 
which means that there is no topological  {$\Theta$-term} that can be added to $\mathcal{L}$ \cite{altland2010condensed,fradkin2013field}.
(Topological {$\Theta$-terms} do not affect the equations of motion, so could not be detected from them.)
In Sec.~\ref{sec:measurement-only} we will need another  NL$\sigma$M, where a {$\Theta$-term} does play a~role.

From now on we rescale time units so that $v=1$, so
\be
    \label{eq:NLSM-lagrangian}
    \mathcal{L}[Q] = \frac{1}{2 g_B} \Tr \left[ \partial_\mu Q^T \partial_\mu Q\right],
\ee
where $\mu=x,t$ is summed over, and again ${Q\in \mathrm{SO}(N)}$.

One final point will be crucial for understanding the phase diagram once we turn on the dimerization parameter. Above we discussed the regime of small $g_B$, where $Q$ can be treated as smooth at the lattice scale. 
However, since the fundamental group of the target space is 
\be
\pi_1\lf \mathrm{SO}(N)\ri=\mathbb{Z}_2
\,\,\, \,\,\,\,\,\,
\text{(for $N>2$),}
\ee
we can in principle have pointlike $\mathbb{Z}_2$ vortices where the order parameter $Q$ is singular 
\cite{fu2012topology, nahum2013loop}. 
These vortices are irrelevant at small $g$
--- this is analogous to the  XY model at low temperature ---
so they do not affect the universal physics of the 
stable nontrivial phase, which we discuss in Sec.~\ref{sec:entanglement}.
However, vortices are responsible for the transition into a disentangled phase. We discuss this in Sec.~\ref{sec:dimerization-and-vortices}.

Remarkably, the Lagrangian in Eq.~\ref{eq:NLSM-lagrangian} gives access to exact universal results for the entanglement structure of the physical state. 
This is due to a special feature of the replica limit: the coupling $g$ flows to smaller and smaller values at large scales, so that semiclassical calculations become better and better controlled.
This flow gives rise to nontrivial scaling forms that can be computed exactly by combining the perturbative beta function with a saddle-point analysis of the field theory.

The two-dimensional $\mathrm{SO}(N)$ NL$\sigma$M appears in the context of Anderson localization, 
where  the two dimensions are two spatial directions (i.e. there is no time coordinate).
There, it describes Bogoliubov-de Gennes excitations of disordered superconductors in symmetry class DIII \cite{PhysRevB.55.1142}.
The main difference, apart from the physical interpretation, 
is that in the Anderson localization problem one is interested in the limit ${N\rightarrow 0}$, whereas here we need ${N\rightarrow 1}$, cf.~Sec.~\ref{sec:model_def}.

The perturbative beta function has been computed for arbitrary $N$ \cite{MCKANE1980169,HIKAMI1981208,WEGNER1989663}:
\be\label{eq:perturbativebetafunctionQtheory}
\f{\dd g_R}{\dd \ln l} = \f{1}{8\pi} (N-2) g_R^2+O(g_R^3)\,.
\ee
The result is available up to $O(g_R^5)$, but the form above will suffice for now. The key point is the sign { change} at ${N=2}$, which means that for each ${N<2}$ there is a stable phase governed by the ${g_R=0}$ fixed point.\footnote{In the Anderson localization problem
(and with ${N=0}$)
 this describes a ``supermetal'' in which the thermal conductance of the Bogoliubov excitations increases with lengthscale (in the NL$\sigma$Ms for Anderson localization, the coupling $g$ is proportional to the conductivity \cite{hikami1980spin,evers2008anderson}).} 

Within this phase for ${N=1}$, solving~\eqref{eq:perturbativebetafunctionQtheory} gives
\be
\label{eq:renormalized-stiffness}
g_R(l)^{-1} = g_B^{-1} +  \f{1}{8\pi}\ln l + \cdots.
\ee
The universal constant of $(8\pi)^{-1}$ in front of the logarithm will appear in the scaling of entanglement entropies. 
The bare value $g_B$ is nonuniversal, but we have fixed its value for  large $N_F$ (Eq.~\ref{eq:gandv}).

The explicit derivation of the sigma model in the previous section guarantees that for large $N_F$ we are in the basin of attraction of the $g_R=0$ fixed point.
But in fact, we can argue on
symmetry grounds that if $N_F$
is odd, 
and if the measurement rates are not dimerized, 
it is impossible for the NL$\sigma$M to be in a disordered phase.
The simplest scenario is that (in the absence of dimerization)  the models with odd $N_F$ 
flow to the $g_R=0$ fixed point for \textit{any} nonzero values of $J^2$ and $\Gamma>0$.
This is consistent with the picture for the RG flows that is proposed in the following sections.
For $N_F$ even we expect flow to this fixed point for large enough $J^2/\Gamma$ , as we discuss briefly in Sec.~\ref{sec:firstlookatphasediagram}.

The fixed point is also completely stable (so long as  replica symmetry is retained)
since the only possible perturbations have larger numbers of derivatives.
Therefore $g_R=0$ governs a stable phase of the monitored dynamics, 
whose properties we will discuss in Sec.~\ref{sec:entanglement}.
In particular, this phase is stable to perturbation by sufficiently weak dimerization $\Delta$.

Of course it is possible to drive the dynamics into a disentangled phase with an appropriate perturbation that is large enough. 
In our model this can be done,
by sufficiently strong and sufficiently strongly dimerized   measurement 
(at least for $N_F=1$, and according to our proposed RG flows, for all $N_F$).
The corresponding  critical fixed point, driven by proliferation of vortices,
is outside the range of validity of the perturbative beta function in Eq.~\ref{eq:perturbativebetafunctionQtheory},
and will be discussed in Sec.~\ref{sec:dimerization-and-vortices}.

We also emphasize that the measurement-only line $J^2=0$ is described by a different field theory, as a result of the enhanced replica symmetry \cite{bao2021symmetry} there: we defer discussing this theory to Sec.~\ref{sec:measurement-only}.

Next --- before describing the physical consequences of the above NL$\sigma$M for the stable phase, or phase transitions out of this phase ---
let us preview the broader phase diagram of the monitored model. 

\section{Phase diagram for  $N_F=1$}
\label{sec:firstlookatphasediagram}

\begin{figure}
    \centering
    \includegraphics[width=\linewidth]{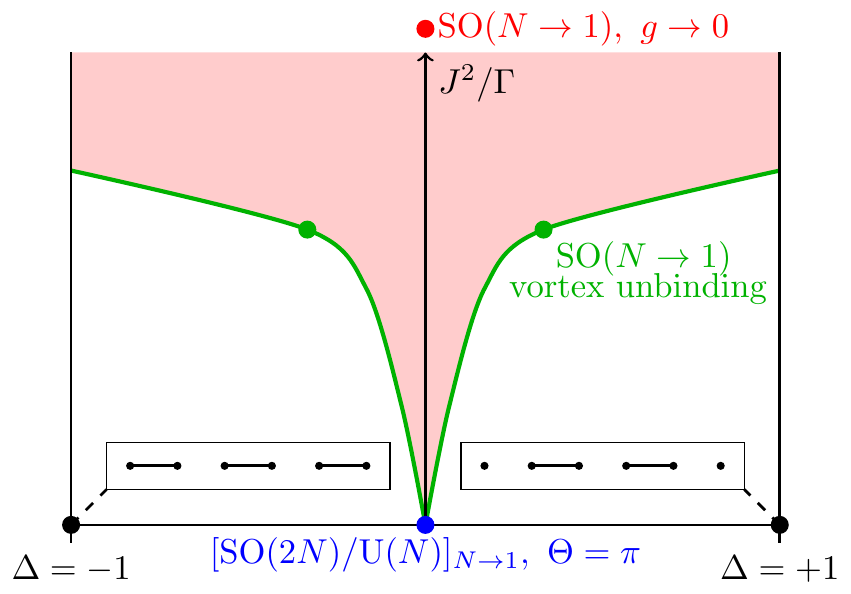}
    \caption{Schematic phase diagram for $N_F=1$ (see also Fig.~\ref{fig:phase-diag-with-flows}) 
    The RG fixed points governing the phases and transitions are  indicated 
    by red, green and blue points, 
    with a description of the associated field theory. 
    Furthermore, two special points in the phase diagram are at $(\Delta, J)=(\pm 1, 0)$, denoted by black dots. 
    Here unitary dynamics is absent and the measurements are fully dimerized. For $N_F=1$, the wavefunction then tends to an eigenstate of $i\gamma_{j}\gamma_{j+1}$ for odd (even) $j$ at $\Delta=-1$ ($\Delta=+1$).}
    \label{fig:phase-diag-sketch}
\end{figure}

In Fig.~\ref{fig:phase-diag-sketch} we sketched our proposed phase diagram for the model with ${N_F=1}$.
In this section we give a schematic overview of the basic features.
The { full justification for various features of this phase diagram will be given in later sections: 
the  phase transition lines at ${J>0}$ are} discussed in terms of the $\mathrm{SO}({N\to 1})$ NL$\sigma$M in Sec.~\ref{sec:dimerization-and-vortices},
while the line ${J=0}$  is related to the ${\mathrm{SO}(2N)/\mathrm{U}(N)}|_{N\to 1}$ NL$\sigma$M
in Sec.~\ref{sec:measurement-only}, which also contains a conjecture about RG flows in that model.
Fig.~\ref{fig:phase-diag-sketch} is also consistent with symmetry considerations for the lattice model.
 
{ For most of this Section we focus on the  simplest case ${N_F=1}$, which corresponds to a chain with a single physical Majorana at each site.
When ${N_F>1}$, so that each site hosts multiple Majoranas, 
 we expect that the stable (pink) phase ``touches'' the ${J=0}$ axis
at a larger number of points:
we comment on this briefly in Sec.~\ref{sec:commentgeneralNF}
and Sec.~\ref{sec:measurement-only}.  
But while the phase diagram for ${N_F>1}$ is slightly more intricate, all the phase transition lines (at $J>0$) remain in the same universality class 
(described by the same bulk CFT) 
as the phase transition lines in Fig.~\ref{fig:phase-diag-sketch}, 
and all critical points (at ${J=0}$)  remain in the same universality class as the ${J=0}$ critical point in Fig.~\ref{fig:phase-diag-sketch}.  
The phase boundaries are symmetric under ${\Delta\rightarrow -\Delta}$ for any value of $N_F$. }
For $N_F$ odd, symmetry ensures that the model is nontrivially entangled everywhere on the vertical axis ${\Delta=0}$.

\subsection{Disentangled phases}

A basic feature of the phase diagram is the existence of two distinct
--- stable --- disentangled phases at large $\Gamma$,
one for each sign of the dimerization $\Delta$.
 These phases also appear in a Majorana model with discrete measurements \cite{nahum2020entanglement}
 (or its  Ising dual \cite{sang2021measurement,lavasani2021measurement,sang2021entanglement,lang2020entanglement})
and in Refs.~\cite{bao2021symmetry, merritt2022entanglement} where more general circuits are studied.
 
The extreme limits of these phases at 
${J=0}$, $\Delta=\pm 1$
have Majoranas { dimerized} in pairs, as shown in Fig.~\ref{fig:phase-diag-sketch}. 
In other words, at late times the state is a 
(random) eigenstate of the operators
 ${i\gamma_j\gamma_{j+1}}$
 for  \textit{either} even $j$ \textit{or} odd $j$, depending on whether ${\Delta=1}$ or ${\Delta = -1}$. 
 On a finite chain, the former case has unpaired boundary Majoranas,
 as shown in Fig.~\ref{fig:phase-diag-sketch}.
The  fermion parity $i\gamma_1\gamma_L$ of these two modes remains ``hidden'' from measurements, giving one ``bit'' that is not purified.
These disentangled phases are stable (we checked this numerically using the techique described in Sec.~\ref{sec:numerics}).

In terms of the NL$\sigma$M, { the disentangled phases} are 
disordered phases,
obtained by proliferating vortices \cite{fu2012topology,nahum2013loop,konig2012metal}. 
Anticipating Sec.~\ref{sec:dimerization-and-vortices}, 
we find that the two disordered phases are distinguished by the sign of the vortex fugacity 
(and the line 
$\Delta=0$ is forced to be nontrivially entangled because the vortex fugacity vanishes there).

The picture of paired Majoranas has a close analog in the effective spin chain $\Hs$ \cite{jian2022criticality}: 
the spins are dimerized, and at  $\Delta=1$ there are gapless boundary spins that are not involved in any dimer.\footnote{Nonrigorous extensions of standard ideas from spin chains  
may be used to argue that these two gapped phases of the spin chain cannot be connected without a phase transition (using the fact that the boundary spins transform projectively under global symmetry \cite{turner2013beyond}),
and that the line $\Delta=0$ is gapless (in the spirit of the Lieb-Schultz-Mattis theorem \cite{lieb1961two}).}

\subsection{Stable nontrivial phase and transition lines}

Next, the phase diagram in Fig.~\ref{fig:phase-diag-sketch} features a stable phase (shaded in pink) 
{ which, in the sigma model language, flows} to ${g_R=0}$. Scaling properties of the entanglement inside this phase are discussed { below} in Sec.~\ref{sec:entanglement}.
The logarithmic flow of the NL$\sigma$M coupling implies that entanglement entropies are larger at large scales than they would be at a conformal fixed point (Sec.~\ref{sec:entanglement}).

Since the line $\Delta=0$ is necessarily  gapless, 
the simplest hypothesis is that this entire vertical axis, for $J>0$, is in this phase.
We  expect the boundary of the gapless phase to meet the lines $\Delta=\pm 1$ at a finite value of $J$:
that is,
 we expect
that the dynamics is entangling whenever the measurements are sufficiently weak, 
regardless of whether they are dimerized.
(There is nothing special about the lines $\Delta=-1,1$ from the point of view of replica symmetry.)

The phase boundary lines between the nontrivially entangled phase 
and the disentangled phases are governed by an RG fixed point that we discuss in Sec.~\ref{sec:dimerization-and-vortices}. Fig.~\ref{fig:phase-diag-with-flows} above { showed the schematic RG flow, which involves} two copies of this fixed point, one for positive and one for negative dimerization (green dots). 
This fixed point is straightforwardly scale-invariant, so that for example the bipartite entanglement entropy { is expected to} scale as ${\ln L}$ on the phase boundary lines, 
as opposed to the $(\ln L)^2$ in the stable nontrivial phase (see discussion just above Sec.~\ref{sec:BCseffectivetheory}).

\subsection{Measurement only axis (${J=0}$)}

The final aspect of the phase diagram is what happens to the transition lines as ${J^2/\Gamma \rightarrow 0}$.
To answer this it is necessary to consider the \textit{distinct}
NL$\sigma$M, with target space ${\mathrm{SO}(2N)/\mathrm{U}(N)}$,
that applies on the ${J=0}$ axis, as a result of  higher replica symmetry there.
This is discussed in Sec.~\ref{sec:measurement-only}.
There, we conjecture that the 
${{\mathrm{SO}(2N)/\mathrm{U}(N)}|_{N\to 1}}$ sigma model (which admits a $\Theta$ term)  has an unstable fixed point with ${\Theta=\pi}$
that gives rise to a critical point 
at the origin of the phase diagram. { We note that this critical point was already studied numerically in Ref.~\cite{kells2021topological}.
The above conjecture for the sigma model} gives the RG flow topology shown in Fig.~\ref{fig:phase-diag-with-flows}.
This fixed point is again scale-invariant, so that bipartite entanglement scales as ${\ln L}$.

{
\subsection{Phase diagram for general $N_F$}
\label{sec:commentgeneralNF}

\begin{figure}
    \centering
    \includegraphics[width=\linewidth]{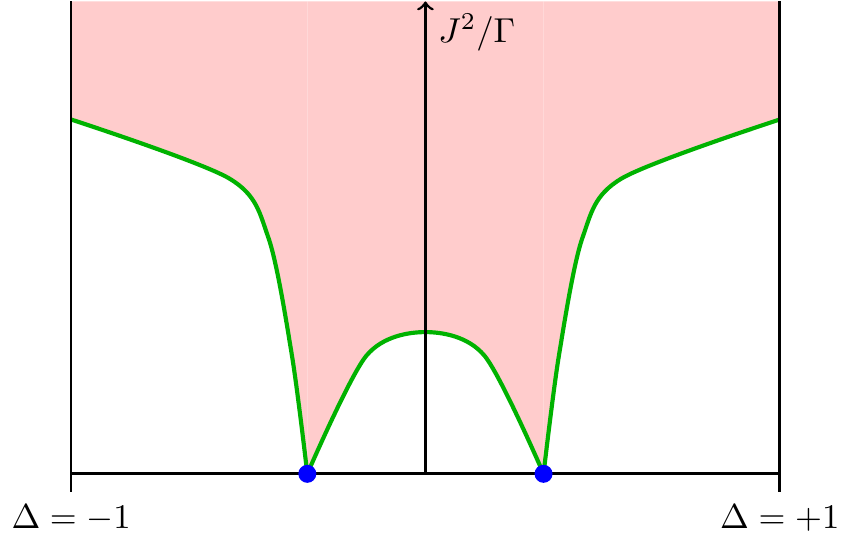}
    \caption{Schematic phase diagram for $N_F=2$ with the same color coding as Fig.~\ref{fig:phase-diag-sketch}, i.e. areas, lines, and points with the same colors are represented by the same infra-red field theory in the bulk.}
    \label{fig:phase-diag-NF2}
\end{figure}

The discussion of the $\Theta$ angle in Sec.~\ref{sec:measurement-only} leads to the conjecture that, for a general number of flavors, $N_F$ critical points appear on the ${J=0}$ axis.
These critical points are due to $\Theta$ cycling $N_F$ times through ${\pi\,(\operatorname{mod}\,2\pi)}$ as $\Delta$ varies from $-1$ to $1$. 
This is in close analogy to the sequence of phase transitions between dimerized and Haldane-like ground states in an antiferromagnetic chain with spin ${S=2N_F}$~\cite{affleck1987critical,
oshikawa1992hidden,
kitazawa1996phase,
kitazawa1997phase}.

Each of the $N_F$ critical points is in the same (bulk) universality class as the ${J=0}$ critical point of Fig.~\ref{fig:phase-diag-sketch}, so that the local topology of the phase boundaries is also the same near each critical point.
As in the ${N_F=1}$ case, we expect that the chain is in the nontrivial phase when ${J^2/\Gamma}$ is sufficiently large, regardless of the value of $\Delta$. 
These considerations give a simple conjecture for the phase diagram, shown in Fig.~\ref{fig:phase-diag-NF2} for the case ${N_F=2}$.
}

\subsection{Aside: Majorana loop model}
\label{sec:majoranaloopphasediagramcomment}

The topology of the phase diagram is very similar to that of a Majorana quantum circuit model with discrete measurements and discrete unitary ``swap'' operations which maps to a classical loop model \cite{nahum2020entanglement, nahum2013loop}
(let us call this the ``Majorana loop model'').
{ Like Fig.~\ref{fig:phase-diag-sketch}, the Majorana loop model shows a stable nontrivial phase where the entanglement scales as ${\log^2 L}$ \cite{nahum2020entanglement} (see Ref.~\cite{sang2021entanglement} for an in-depth discussion of this phase).
The similarity of the two phase diagrams is} consistent with the numerical results of 
Ref.~\cite{merritt2022entanglement}, who found previously  that the phase diagram topology of the Majorana loop model was robust when more general unitary gates were allowed (and
we expect that the gapless phase found here for $J>0$ is the same as that found in Ref.~\cite{merritt2022entanglement} when generic unitaries are allowed).

{ But despite the similar phase diagram topology, the relevant universality classes are different in the loop model and in the ``generic'' models studied here.}
The similarity of the phase diagrams may be understood in terms of close similarities between the relevant field theories 
(which we discuss briefly in Sec.~\ref{sec:measurement-only}).
The RG flows have a similar topology, and a similar role is played by both  vortex defects in the bulk of the phase diagram and a $\Theta$ term on the lower axis.\footnote{See Refs.~\cite{gruzberg1999exact,beamond2002quantum} for an unexpected exact correspondence between an Anderson localization   problem (in symmetry class C) and a classical model of loops.}
However, there are also significant differences between the Majorana loop model and more general models.
In the former, 
Majoranas are only ever entangled in pairs: 
this is an additional structure that is 
more constraining than Gaussianity. This results in a larger replica symmetry and a different NL$\sigma$M. 

\section{Entanglement in the stable non-trivial phase}
\label{sec:entanglement}

\begin{figure}
    \centering
    \begin{tikzpicture}
        \begin{scope}
            \draw[thick,->] (0,-0.1) -- (0,4.3) node[anchor=south west] {time};
            \draw[thick,->] (-0.1,0) -- (3.2, 0) node[anchor=west] {$x$};
            \node at (-0.3,0) {$0$};
            \node at (0,-0.3) {$0$};
            \draw[thick] (3,4) -- (3,-0.1) node[anchor=north] {$L$};
            \draw[thick] (3,4) -- (-0.1,4) node[anchor=east] {$t$};
            \draw[ultra thick, red] (3,4) -- (0,4);
            \node at (1.5, 3.7) {\red $Q_{k,n}$};
            \draw[ultra thick, blue] (3,0) -- (0,0);
            \node at (1.5, 0.3) {\blue $\mathds{1}$};
            \draw[dashed, -{Stealth[length=10pt, width=7pt]}] (0.75,0.1) -- (0.75, 2); 
            \draw[dashed] (0.75, 2)-- (0.75,3.9);
            \draw[dashed, -{Stealth[length=10pt, width=7pt]}] (2.25,0.1) -- (2.25, 2); 
            \draw[dashed] (2.25, 2)-- (2.25,3.9);
        \end{scope}

        \begin{scope}[shift={(4.5,0)}]
            \draw[thick,->] (0,1) -- (0,4.3);
            \draw[thick,dashed] (0,0) -- (0,1);
            \draw[thick] (3,1) -- (3,4.1) node[anchor=south] {$L$};
            \draw[thick] (1.5,4) -- (1.5,4.1) node[anchor=south] {$L/2$};
            \draw[thick, dashed] (3,0) -- (3,1);
            \draw[thick] (3,4) -- (-0.1,4) node[anchor=east] {$t$};
            \draw[ultra thick, red] (1.5,4) -- (3,4);
            \node at (2.25, 3.7) {\red $Q_{k,n}$};
            \draw[ultra thick, blue] (0,4) -- (1.5,4);
            \node at (0.75, 3.7) {\blue $\mathds{1}$};
            \draw[dashed, -{Stealth[length=10pt, width=7pt]}] (1.0,4) arc (-180:-90:0.5);
            \draw[dashed] (1.5,3.5) arc (-90:0:0.5);
            \draw[dashed, -{Stealth[length=10pt, width=7pt]}] (0.5,4) arc (-180:-90:1.0);
            \draw[dashed] (1.5,3.0) arc (-90:0:1.0);
        \end{scope}
    \end{tikzpicture}
    \caption{Boundary conditions for the computation of the entropy of the whole state { at finite time} (left) and the bipartite { entanglement entropy at asymptotically late times} (right). Dashed lines denotes the direction along which $\nabla Q^{\rm cl}$ is non-zero, with $Q^{\rm cl}$ denoting the classical configuration minimizing the energy.}
    \label{fig:boundary-conditions}
\end{figure}
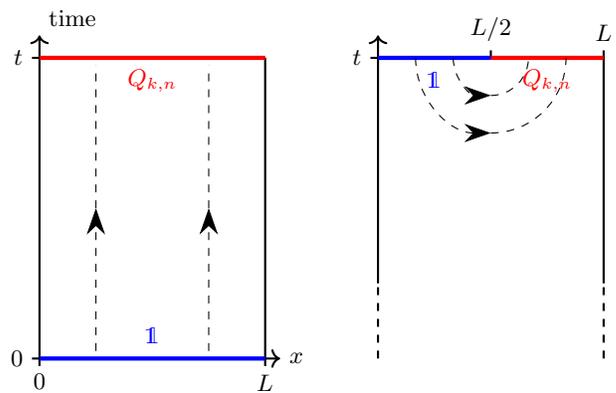

We now turn to the  universal properties of the stable nontrivial phase (shaded pink in Fig.~\ref{fig:phase-diag-sketch}), where
entanglement entropies have non-trivial scaling with time and system size $L$.
Asymptotically, the scaling depends only on the RG flow near the ${g_R=0}$ fixed point discussed in Sec.~\ref{sec:firstlookatphasediagram}, so is independent of the value of $N_F$.
In the following subsections we discuss both the purification of the state over time 
in a given quantum trajectory \cite{gullans2020dynamical},
and the entanglement of the pure states that arise at very long times.
We first summarize the main results of this analysis.

Our initial state ${\rho(0)\varpropto \mathds{1}}$ has maximal entropy,
\be
    S_n(t=0) = \f{N_F L}{2} \ln 2,
\ee
for all R\'enyis. 
However, measurements tend to purify the state as $t$ increases. 
We find that when  $L$ and $t$ are both asymptotically large,
and if ${t\lesssim L}$,
\be\label{eq:purity-scaling}
S_n \sim \frac{\pi (n+1)}{48 n} \f{L\ln t}{v t},
\ee
with the von Neumann entropy obtained as the ${n\rightarrow 1}$ limit.
The velocity $v$ appearing here is  nonuniversal: 
in Sec.~\ref{sec:mapping-to-NLSM} we have calculated it for large $N_F$  and ${\Delta = 0}$.

The above formula holds for $S_n$, and not only for its mean $\overline{S_n}$: we find that the entropies and entanglement entropies are self-averaging (fluctuations  are subleading compared to the mean).
{ Note that the numerical prefactors in Eq.~\ref{eq:purity-scaling} and  in Eq.~\ref{eq:renyi-entropy-result} below are universal (ultimately arising from the RG flow in Eq.~\ref{eq:renormalized-stiffness})
and so apply to a larger class of monitored systems with the same symmetries. 
 }

At times $t\gtrsim L$, we find a crossover after which $S_n$ decays exponentially. The typical time of decay gives us a ``purification timescale'', 
\be \label{eq:tauP}
\tau_P \sim L \ln L,
\ee
after which the entropy becomes of order 1. 
The logarithms in this formula and  the previous one are due to the marginal flow of the coupling $g$.
In a model where the flow is to a fixed point at finite coupling they are absent, cf. Sec.~\ref{sec:measurement-only}
(the two kinds of flow can also be found in the Majorana loop model \cite{nahum2020entanglement,nahum2021measurement,merritt2022entanglement,sang2021entanglement}).

When $t/\tau_P\rightarrow\infty$ we have a pure state, which we may characterize by its bipartite entanglement.
For a region $A$ including the leftmost $L/2$ sites 
(with non-periodic boundary conditions)
we find that 
\ba
\label{eq:renyi-entropy-result}
  S_{n,A} = \f{n+1}{96 n} (\ln L)^2 + o\left[(\ln L)^2\right].
\end{align}
Again the additional logarithm, compared to the single power of $\ln L$ that would be dictated by scale invariance~\cite{vasseur2019entanglement,Calabrese_2009}, is due to the marginal flow. Fluctuations are again subleading.

The above formulas are asymptotic formulas that hold for any fixed $N_F$ in the limit that $\operatorname{min}(t,L)$  becomes large.
A factor $(8\pi)^{-1} \ln \operatorname{min}(t,L)$ that contributes to  these formulas arises from the  asymptotic 
form of the inverse coupling $1/g_R(L)$.
If  the number of flavors $N_F$ is large, we see from 
 Eq.~\ref{eq:renormalized-stiffness} and Eq.~\ref{eq:gandv}
 that  there is a very large lengthscale below  which the ``bare'' term in $1/g_R(L)$ is larger than the logarithmic term. 
Consequently for large $N_F$ a much better formula is given by replacing $(8\pi)^{-1}\ln L$ by the flowing inverse coupling $1/g_R(L)$ --- for example for the purification protocol
\be
S_n(t) \approx \frac{\pi^2 (n+1)}{6n} \f{L}{v t} g_R^{-1}(t)
\ee
with $g_R^{-1}$ given by Eq.~\ref{eq:renormalized-stiffness}
(subleading terms in the beta function could also be taken into account to obtain more precise fits to data).

As mentioned in the introduction, the flow of $1/g$ to larger values as a function of lengthscale means that the entanglement structure of the state is not scale invariant (at least in the usual sense\footnote{ The Goldstone modes become more and more weakly interacting as $g\to 0$, but while we can rescale these fields so that they are conventional free fields in the limit $g\to 0$, the violation of scale invariance reappears as the indefinite growth of the size of the target space with scale.}). In a scale or conformally invariant theory \cite{Calabrese_2009,vasseur2019entanglement}, the bipartite entanglement scales as $\ln L$, which may be thought of heuristically (in the spirit of real space RG) as a sum of equal contributions from all logarithmically-spaced scales up to $\ln L$.
Here, the contribution instead increases with scale.

While here we discuss the NL$\sigma$M with ${N\to 1}$
relevant to measurements, the considerations in this section 
carry over to the metallic phase of the theory with ${N\to0}$ which was proposed in Ref.~\cite{jian2022criticality} to describe Gaussian random tensor networks.
Applied to this limit, our calculation gives the same $(\ln L)^2$ entanglement scaling as in Eq.~\ref{eq:renyi-entropy-result}, but with a universal coefficient larger by a factor 2 as a result of Eq.~\ref{eq:perturbativebetafunctionQtheory}. This { result differs from the form proposed in Ref.~\cite{jian2022criticality}, where  scale invariance was assumed (cf. the discussion above about scale invariance).}

We now describe how the results above may be obtained by minimizing the effective action $\mathcal{S}[Q]$ with the appropriate boundary conditions.
This approach could be straightforwardly generalized to many other settings to study for example the entanglement of multiple intervals, or, say, other spatial boundary conditions.\footnote{For example, we could include decoherence at the spatial boundaries~\cite{PhysRevX.9.021007,PhysRevLett.123.110601}  that prevents the complete purification of the state (this would lead to  Dirichlet boundary conditions on  $Q$ field at these boundaries).}

\subsection{Boundary conditions in the effective theory}
\label{sec:BCseffectivetheory}

We sketched in Secs.~\ref{sec:replicareviewmaintext}  and Sec.~\ref{sec:effectivereplicahamiltonian} how the generating function for any R\'enyi entropy $S_n(t)$ may be written as a transition amplitude in the spin chain:
 \be\label{eq:genfunamplitude}
\mathbb{E} \left[ e^{-k(n-1)S_n(t)} \right]
= 
\big\langle \mathcal{C}_{k,n} \big|
\exp\lf - t N_F \Hs \ri
\big|\rho^{(N)}(0) \big\rangle.
\ee
Recall that in the replica approach, as we have formulated it, we initially take ${N>kn}$ and then continue to ${N=1}$. We have left the limit ${N\rightarrow 1}$ implicit in Eq.~\ref{eq:genfunamplitude}: since we will always be interested in the limit ${N\to 1}$ we will often simplify the notation by leaving the $N$ dependence implict, for example in the state $\ket{\mathcal{C}_{k,n}}$ above. 

The initial and final states appearing above are discussed in Appendix~\ref{app:sec:boundary-stabilizers}.
They are precisely (products of) coherent states $\ket{S}$ of the form mentioned in Sec.~\ref{subsec:path-integral-deg}, which are labelled by a choice of expectation value for the matrix $S$. Since the states of interest have vanishing expectation values for $L$ and $R$,
they are parameterized by a value for the $Q$ matrix: 
\be
\mathbb{E} \left[ e^{-k(n-1)S_n(t)} \right]
= 
\big\langle Q_{k,n} \big|
\exp\lf - t N_F \Hs \ri
\big|Q = \mathds{1}_{N} \big\rangle.
\ee
We have now labelled the initial and final states by their expectation values of $Q$ (which are translation invariant in the present setting).
In Appendix~\ref{app:sec:boundary-stabilizers} we show that the initial
maximally mixed state 
corresponds (after replication)
to   $Q=\mathds{1}_{N}$ (the ${N\times N}$ identity matrix), 
as indicated above, and that
the final state  ${\ket{\mathcal{C}_{k,n}}=\ket{Q_{k,n}}}$
corresponds to an $N\times N$ matrix with a block structure: 
\be
Q_{k,n} \equiv \overset{\text{$k$ times}}{\overbrace{q_n \oplus \cdots \oplus q_n}} \oplus \mathds{1}_{N-nk}.
\ee
The nontrivial $n\times n$ block has a cyclic form,
\be
\label{eq:qn-boundary}
q_n \equiv 
\lf \begin{array}{ccccc}
    \phantom{+} 0 & \phantom{+} 0 & \cdots &  \phantom{+} 0 &  + 1  \\
    -1 & \phantom{+} 0 & \cdots &  \phantom{+} 0 &  \phantom{+} 0  \\
    \phantom{+} 0 & -1 & \cdots &  \phantom{+} 0 & \phantom{+}  0  \\
     &   & \cdots & &     \\
     \phantom{+} 0 & \phantom{+} 0 & \cdots & -1 &\phantom{+}  0
\end{array}\ri
\ee
with $-1$s on the subleading diagonal and $+1$ in the top right entry.

These $Q$ matrices are  closely related to the permutation matrices that can be used to express the pattern of index contractions at the boundary (in the language of a bosonic tensor network). 
However  they are not equivalent to these objects due to their sign structure (see Appendix~\ref{app:sec:boundary-stabilizers}). The signs (which would be absent in a permutation matrix) are crucial to ensure that ${\operatorname{det} Q = 1}$ and ${Q\in \mathrm{SO}(N)}$. 

When we pass to the field theory, coarse-graining over microscopic scales, the initial and final states above set the boundary conditions for the field:
\ba\label{eq:entropytofieldtheory}
\mathbb{E} \left[ {e^{-k(n-1)S_n(t)}} \right]
    &= \int_{Q(x,0)=\mathds{1}}^{Q(x,t)=Q_{k,n}} \mathcal{D}Q\, e^{- S[Q]}.
\end{align}
These boundary conditions (BCs) are shown in Fig.~\ref{fig:boundary-conditions} (Left). 
We have omitted a normalization constant in \eqref{eq:entropytofieldtheory} since it will drop out automatically in the saddle-point calculation.

In the simplest case, we will be able to approximate the right-hand-side of \eqref{eq:entropytofieldtheory} simply by the exponential of minus the
action for the saddle-point solution $Q^{\text{cl}(x,t)}$.
In cases where the saddle-point configuration has variation on many lengthscales, we must separate out contributions
to the free energy
from different lengthscales since the value of $g_R$ renormalizes nontrivially as a function of scale.
Crucially, these calculations are asymptotically exact since the theory flows to weak coupling ($g_R\to0$).

Note that if the resulting  action is simply proportional to $k$, 
then the generating function
becomes that of a deterministic (nonrandom) variable. This is what we find in our leading-order calculation, indicating that fluctuations in the entanglement entropy are parametrically smaller than the mean.

The BCs above apply for the entropy of the entire system. 
A simple generalization gives the entropy of a subsystem $A$.
Schematically, 
we must  trace out region $\bar A$
from the physical density matrix:
this means that the needed ``index contractions'' in region $\bar A$ at the final time are the same as at the initial time 
(here we are using the language of the equivalent bosonic system).
As a result, the final time boundary condition becomes $Q=\mathds{1}_N$ in region $\bar A$, and $Q=Q_{k,n}$ in region $A$.
These BCs are shown in Fig.~\ref{fig:boundary-conditions} (Right).
We will focus on the case where $t\rightarrow \infty$, so that we are computing the entanglement entropy  of a pure state.

Next we apply the above to compute the entropy of the full state: 
due to the uniform boundary condition at time $t$ 
this is the simplest case.
Then we build on these results to compute the bipartite entanglement in Subsec.~\ref{subsec:bipartite}.

\subsection{Purification of mixed state}
\label{subsec:purity:tlessL}

To compute the entropy of the final state we need (by Eq.~\ref{eq:entropytofieldtheory})  the minimal action configuration for the boundary conditions shown in Fig.~\ref{fig:boundary-conditions} (Left), 
with $Q=\mathds{1}_N$ at the initial time and $Q=Q_{k,n}$ at the final time. Let us assume to begin with that $t\lesssim L$, and for simplicity let us choose units so that the nonuniversal velocity is  $v=1$.

We first run the RG up to a scale that is comparable with
(but somewhat less than) $t$,  integrating out modes with shorter wavelengths.
We expect the RG up to this scale to be approximately insensitive to the boundary conditions, and to produce a renormalized stiffness $g_R^{-1}(t)$ described by Eq.~\ref{eq:renormalized-stiffness}.\footnote{For general $N$, integrating out  high-frequency modes  also contributes an  additive constant to the free energy that is independent of boundary conditions. This is  subleading at small $g$, but it also vanishes when ${N\to 1}$, because the number ${N(N-1)/2}$ of Goldstone modes vanishes. This is consistent with the fact that the transition amplitude with equal initial and final boundary conditions, ${\bra{\mathds{1}_N} e^{-tN_F \Hs} \ket{\mathds{1}_N}}$, is trivial when ${N\rightarrow 1}$. Up to a trivial constant, it just computes ${\mathbb{E}_G [\tr \check \rho(t)]^N \to \mathbb{E}_G [\tr \check \rho(t)]}$, which is the normalized sum of trajectory probabilities and is equal to 1 (Eq.~\ref{eq:expvalwithBorn}).}
We now have a system { which, measured in units of the new UV cutoff, is} of length { $\sim L/t$} in the spatial direction and { $\sim 1$} in the time direction. 
Since the stiffness is large and the cutoff is comparable with the temporal extent, 
we can neglect further renormalization effects and simply compute the path integral using saddle point. 

We need the lowest-action configuration ${Q_{k,n}^\text{cl}}$ which interpolates between $Q=\mathds{1}_N$ and $Q=Q_{k,n}$. 
Denoting the rescaled coordinates by $\tilde x$, $\tilde t$ (so that ${\tilde t \in [0,1]}$),
we assume that the optimal $Q_{k,n}^{\text{cl}}(\tilde x, \tilde t)$ is $\tilde x$-independent. The equation of motion then implies that $Q_{k,n}^\text{cl}$ has the form
\ba
    Q_{k,n}^\text{cl}(\tilde x, \tilde t) &= \exp\left( \, \tilde t A_{k,n} \right),
\end{align}
where the antisymmetric matrix $A_{k,n}$ 
(the logarithm\footnote{While $\ln Q_{k,n}$ is not unique, here we intend the branch of the $\ln$ that minimizes the Frobenius norm of $A_{k,n}$, as this is the branch minimizing the energy: see Appendix~\ref{app:sec:bipartite-entanglement}.} of $Q_{k,n}$)
also has a block structure,
\be
A_{k,n} = \ln Q_{k,n} = \overset{\text{$k$ times}}{\overbrace{a_n \oplus \cdots \oplus a_n}} \oplus \mymathbb{0}_{N-nk},
\ee
where, as usual, we leave the $N$-dependence of $A_{k,n}$ implicit. The gradient term in the renormalized action is then
\ba
\tr (\partial_{\tilde t}Q_{k,n}^\mathrm{cl})^T
(\partial_{\tilde t}Q_{k,n}^\mathrm{cl})
=  k \tr a_n^T a_n 
& = \f{\pi^2k(n^2-1)}{3n},
\end{align}
where the second equality is explained in Appendix~\ref{app:sec:bipartite-entanglement}.
Therefore, the minimal action in the renormalized theory is 
\be
\mathcal{S}[ Q_{k,n}^\text{cl}] =
\f{\pi^2}{6}
\f{k(n^2-1)}{n}
\f{L}{g_R(t) t}.
\ee
Comparing with Eq.~\ref{eq:entropytofieldtheory} for the generating function then gives us the result for purification in Eq.~\ref{eq:purity-scaling} (where we have restored the nonuniversal velocity).

Finally, for $t> L$ we can proceed in a similar fashion. 
We first run the RG up to scale $L$, which renormalizes the coupling constant to $g_R^{-1}(L)\sim \log L$. 
{ At this stage it is convenient to  use coordinates that are rescaled by a factor of $L$, so that} the system has size $1$ in the spatial direction and size $t/L$ in the time direction
(and a UV cutoff on the frequencies of order 1 in the rescaled units). 
The RG from this point onward is very different, as the system has become effectively one-dimensional. In 1D, the stiffness $g_R^{-1}$ is not dimensionless anymore, 
but rather it defines a correlation time $\tilde t_* \varpropto g_R^{-1}$.
Returning to the original units, this gives us the purification timescale in Eq.~\ref{eq:tauP},
\be
\tau_P \sim \f{L}{g_R(L)}. 
\ee
The action cost  $\mathcal{S}$, imposed by boundary conditions, scales as $ \exp(-t/\tau_P)$  (to leading  order in the exponential), implying exponential decay of the entropies at very long times.
Note that once $t$ is larger than  $\tau_P$, so that fluctuations become important in the effective 1D model, 
the (small) value of $S_n$ in a realization will no longer be close to the mean.

\subsection{Bipartite entanglement in the pure state}
\label{subsec:bipartite}

At times $t\gg \tau_P$ the dynamics generates an ensemble of random pure states that is independent of the initial state (pure or mixed).
As discussed in Sec.~\ref{sec:BCseffectivetheory}, the bipartite entanglement in this ensemble can be computed  using the BCs shown in Fig.~\ref{fig:boundary-conditions} (Right).  
We take ${t\rightarrow \infty}$, so only the final-time BCs are important.

In contrast to the previous section, these BCs have a discontinuity.
Taking  $A$ to be the right half of the chain and  $\bar A$ to be the left,
\ba\notag    Q(x,t) &= \mathds{1} 
    \text{ for $x<L/2$},
    &
      Q(x,t) &= Q_{k,n} 
    \text{ for $x>L/2$}.
\end{align}
It is convenient to employ polar coordinates $(r,\theta)$ centered at the discontinuity, with $\theta\in [0,\pi]$.
It is possible to find a vortex-like solution of the saddle-point equations that depends only on $\theta$:
\ba
\label{eq:bipartite-entanglement-Q-min}
    Q^\mathrm{cl}_{k,n}(r,\theta) &= e^{\f{\theta}{\pi} A_{k,n}},
\end{align}
where $A_{k,n}$ has been described above.
We have neglected the spatial boundary conditions at large ${|x-L/2|}$, which will affect the solution on the largest scales but will not affect the leading term in the free energy.

The above solution has variations on all scales up to $L$. Since the effective coupling varies (albeit slowly) with scale, 
it would not be correct simply to approximate the path integral by the saddle-point action.
However, this is simply remedied.

We split the free energy $F$ into contributions  from nested annuli at sequentially larger radius  \cite{Kosterlitz_1973, Young_1978}.
Consider the additional contribution $F_{[l, l e^s]}$ from an annulus
$r\in [l, l e^s]$.
Within this region, the typical lengthscale for variation of the expectation value of $Q$ is $l$.
A simple coarse-graining argument shows that  $F_{[l, l e^s]}$ may be approximated by the part of the saddle-point action for this annulus, but weighted by the coupling for this scale, $g_R(l)$.\footnote{This approximation is valid since $g_R(l)$ varies sufficiently slowly with $l$:   $g_R(l)/g_R(e^s l)\simeq 1$ when $g_R$ is small.}

Taking a small enough $s$, we obtain a differential form that may be integrated: 
\ba\label{eq:Fboundary}
F & =
 \f{k \tr a_n^Ta_n }{2\pi}
 \int_{\ln a}^{\ln L} 
 \f{\dd \ln l}{g_R(l)}.
 \end{align}
(We may also think of $e^{-F}$ as a corrrelation function involving a boundary-condition-changing operator~\cite{francesco2012conformal} at ${r=0}$. In this language, Eq.~\ref{eq:Fboundary} comes from the Callan-Symanzik equation for this correlator).

Using the form of the coupling (Eq.~\ref{eq:renormalized-stiffness}) as ${L\rightarrow\infty}$, the relation in Eq.~\ref{eq:entropytofieldtheory} between the partition function ${Z=e^{-F}}$ and the entanglement entropy, and the value of $\tr a_n^Ta_n$ in Appendix~\ref{app:sec:bipartite-entanglement},
gives the $(\ln L)^2$ scaling stated in Eq.~\ref{eq:renyi-entropy-result}.
However if $N_F$ is large, then there will be a very large range of scales where the bare term in $1/g_R$ dominates, 
and in this intermediate range the entanglement entropies are proportional to~${N_F\ln L}$. 

\subsection{Numerical test of the entanglement scaling}
\label{sec:numerics}

\begin{figure*}
    \centering
    \includegraphics[width=0.48\linewidth]{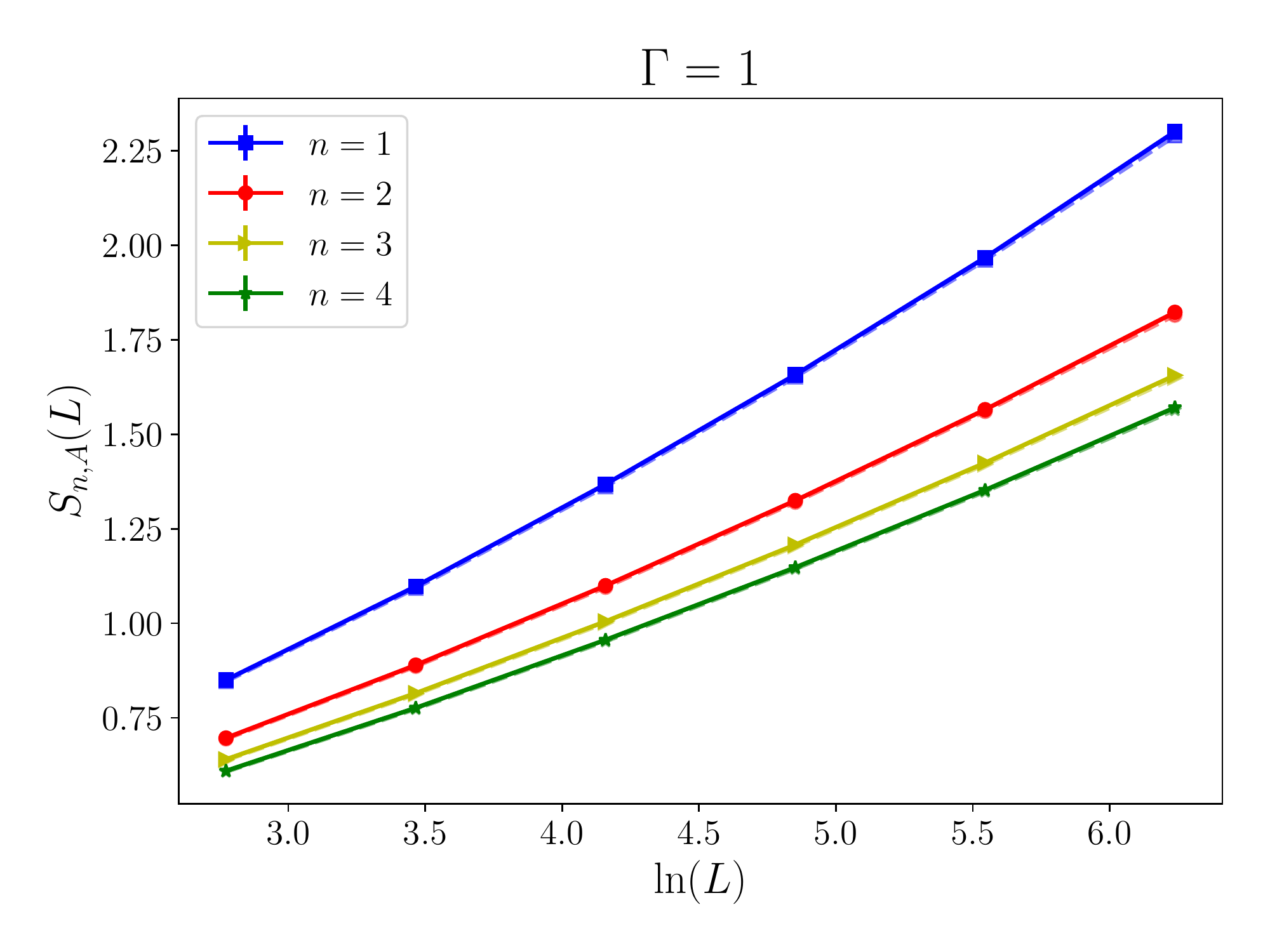}
    \includegraphics[width=0.48\linewidth]{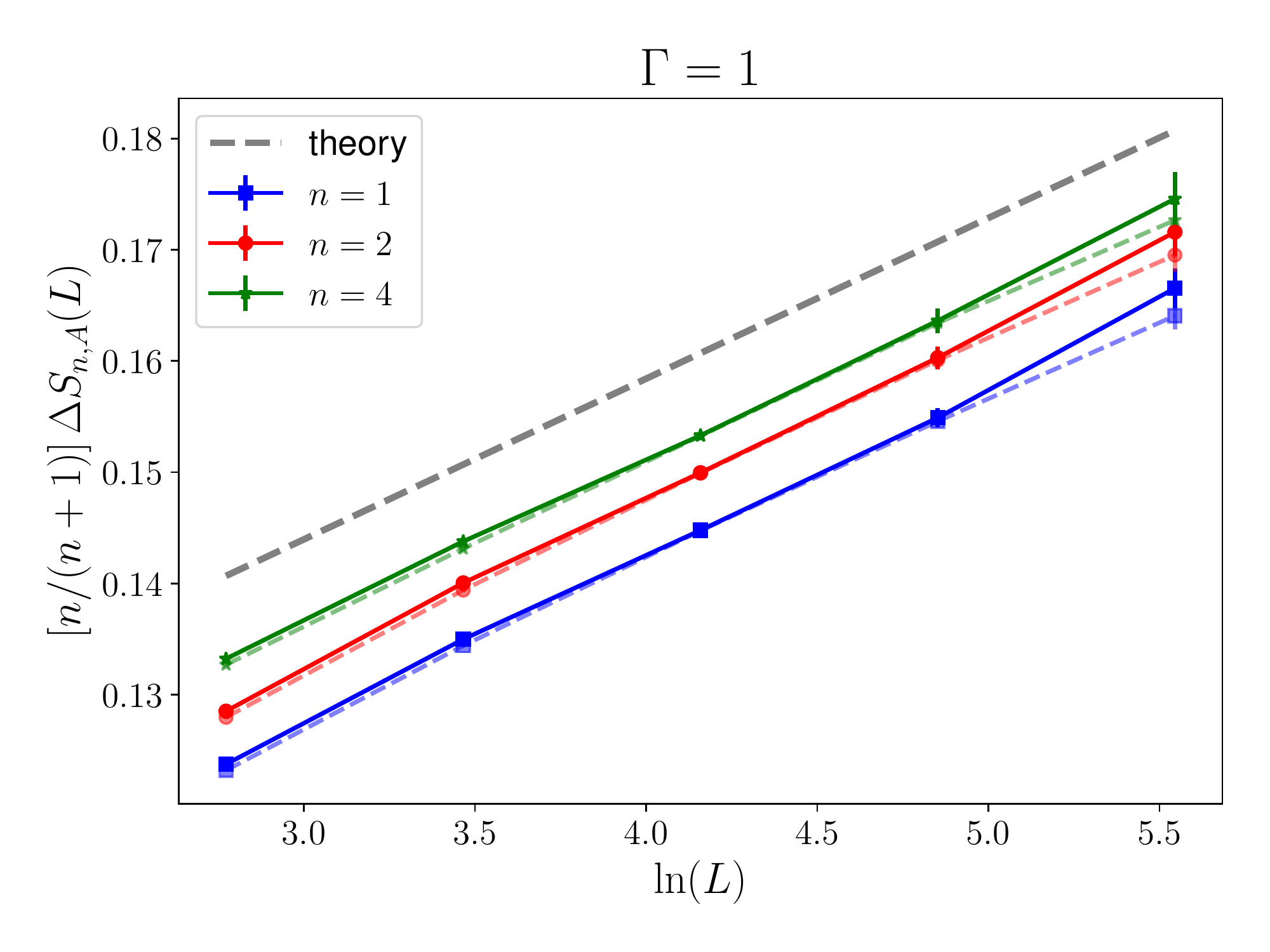}
    \caption{Simulations of Eq.~\ref{eq:quantum-state-diffusion} for $\Gamma=1$, $\Delta=0$,  $J=2$. For a given quantum trajectory we average $S_{n,A}$  over the time-interval $t\in[40,80]$, and we also average over more than $400$ independent quantum trajectories. The errorbar is the standard error over the set of distinct quantum trajectories.  Dashed faded lines and solid lines  are  data obtained with timestep ${\dd t = 5\cdot 10^{-3}}$ and ${\dd t = 2.5\cdot 10^{-3}}$ respectively. {\bf Left:} Steady-state entanglement $S_{n,A}(L)$ vs. $\log L$. {\bf Right:} $[n/(n+1)]\Delta S_{n,A}$ vs. $\log L$. The dashed gray line shows the predicted universal slope, Eq.~\ref{eq:theory-slope}.}
    \label{fig:entanglement-scaling-Gamma-1}
\end{figure*}

We now proceed to numerically test the theory and in particular Eq.~\ref{eq:renyi-entropy-result} for the bipartite entanglement.
For numerical convenience we restrict ourselves to $N_F=1$, so that larger system sizes can be reached.
For the purpose of the simulation it is convenient to re-express the dynamics as a quantum state-diffusion equation~\cite{caves1987quantum,diosi1998non,gisin1992quantum} 
\ba
\label{eq:quantum-state-diffusion}
    \dd\ket{\psi(t)} = & -i \, H(t)\dd t \ket{\psi(t)} \nn\\
    &+{\sum}_j \lf i\gamma_j \gamma_{j+1} - B_j(t) \ri \dd \xi_j \ket{\psi(t)} \nn\\
    &- \f{\dd t}{2} {\sum}_j \Gamma_j \lf i\gamma_j\gamma_{j+1} - B_{j}(t) \ri^2\ket{\psi(t)},
\end{align}
where $H(t)$ is the Hamiltonian generating the unitary part of the evolution,
\be
    H(t) = i \sum_{j=1}^{L-1} J_j(t) \gamma_j \gamma_{j+1},
\ee
(we use open { boundary} conditions for our numerics) and $B_j(t)$ depends on $\ket{\psi(t)}$ itself:
\be
    B_j(t) = \braket{\psi(t)|i\gamma_j \gamma_{j+1}|\psi(t)}.
\ee
The big advantage of this approach is that $\dd \xi_j$ is now simply the differential of a Brownian motion, i.e. $\overline{\dd \xi_j}=0$ and $\overline{(\dd \xi_j)^2} = \Gamma_j \dd t$, while higher-order cumulants vanish in the $\dd t\to 0$ limit. 
Note that therefore we do not need to sample measurement outcomes explicitly --- computing $B_j(t)$ at each time-step automatically takes Born's rule into account.

We discuss how Eq.~\ref{eq:quantum-state-diffusion} can be efficiently implemented for Gaussian states in Appendix~\ref{app:sec:numerics}. An important aspect is that Eq.~\ref{eq:quantum-state-diffusion} reproduces Born-rule sampling only in the $\dd t\to 0$ limit, whereas the numerical simulations are necessarily performed using a finite time step $\dd t$.  However, the deviation will tend to zero as $\dd t\to 0$. In the following we will show the convergence of our results w.r.t. $\dd t$.

Initializing the state in the vacuum state $\ket{\boldsymbol{0}}$ associated with the fermions $c_j=(\gamma_{2j-1}-i\gamma_{2j})/2$, we consider the evolution of the bipartite R\'enyi entropies  $S_{n,A}$ of the evolved state as a function of time --- see Appendix~\ref{app:sec:further-data} for a plot as a function of time. After $S_{n,A}$ plateaus at long-enough times, the prediction Eq.~\ref{eq:renyi-entropy-result} is expected to hold. 
Averaging over a time window after the plateau, and over quantum trajectories, we can study the dependence of $S_{n,A}$ on the system size $L$.

Fig.~\ref{fig:entanglement-scaling-Gamma-1} shows the result for $S_{n,A}$ at $\Gamma=1$ and $\Delta=0$ (we set $J=2$). 
In the left-hand panel we see that $S_{n,A}$ appears to have a positive curvature as a function of $\log L$. Eq.~\ref{eq:renyi-entropy-result} on its own is not a good fit, because subleading terms (neglected there) 
are still large at the  system sizes we have access to.
However, 
 taking the discrete logarithmic derivative
\be
    \Delta S_{n,A}(L) = S_{n,A}(2L) - S_{n,A}(L),
\ee
suppresses some of the contributions from short scales (for example a constant term would cancel out).
Eq.~\ref{eq:renyi-entropy-result} gives the scaling
\ba
\label{eq:theory-slope}
  \f{n}{n+1}\Delta S_{n,A} = \f{\ln 2}{48}  \ln L + o\left(\ln L\right).
\end{align}
Plotting the LHS against $\log L$ (Fig.~\ref{fig:entanglement-scaling-Gamma-1} right), we indeed see a linear scaling with a universal slope  compatible with that by predicted by the analysis in the previous sections.

\begin{figure}
    \centering
    \includegraphics[width=\linewidth]{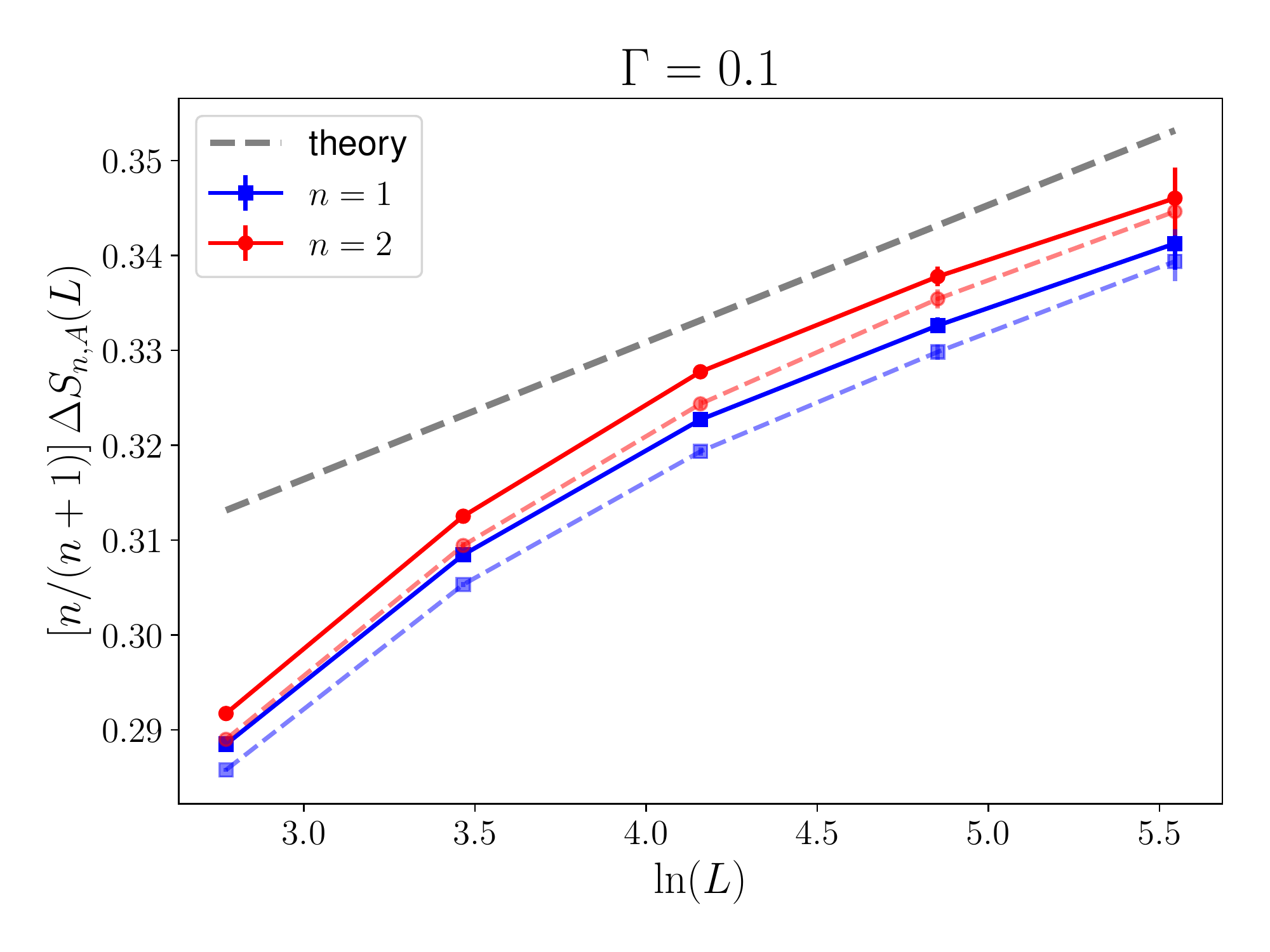}
    \caption{The analog of the Right panel of Fig.~\ref{fig:entanglement-scaling-Gamma-1}, but for the smaller measurement rate $\Gamma=0.1$ (with $\Delta=0$, $J=2$). Averages are performed in the time interval $t\in[100,150]$. See caption of Fig.~\ref{fig:entanglement-scaling-Gamma-1} for further details.}
    \label{fig:entanglement-scaling-Gamma-01}
\end{figure}

In Fig.~\ref{fig:entanglement-scaling-Gamma-01} we check that for a much smaller measurement rate, ${\Gamma=0.1}$, the slope of $\Delta S_{n,A}$ versus $\log L$ remains compatible with the predictions, though with larger finite-size effects, presumably due to crossover from the unitary behavior \cite{unitarymajorana}.
Finally, in Appendix~\ref{app:sec:further-data} we show data testing the univeral dependence on the R\'enyi index,  ${S_{n,A}\varpropto (n+1)/n}$, finding (at these sizes) agreement at the level of a few percent.

\section{Dimerization and vortices}
\label{sec:dimerization-and-vortices}

Having discussed the stable phase, it remains to analyze the transitions out of this phase (for $J>0$), and also the separate universal behavior in the measurement  only model (${J=0}$). We discuss the first of these issues here and the second in the next section.

To capture the transition into the area-law phase, which is a disordered phase for the $Q$ field, we need to take into account vortices (which proliferate in the disordered phase).\footnote{Recall that a vortex is a pointlike singularity in spacetime (instanton). Encircling the vortex, $Q$ describes a nontrivial path in ${\rm SO}(N)$ associated with the nontrivial element of $\pi_1({\rm SO}(N))=\mathbb{Z}_2$.
We may worry about how to make sense of a vortex in the replica limit: one way to make the idea more precise may be via an equivalent supersymmetric formulation of the NL$\sigma$M \cite{evers2008anderson}. 
Various kinds of topological defects arise in replica NL$\sigma$Ms including domain walls \cite{bocquet2000disordered,
gruzberg2005localization} and vortices \cite{fu2012topology,konig2012metal,nahum2013loop}.}
Vortices introduce an additional coupling in the (crudest approximation to the) RG flows, which is the vortex fugacity.
We begin by discussing the symmetry properties of this coupling.
In this section and the next, we draw heavily on the analogy with instantons in the XXZ spin chain and related systems \cite{auerbach1998interacting,affleck1986mass,senthil2006competing}.

Following a standard heuristic picture (see e.g. Ref.~\cite{Kosterlitz_1973}),
we may  imagine an expansion of the partition function in terms of the number $n$ of vortices. Schematically,
\be
\label{eq:partition-fun-with-vortices}
    Z \simeq \sum_n \frac{1}{n!} \sum_{x_1,\dots,x_n} y_{x_1}\dots y_{x_n} \int \dd t_1\dots \dd t_n \int \mathcal{D}Q\, e^{-S[Q]}.
\ee
The $k$th vortex is located at position $(x_k, t_k)$:
we take $x$ to be a lattice coordinate taking values $x\in \mathbb{Z}$.\footnote{Depending on microscopic considerations it may be more natural to think of $x$ as living on bonds of the lattice \cite{haldane19883}, or alternately on sites, but it will  will not matter which for the present discussion.}
Each vortex costs a fugacity $y_x$.
The magnitude of this fugacity   will be sensitive to the way the theory is cut off near the vortices
(this magnitude will be independent of $x$, by translation invariance, if $\Delta=0$).
However, the sign structure of $y_x$ carries universal information.
We argue that 
\be\label{eq:fugacityalternation}
y_x \propto (-1)^{x N_F}.
\ee
The same kind of alternation holds for vortices (instantons) in the easy-plane XXZ chain, with the sign factor being~$(-1)^{2 S x}$.
The minus sign arises from the Berry phase terms in the lattice action (for the  coherent states path integral)  
when we consider configurations with vortex singularities. 
The argument 
is of a standard kind and is given in Appendix~\ref{app:sec:vortices}.
Importantly,  the  argument applies independently of the microscopic structure  near the core of the vortex. 

Next (in a standard argument \cite{haldane19883}) we may imagine coarse-graining pairs of sites to define a coarse-grained vortex fugacity $y$. Heuristically,  ${y\propto y_x+y_{x+1}}$.
In the case where $N_F$ is odd, 
this leads to a cancellation.
The cancellation is perfect if 
$\Delta=0$, but if  $\Delta\neq 0$
(so that translational symmetry by one lattice spacing  is broken, and $y_x\neq y_{x+1}$)  
it is not perfect. At small $\Delta$ we therefore expect
\ba\label{eq:ydelta}
    y& = b \Delta + \ldots 
    & 
    & (N_F\text{ odd}),
\end{align}
where $b$ is a constant.
Conversely, for even $N_F$, the fugacity of a vortex is independent of its lattice position $j$. Therefore, in this case $y\neq 0$ even when $\Delta=0$, and by symmetry
\ba\label{eq:ydelta2}
    y =  c + d \Delta^2 + \ldots
    & 
    & (N_F\text{ even}).
\end{align}
(where the discussion in Sec.~\ref{sec:measurement-only} suggests that $c$ is negative and $d$ is positive for $N_F=2$).  The expansions above are for small $\Delta$ and do not rule out a nonmonotonic dependence at larger $\Delta$.

We see that in the $N_F$-odd models with $\Delta \neq 0$
and in the $N_F$-even models with generic $\Delta$, it is possible for the bare vortex fugacity to be nonzero.
In order to address the transition out of the stable nontrivial phase,  let us now consider an approximate RG involving $y$ and $g$.

For $N=1$ the critical fixed point is not at small $g$ or small $y$, preventing a controlled perturbative calculation.
However, the topology of the flows can be determined in a limit where  ${2-N}$ is treated as a small parameter, in the spirit of the epsilon expansion.
This was discussed for 
the $S^{N-1}$ sigma model
(the classical ferromagnet with ${\rm O}(N)$ symmetry)
near $N=2$ and $d=2$ in~\cite{PhysRevLett.45.499} and for  replica NL$\sigma$Ms in \cite{fu2012topology,nahum2013loop} (where sign effects for $y$ like those discussed above are relevant).
The key point is that in all these 
hierarchies of models, the point $N=2$ is an XY model, 
with the well-known Kosterlitz-Thouless RG equations in which $y=0$ is a fixed line. 
Assuming analyticity in $N$,
the beta functions can then be expanded in ${\varepsilon=2-N}$ \cite{PhysRevLett.45.499}:
\ba
\label{eq:vortices-RG-flow}\notag
    \f{\dd g_R}{\dd \log \ell} &=
    y^2_R + \varepsilon F(g_R) + O\lf y_R^4, \, \varepsilon^2\ri,\\
    \f{\dd y_R}{\dd \log \ell} &= (2-\pi g^{-1}_R) y_R + O\lf  \varepsilon y_R, \, y_R^3 \ri.
\end{align}
Here $F(g_R)$ is the derivative of the beta function for $g$ with respect to $\varepsilon$ at $\varepsilon=0$ and $y=0$.
At small $g_R$ it
is given by the perturbative result quoted in Eq.~\ref{eq:perturbativebetafunctionQtheory}. The values of $g_R$ of  interest now are not necessarily small, but for a qualitative picture of the flows what matters is the sign structure of 
$F(g_R)$ which we take to be as  suggested by the perturbative beta function \cite{PhysRevLett.45.499}.

\begin{figure}
    \centering
    \includegraphics[width=0.9\linewidth]{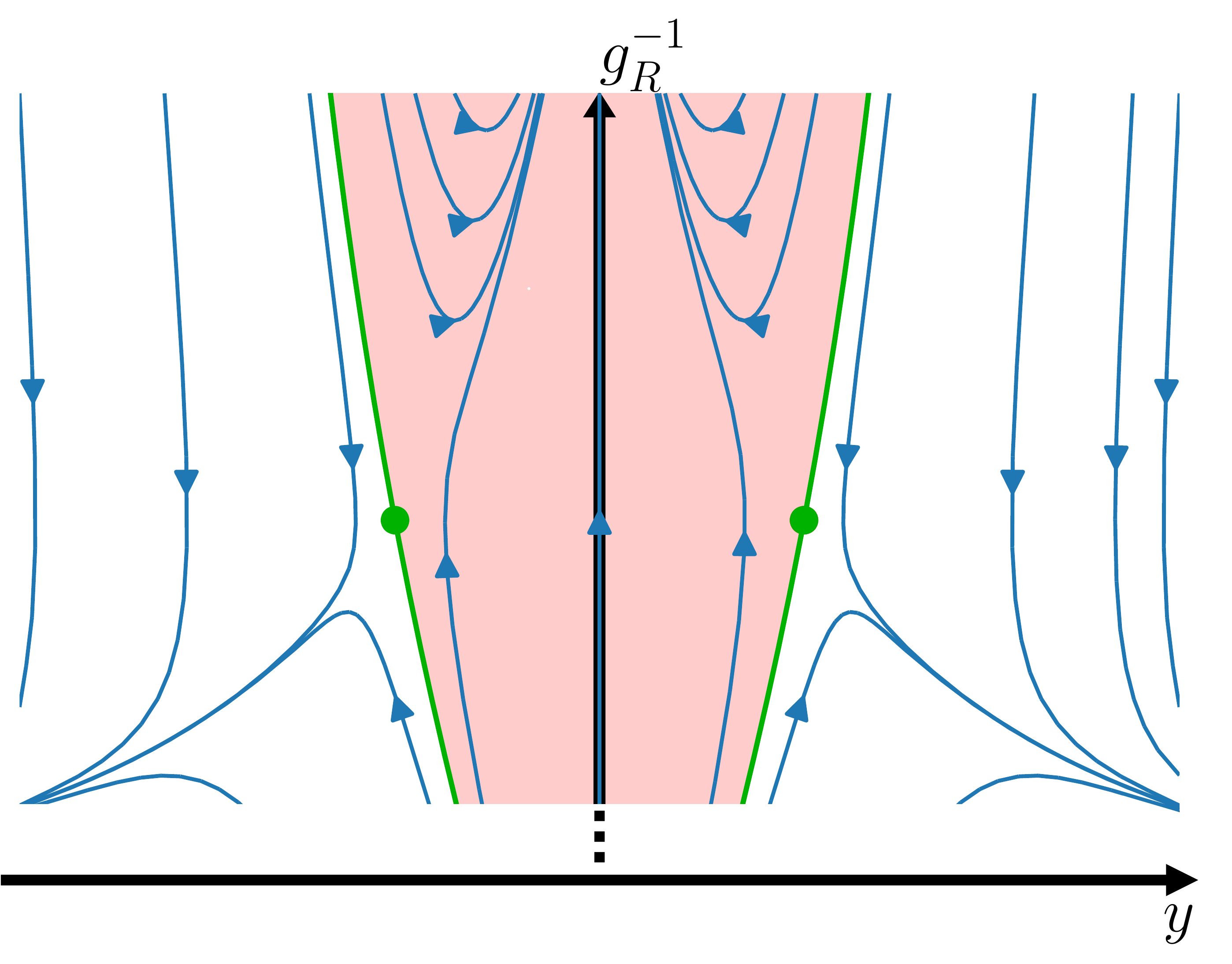}
    \caption{
    Schematic RG flow for $N=2-\varepsilon$, cf. Eq.~\ref{eq:vortices-RG-flow}. Two green points denote the fixed points (unstable in one direction) which control the transitions from the nontrivial phase
    (the pink region, which flows to $y=0$, $g_R^{-1}=\infty$) to the disordered phases (note that the lower boundary of this figure corresponds to an arbitrary value of $g_R^{-1}$, and does not correspond to the ${J=0}$ boundary of the fermion model phase diagram).}
    \label{fig:SON-RG-flow}
\end{figure}

Fig.~\ref{fig:SON-RG-flow} shows the schematic RG flow obtained from these equations. 
The structure of these flows matches the ``upper'' part of the phase diagram in Fig.~\ref{fig:phase-diag-sketch} (the limit $J\rightarrow 0$ will be discussed in the next section).
The axes are in heuristic correspondence between the two figures: recall from Eq.~\ref{eq:gandv} that $g_B^{-1}$ increases with $J^2/\Gamma$, and from Eq.~\ref{eq:ydelta} that the sign of $y$ is equal to that of $\Delta$ { (at least for small enough $\Delta$).}

For large $g_R^{-1}$, a small vortex fugacity $y$ is irrelevant and the couplings flow towards the 
stable fixed point at 
$(g^{-1}_R,y_R)=(+\infty,0)$ 
that we have already discussed. 
There are phases where $g_R^{-1}$ flows to small values and $y$ flows to large positive or negative values, which we identify with disordered phases. The analysis of   Sec.~\ref{sec:entanglement} makes clear that a disordered phase corresponds to an area-law entangled phase.

The transitions between the nontrivial and trivial phases are controlled by a pair of unstable fixed points (at positive and negative $y$), 
shown in red in Fig.~\ref{fig:SON-RG-flow}. 
When  ${N\to 2}$, these fixed points approach the $y=0$ line, 
giving  the conventional BKT flow diagram. 
For $N<2$ the flow structure is different, and for example the fixed points at finite $y$ have a finite value for the correlation length exponent $\nu$.\footnote{This exponent can be crudely estimated to leading order in $1/(2-N)$  using the perturbative expression for $F(g)$. However, the relevant coupling value $g\approx 2/\pi$ is not parametrically small.} 
By contrast, an ``annealed'' aproximation to the Majorana dynamics gives a Kosterlitz Thouless transition with $\nu=\infty$ \cite{bao2021symmetry}.

Finally, given that at this fixed point there are no marginal flows, we expect a model that is on the phase transition line to exhibit the entanglement scaling dictated by conformal invariance \cite{Calabrese_2009,vasseur2019entanglement}, i.e.  $S_{n,A}\varpropto \log L$, rather than $(\log L)^2$, cf. Sec.~\ref{sec:entanglement}.

\section{Measurement-only model (${J=0}$)}
\label{sec:measurement-only}

The ${{\rm so}(N)\times {\rm so}(N)}$ replica symmetry of the 
Majorana chain is generic, in the sense that it is the 
minimal continuous symmetry  shared by  any model of free fermions.\footnote{This includes models written in terms of complex fermions $c$ and $c^\dag$, which can always be rewritten in terms of Majoranas. However, charge conservation will increase the replica symmetry group further, and this will change the NL$\sigma$M description \cite{jian2022criticality}.}
 However, a larger ${\rm so}(2N)$ replica symmetry is present  both in the  measurement-only limit 
${\Gamma=0}$  and in the unitary limit ${J=0}$  \cite{bao2021symmetry}.
In this section we explore the consequences of this  symmetry for the measurement-only line.\footnote{The  unitary model is discussed in Ref.~\cite{unitarymajorana}.}

The feature responsible for the additional symmetry on the ${J=0}$ line is not the absence of unitaries, but rather a bipartite structure: 
by  examining the replicated Hamiltonian, one may check that ${\rm so}(2N)$ is present whenever it is possible to group the physical Majoranas into two subsets, $A$ and $B$, such that the unitary hopping only involves pairs $i\gamma \gamma$ from the same set, and measurements are only of pairs $i\gamma\gamma$ from opposite sets { (see endnote for examples\footnote{The ${J=0}$ case obeys this condition with the labelling ${\rm ABAB\cdots}$ as we go along the chain,
and the $\Gamma=0$ case obeys it with the labelling ${\rm AAAA\cdots}$, and a model studied in Ref.~\cite{turkeshi2021measurementinduced} obeys the condition with the labelling ${\rm ABBAABBA\cdots}$.}).}

In the present model, the symmetry is made manifest by a sign change 
$\gamma^{(\sigma a)}_{j,\mu} \mapsto (\sigma)^j \gamma^{(\sigma a)}_{j,\mu}$,
so that, in the block decomposition   of the local ${2N\times 2N}$ matrix degree of freedom $S_j$ (Eq.~\ref{eq:blocks}),
the off-diagonal blocks are redefined as $Q_j\mapsto (-1)^j Q_j$ while $L_j$ and $R_j$ are unaffected. The Hamiltonian is then
\be
    \Hs = \sum_j \f{\Gamma_j}{2} \Tr\left( S^T_j S_{j+1} \right).
\ee
$\mathrm{SO}(2N)$ transformations act on $S$ by conjugation.
This is an antiferromagnet.
For $N=2$ one can check 
that it reduces to the conventional $\mathrm{SU}(2)$ Heisenberg  chain with spin $S=2N_F$. We have allowed for dimerization in the measurement rates, ${\Gamma_j=\Gamma(1+(-1)^j\Delta)}$.

In the semiclassical limit, or in a coherent-states representation of the path integral,  $S$  
becomes an antisymmetric orthogonal matrix that lives in the manifold 
${{\rm SO}(2N)/{\rm U}(N)}$ 
(see  Sec.~\ref{subsec:path-integral-deg}).
In the model with ${J>0}$ discussed in Sec.~\ref{sec:mapping-to-NLSM}, the
 Hamiltonian favored ordered states in which only $Q$ was nonzero,
 and we integrated out $L$ and $R$ to obtain the theory for $Q$ only. 
Here 
the larger symmetry means 
we must retain the full target space for the low-energy theory, 
with is therefore an NL$\sigma$M on ${{\rm SO}(2N)/{\rm U}(N)}$~\cite{senthil2000quasiparticle,
read2000paired,bocquet2000disordered}.

By symmetry,
the  Lagrangian for such a theory is expected to take the standard form
with two couplings, namely the inverse stiffness  $g$ and a $\Theta$ angle.
The bulk physics is invariant under ${\Theta\rightarrow \Theta+2\pi}$.\footnote{Recall that a $\Theta$ term can be added to a two-dimensional NL$\sigma$M whenever the target space $\mathcal{M}$ has $\pi_2(\mathcal{M})=\mathbb{Z}$ \cite{altland2010condensed}, as in the present case. The $\Theta$ term weights the partition sum by $e^{i\Theta n}$, where for a closed system $n$ is a (signed) integer that counts the number of skyrmion textures in the configuration.}
The values of these couplings should in principle be obtained at large $N_F$ 
by taking a continuum limit of the coherent states path integral  \cite{auerbach1998interacting,altland2010condensed}. 
We do not perform this calculation here,
instead conjecturing the  form of $\Theta$ 
on the basis of the ${N=2}$ special case.
 The latter reduces to the   ${\rm SU}(2)$ chain with spin $N_F/2$, for which the NL$\sigma$M mapping at large spin is well known and gives the sphere sigma model with a $\Theta$ term.
 
The simplest assumption that matches with this case 
is that  ${\Theta=N_F \pi}$
in the undimerized model, 
with   ${\Theta=N_F \lf \pi  + O(\Delta) \ri }$ for weak dimerization $\Delta$. 
Finally  if $\Delta$ is varied all the way from $-1$ to $1$ the topological angle varies monotonically from $0$ to $2\pi N_F$ \cite{affleck1987critical}.

The ${{\rm SO}(2N)/{\rm U}(N)}$ model with ${N\to 0}$
arises as a description of Anderson localization problems in symmetry class D \cite{senthil2000quasiparticle,
read2000paired,bocquet2000disordered,chalker2001thermal,read2000absence,gruzberg2001random,gruzberg2005localization,mildenberger2007density,cho1997criticality,merz2002two}. 
Localization models in this class have a rich phase diagram  with various kinds of transition \cite{evers2008anderson}: 
in addition to a nontrivial dependence on the $\Theta$ angle
 there is a stable metallic phase when ${N\to 0}$
(also, two slightly different models arise\footnote{On  ${{\rm SO}(2N)/{\rm U}(N)}$  or on  ${{\rm O}(2N)/{\rm U}(N)}$ \cite{evers2008anderson}.}).
Here we expect a simpler phase diagram, in particular because the
perturbative beta function  \cite{evers2008anderson} shows that unlike the ${N\to 0}$ case the ${N\to 1}$ case does not have  a metallic phase.

Instead, the simplest conjecture, 
given the instability of the ${g=0}$ fixed point,
is that the flows in the $(\Theta, g)$ plane at ${N\to 1}$ resemble those of 
various other replica NL$\sigma$Ms, including the Pruisken sigma model for the Integer Quantum Hall transition (a celebrated example of the effect of a $\Theta$ term  
\cite{pruisken1984localization,evers2008anderson}).
If so, then  the theory is gapped for all $g$, 
except on the lines with  
${\Theta=\pi\,\text{mod } 2\pi}$
(which are preserved under RG, by parity symmetry).
On these lines the model flows, 
for any initial $g$, 
 to a critical fixed point
 that is unstable in the $\Theta$ direction.

 At ${N_F=1}$ this leads to a single critical point on the ${J=0}$ axis, at ${\Delta=0}$. { This is consistent with the numerical results of Ref.~\cite{kells2021topological}. 
In general it leads} to $N_F$ critical points on this axis, as varying $\Delta$ causes the $\Theta$ angle to cycle $N_F$ times through the value ${\pi\,\operatorname{mod }2\pi}$ 
 (in analogy to the spin-$N_F/2$ chain \cite{affleck1987critical,oshikawa1992hidden,
kitazawa1996phase,
kitazawa1997phase}).
 
 We also conjecture that the ${\Theta=\pi}$ fixed point is unstable to the symmetry-breaking perturbation induced by turning on $J$.
 The considerations above then yield the phase diagram structure for small $J$ that was proposed in Sec.~\ref{sec:firstlookatphasediagram}.\footnote{In a semiclassical limit, the   the vortex fugacity in the resulting reduced-symmetry model is proportional to $y\propto \cos(\Theta/2)$ \cite{affleck1986mass,senthil2006competing}, with a positive prefactor, consistent with Eqs.~\ref{eq:ydelta},~\ref{eq:ydelta2} (with $c<0$ and $d>0$).}
However, we emphasize that these are conjectures: in particular it is not ruled out that the stability properties of the $\Theta=\pi$ fixed point could be different (we will return to this in the future).

Finally
let us briefly note the analogy between the RG flows described above and those in the Majorana loop model  \cite{nahum2020entanglement, merritt2022entanglement,sang2021entanglement,lavasani2022monitored}  (see Sec.~\ref{sec:firstlookatphasediagram}).
The Majorana loop model has a measurement-only line  described by the $\mathbb{CP}^{M-1}$ NL$\sigma$M  
with a  $\Theta$ term in the limit ${M\to 1}$  \cite{read2001exact,candu2010universality}. If the measurements are un-dimerized, then ${\Theta}$ is equal to  ${\pi}$, and this is a critical point between two disentangled phases.
Adding SWAP unitaries gives a symmetry-breaking perturbation in the NL$\sigma$M, which induces a flow to the $\mathbb{RP}^{M-1}$ NL$\sigma$M, which has a metallic phase \cite{nahum2013loop,jacobsen2003dense}. 
The $\mathbb{RP}^{M-1}$ NL$\sigma$M allows $\mathbb{Z}_2$ vortex defects, and the sign of the vortex fugacity is inherited from the sign of ${\Theta-\pi}$ in the parent $\mathbb{CP}^{M-1}$ model. These vortices can drive a phase  transition to a trivial phase.
This structure of flows is similar to that proposed above, with the $\mathrm{SO}(2N)/\mathrm{U}(N)$ NL$\sigma$M at $\Theta=\pi$ playing the role of the $\mathbb{CP}^{M-1}$ NL$\sigma$M, and the $\mathrm{SO}(N)$ NL$\sigma$M playing the role of the $\mathbb{RP}^{M-1}$ NL$\sigma$M. However, there are also basic differences between the two kinds of measurement model, cf. Sec.~\ref{sec:majoranaloopphasediagramcomment}.

\section{Conclusions}
\label{sec:conclusions}

In the context of a simple Majorana Hamiltonian,
we have argued that monitored free fermions 
give rise to problems in critical phenomena
 that can be viewed formally as sitting (at replica number ``$N$'' equal to 1) 
 in between two well-studied classes of critical points: 
zero-temperature phase transitions in spin chains, described by NL$\sigma$Ms 
 with values of ``$N$'' that are greater than one, 
 and Anderson transitions in eigenstates of disordered Hamiltonians, 
 described by the limit ${N\to 0}$.
Our aim in this paper has been to give an analytically controlled derivation of an effective field theory for a generic model and to use it to analyze the entanglement properties of the phases and transitions.

Concretely, our approach was to map the generator of the dynamics for moments of  
the density matrix (prior to normalization) to an ${\rm so}(2N)$ spin chain, which at large $N_F$ is in a semiclassical limit, allowing a controlled reduction to a long-wavelength theory.

The nonlinear sigma model was particularly useful in the stable nontrivial phase, where the calculation of the entanglement is asymptotically exact, and agrees with with numerical simulations made with the quantum-state-diffusion method. 
The calculation illustrates that the nonlinear sigma model stiffness $1/g$ 
(analogous to a conductivity in a localization problem)
functions as a scale-dependent strength of entanglement. 
The NL$\sigma$M also gives a picture for the structure of critical points,
via an ${\epsilon=2-N}$ expansion in the generic case, or via transitions between 
different $\Theta$-vacua in Majorana models with a certain bipartite structure for interactions and measurements,
in particular in a model in which the dynamics involves only noncommuting measurements~\cite{PhysRevX.11.011030,nahum2020entanglement}.
This structure of RG flows for these sigma models explains the previously observed~\cite{merritt2022entanglement} similarity between the phase diagram of the Majorana loop model and more generic models, 
despite these problems being in distinct universality classes.

We emphasize that various kinds of connection between either free monitored  \cite{buchold2021effective}
or free non-Hermitian  \cite{jian2022criticality,2021Quant...5..579Z} dynamics and replica Lagrangians have been developed previously.
Ref.~\cite{buchold2021effective} mapped a regime of monitored Dirac fermions in the spatial continuum onto a non-Hermitian Sine-Gordon theory. 
A Majorana model closer to ours was considered in Ref.~\cite{2021Quant...5..579Z} in the context of  the SYK$_2$ model, with a non-Hermitian Hamiltonian rather than measurements.
There a different regime was considered, where fluctuations were suppressed, so that the effective description was different.
Most relevant to our work is Ref.~\cite{jian2022criticality}, 
which  provided a symmetry classification of Gaussian random tensor networks (observing the  connection with nonunitary time evolution operators) and proposed the DIII sigma model with ${N\to 0}$ as a description of generic Gaussian tensor networks on grounds of symmetry.

The  relevant limit of the target space for NL$\sigma$Ms for monitored systems is $N\to 1$.
It will be interesting to explore the RG fixed points \cite{fendley2001critical} relevant to these ${N\to 1}$ sigma models { (with various symmetries)}, using simulations with Born's rule or with the state diffusion formalism to compute exponents and flow diagrams, and to map out the relation to  different monitored models \cite{cao2019entanglement,PhysRevLett.124.040401, PhysRevB.106.035408, PhysRevLett.109.090404, PhysRevA.86.044103, PhysRevX.11.021037}. 
{ For the specific models studied here, it also remains to perform a fuller numerical analysis of the phase diagrams in order to check the conjectured structure for general~$N_F$. }

In this work we have focussed on a simple case where the Hamiltonian couplings fluctuate as  white noise, since this led to a particularly simple effective model. In the future it will be interesting to study models where the hopping amplitudes either have a deterministic part (nonzero mean), 
or are nonrandom (so that the randomness is only from measurements, as for example in \cite{buchold2021effective}). In these cases the model does not reduce microscopically to a purely bosonic model. However, one could employ tools from Anderson localization to derive NL$\sigma$Ms in the long-wavelength limit \cite{lerner2003nonlinear,altland2010condensed}.\\

\noindent \emph{Note added}: While finalizing this manuscript, a closely related work appeared on the arXiv~\cite{jian2023measurement}, studying monitored free fermions dynamics with the same symmetry, albeit using a different microscopic model.

\acknowledgements

We thank Beno\^it Dou\c{c}ot and Andrea De Luca for useful discussions. MF thanks Sounak Biswas for clarifying some aspects of jackknife resampling.
TS was supported by a James Buckee Scholarship.
DB was  supported in part by CNRS, by the ENS, and by the ANR project “ESQuisses”, contract number ANR-20-CE47-0014-01.
AN was supported by CNRS and the ENS.

\appendix
\section{Weak measurements, continuum limit, and averaging}
\label{app:sec:discrete-time}

In this appendix we derive the effective non-Hermitian Hamiltonian~\eqref{eq:HamiltonianNH} { as the continuum limit of} a discrete time-evolution alternating unitary evolution and weak measurements. { This discretization gives a more precise justification of} the scheme to compute averages presented in the main text.

We start from unitary dynamics of Majorana fermions with the  time-dependent Hamiltonian~\eqref{eq:Hamiltonian}.
It is useful to think of the dynamics generated by~\eqref{eq:Hamiltonian} as the continuous limit of the discrete process
\begin{align}
U(t)&=U(t_n)U(t_{n-1})\ldots U(t_1)\nonumber\\
&=e^{-i \Delta t H(t_{n})} e^{-i \Delta t H(t_{n-1})} \cdots e^{-i \Delta t H(t_{1})}\,,
\label{eq:discrete_steps}
\end{align}
where $\Delta t=t/n$ and $t_j=j\Delta t$, while the delta function in Eq.~\ref{eq:variance} is regularized as
\begin{equation}\label{eq:variance_discrete}
	\mathbb{E}_G[J_j^{\mu_1\nu_1}(t_r)J_k^{\mu_2\nu_2}(t_s)]=\frac{J^2}{N_F \Delta t}\delta_{r,s}\delta_{\mu_1,\mu_2}\delta_{\nu_1,\nu_2}\,.
\end{equation}

To this we add  measurements. After each time step we make weak measurements of all the fermion parity operators of the form $i\gamma_{j,\mu} \gamma_{j+1,\nu}$ for adjacent sites, with the strength of the measurements approaching zero as ${\Delta t \rightarrow 0}$. 
For a given { bond} $j$ and pair of flavors $(\mu,\nu)$, we may describe this process by a continuous family of Kraus operators $\hat k (M_{j}^{\mu\nu})$ indexed by a real number $M^{\mu\nu}_j$
\be
    \hat k (M_{j}^{\mu\nu})  = 
    \mathcal{N}_{\Gamma_j}
    \exp
    \lf    i \Delta t M^{\mu\nu}_j \gamma_{j,\mu} \gamma_{j+1,\nu} \ri,
\ee
and a measure $\dd \mu_G$ over $M^{\mu\nu}_j$, which we choose to be a normalized Gaussian measure with mean $0$ and variance $\Gamma_j / (N_F \Delta t)$, viz. ${\dd \mu_G(M^{\mu\nu}_j)\propto e^{- (M^{\mu\nu}_j)^2 \Delta t N_F /(2 \Gamma_j)} \dd H}$.
The normalization factor $\mathcal{N}$ is chosen such that the Kraus set is properly normalized, i.e.%
\footnote{At the level of the normalization condition, there is an ambiguity as for any function $f(M)$ we can redefine the measure as $\mu(M)\mapsto f(M) \mu(M)$ and the operators as $\hat k(M) \mapsto \hat k(M)/\sqrt{f(M)}$, while preserving the normalization condition. This ambiguity can be resolved e.g. by specifying a discrete Kraus set approaching the continuous one in the limit of dense measurement outcome. Then, the measure $\mu$ is determined by the density of the of the possible measurement outcomes $M$. In practice, this ambiguity affects the definition of average evolution operators for $N>1$ replicas, but is anyway resolved when taking the $N\to 1$ replica limit.}
\begin{equation}
    \int_{-\infty}^{+\infty} \dd \mu_G(M^{\mu\nu}_j)  
    \,
    \hat k(M^{\mu\nu}_j)^\dag \hat k(M^{\mu\nu}_j)
    = \mathds{1}.
\end{equation}
The condition above arises by requiring that, averaging over the measurement outcomes, the quantum channel obtained is trace-preserving.
From the normalization condition we can explicitly find $\mathcal{N}_{\Gamma_j}= e^{-\Gamma_j \Delta t N_F^{-1}}$.

The overall measurement process across the whole chain can then be described as a two step process, where first all odd bonds are measured and then all even ones are measured, viz.
\begin{equation}\label{eq:weak_measurements}
	\tilde{K}(\{M^{\mu\nu}_j \})=
    \prod_{j} \hat{k}(M_{2j}^{\mu\nu})
    \prod_{j} \hat{k}(M_{2j+1}^{\mu\nu}).
\end{equation}

A set of couplings $J_j^{\mu\nu}(t_k)$ and measurement outcomes $M^{\mu\nu}_j(t_k)$ defines a quantum trajectory. Putting everything together we have that in a specific quantum trajectory, for a given time-step, the unnormalized density matrix of the system evolves according to
\ba
    &\check{\rho}_{J,M}(t_k+\Delta t) \nonumber\\
    &~~=\tilde{K}(\{M^{\mu\nu}_j(t_k) \}) U(t_k)
    \check{\rho}_{J,M}(t_k)
    U^\dag(t_k) \tilde{K}^\dag(\{M^{\mu\nu}_j(t_k) \}),
\end{align}
in terms of which the normalized density matrix is
\be
    \rho_{J,M}(t) = \f{\check{\rho}_{J,M}(t)}{\Tr \check{\rho}_{J,M}(t)}\,,
\ee
which can be equivalently expressed in the form of Eq.~\ref{eq:defnrhounnormalized} with
\be
    K_{J,M}(t_n) =  \prod_{k=1}^n \left[ \tilde{K}(\{M^{\mu\nu}_j(t_k) \}) U(t_k) \right]\,.
\ee

In the limit $\Delta t\to 0$, $K_{J,M}$ is given by \eqref{eq:timeorderedexp} with
\ba
    H_{J,M}(t) &=  \sum_{j=1}^L \sum_{\mu, \nu=1}^{N_F} \left[
    J_j^{\mu\nu}(t) 
    +  i M_j^{\mu\nu}(t)   \right]
    \, i \gamma_{j,\mu} \gamma_{j+1,\nu}\nonumber\\
    &- N_F\sum_j \Gamma_j,
\end{align}
which, except for the additive constant arising from the normalization $\mathcal{N}_{\Gamma_j}$, agrees with Eq.~\ref{eq:HamiltonianNH}. The fact that this constant has been neglected has been adjusted through the denominator of Eq.~\ref{eq:expvalwithBorn1}, since if we were to neglect $\mathcal{N}$, as done in the main text, we would ultimately obtain
\be
    \mathbb{E}_G[\check\rho_{J,M}(t)] = \exp(2 N_F t \sum_j \Gamma_j).
\ee
We explicitly checked this in the long-time limit by computing the energy of the classical ground state polarized along $Q$ and seeing that it is $2\sum_j \Gamma_j$ for $N\to1$.%
\footnote{This requires keeping track of the constants neglected in the main text in going from Eq.~\ref{eq:H_N-spin} to Eq.~\ref{eq:Heff-first-expr}.}

Finally, the most elementary objects we are interested in computing are averages of replicas of $\rho^{\otimes n}(t)$. [As the tensor product of fermionic Fock spaces is ill-defined, here we have in mind that the fermionic Fock space is first mapped to a bosonic one, see Appendix~\ref{app:sec:replica-majorana}.]
According to Born rule, these are given by
\ba
    \mathbb{E}[\rho^{\otimes n}(t)] &= \int \dd\mu_{G}(J,M) \f{\check{\rho}^{\otimes n}_{J,M}(t)}{\left(\Tr \check\rho_{J,M}(t)\right)^{n}} \Tr \check\rho_{J,M}(t)=\nn\\
    &=\mathbb{E}_G\left[ \check{\rho}^{\otimes n}_{J,M}(t) \left(\Tr \check\rho_{J,M}(t)\right)^{1-n} \right]    
\end{align}
where $\check{\rho}$ is the unnormalized density matrix along one trajectory~\eqref{eq:defnrhounnormalized}, and, with a slight abuse of notation, we are writing $\dd\mu_{G}(J,M)$ to denote the product of the various Gaussian measures at different positions and time-steps.
Here the last factor of $\Tr \check\rho_{J,M}$ ensures that quantum trajectory are chosen according to Born's rule.
The key of the replica trick lies in the observation that the average over $J$ and $M$ can be most easily performed by computing
\be
\label{app:eq:replica-basic}
    \mathbb{E}_G\left[\check{\rho}^{\otimes n}_{J,M}(t) \left(\Tr \check\rho_{J,M}(t)\right)^{N-n}\right]
\ee
for $N\geq n$ and finally performing an analytic continuation to $N\to 1$. In this approach the fundamental object is $\rho^{(N)}= \mathbb{E}_G \left[ \check{\rho}_{J,M}^{\otimes N}(t)\right]$ (Eq.~\ref{eq:Nreplicastate}),
which through appropriate contractions can yield all terms of the form~\ref{app:eq:replica-basic}.

Averages are most simply computed in the ``folded'' respresentation  (via the Choi–Jamiołkowski isomorphism). Fixing a basis $\{\ket{k}\}$ of the Hilbert space $H$, we can identify any operator
$\hat{O}$ with a state $\ket{O}$  in $H\otimes H$ via
\be
\label{app:eq:operator-to-state}
    \ket{O} = \sum_{k,l} \braket{k|O|l} \ket{k} \otimes \ket{l}.
\ee
In this formalism
\ba
\ket{\rho^{(N)}(t)} &= \mathbb{E}_G\left[(K_{J,M} \otimes K^*_{J,M})^{\otimes N} \right] \ket{\rho^{\otimes N}(0)}.
\end{align}
At this stage the average can simply be taken using the cumulant expansion at each time step.
It is straightforward to see that this yields result consistent with the averaging rules given in the main text Eqs.~\ref{eq:variance} and~\ref{eq:varianceM}. However, in order to write an explicit expression for $K^{(N)}$, we need to  define the tensor product by mapping to a bosonic  Fock space.

\section{Replica formalism for Majorana fermions}
\label{app:sec:replica-majorana}

The convention that we used to define replicated Majorana operators is an extension of the one used in Ref.~\cite{sunderhauf2019quantum}, which for the reader's convenience we summarize here. For simplicity, in our treatment, we will assume that $L N_F$ is a multiple of $8$.

While the tensor product of fermionic Fock spaces is not well-defined, due to the anticommutation relations, we can  construct a bosonic Hilbert space by means of a Jordan-Wigner transformation. For this purpose, we impose some ordering on the indices $(j,\mu)$ of the Majoranas operators and  label them with a single index $k$, e.g. by defining $j_k = \lfloor (k-1)/N_F \rfloor + 1$ and $\mu_k-1 = (k-1)\,(\text{mod }N_F)$. We can then identify the $L N_F$ Majorana fermions with spin operators of a spin-$1/2$ chain with $L N_F/2$ sites:
\begin{align}
\label{app:eq:jordan-wigner}
    \gamma_{2k+1} &= \left(\prod_{k'<k} Z_{k'} \right) X_k,
    &
    \gamma_{2k+2} &= \left(\prod_{k'<k} Z_{k'} \right) Y_k.
\end{align}
In terms of the bosonic (spin) Hilbert space $\mathcal{H}_B$, tensor products are well defined and we can consider replicas $\mathcal{H}_{2N} = (\mathcal{H}_B)^{\otimes 2N}$ and Pauli operators acting on them
\ba
    V^{(+a)}_k &= \mathbb{I}^{\otimes 2 a} \otimes V_k \otimes  \mathbb{I}^{\otimes 2 N - 2 a - 1}\\
    V^{(-a)}_k &= \mathbb{I}^{\otimes 2 a+1} \otimes V_k \otimes  \mathbb{I}^{\otimes 2 (N-a-1)}
\end{align}
with $V= X,\,Y,\,Z$ and $\mathbb{I}$ denoting the identity on the space $\mathcal{H}_B$.
We then proceed to define replicated Majoranas operators in the extended space. As an intermediate step, we introduce the operators $\chi_{k}^{\sigma a}$, where $a$ ($1<a<N$) and $\sigma=\pm$ label the different replicas:
\ba
    \chi^{(+a)}_k &= \mathbb{I}^{\otimes 2 a} \otimes \gamma_k \otimes  \mathbb{I}^{\otimes 2 N - 2 a - 1}\\
    \chi^{(-a)}_k &= \mathbb{I}^{\otimes 2 a+1} \otimes \gamma_k^* \otimes  \mathbb{I}^{\otimes 2 (N-a-1)}
\end{align}
Here $\gamma_k$ denote the Majorana operators, which act on the bosonic Hilbert space according to Eq.~\ref{app:eq:jordan-wigner}. Finally $\gamma^*_k$ denotes the complex conjugate of $\gamma_k$ w.r.t. some basis --- the actual basis choice is inconsequential.

Note that the operators $\chi_{k}^{(\sigma a)}$ cannot be interpreted as Majorana operators in the enlarged space, as $\chi$-operators acting on different replicas commute. This can however be fixed, by adding a Klein factor among different replicas. We define the Klein factors as
\begin{equation}
    F^{(\sigma a)} = \prod_{k=1}^{L N_F} \chi_{k}^{(\sigma a)} = \prod_{k=1}^{L N_F/2} Z_k^{(\sigma a)},
\end{equation}
and the replicated Majoranas
\begin{align}
    \gamma^{(+a)}_k &= i \left( \prod_{a'<a} F^{(+a')} F^{(-a')} \right) F^{(+a)} \chi^{(+a)}_k\\
    \gamma^{(-a)}_k &= \left( \prod_{a'<a} F^{(+a')} F^{(-a')} \right) F^{(+a)} \chi^{(-a)}_k.
\end{align}
One can then verify that the operators $\gamma^{(\alpha)}_k$ with $\alpha=\sigma a$ indeed form  a set of Majorana operators, i.e. ${(\gamma^{(\alpha)}_k)^\dag = \gamma^{(\alpha)}_k}$ and ${\{ \gamma^{(\alpha)}_k, \gamma^{(\alpha')}_{k'}\}}={\delta_{k,k'} \delta_{\alpha,\alpha'}}$.
The specific choice of Majorana operators above turns out to be very convenient both to compute averages of $(K\otimes K^*)^N$, as we now discuss, and to  characterize the boundary states needed to  write the entanglement entropies (see Appendix~\ref{app:sec:boundary-stabilizers}).

We are now set to compute the average evolution for $N$ replicas $\mathbb{E}_G\left[(K_{J,M} \otimes K^*_{J,M})^{\otimes N} \right]$.
We can start by observing that in the replicated space the time-evolution along a quantum trajectory can be expressed in terms of a new continuum Hamiltonian playing a role analogous to~\eqref{eq:HamiltonianNH}
\ba
     (K_{J,M}\otimes K_{J,M}^*)^{\otimes N} & = \mathcal{T}
     \exp \lf   - i
     \int^t_0 \dd t' H_{J,M}^{(N)}(t')
     \ri.
\end{align}
In terms of the $\chi$-operators this takes the form
\ba
H_{J,M}^{(N)}(t) & =  i \sum_{j}\sum_{\sigma=\pm} \left[ J_j^{\mu\nu}(t) + i \sigma  M_j^{\mu\nu}(t)  \right]  \chi_{j,\mu}^{(\sigma a)} \chi_{j+1,\nu}^{(\sigma a)}.
\end{align}
Note that $\sigma$ multiplies { $M_j^{\mu\nu}(t)$} and not $J_j^{\mu\nu}(t)$, viz. the part of the Hamiltonian produced by weak measurements and not the one produced by unitary evolution. 
This is unlike what happens in replicas of real bosonic Hamiltonians, and it happens because the complex conjugation of the Majorana operators is absorbed in the definition of the operators $\chi$. 
(However, it is possible to move the $\sigma$ from one term to the other in the replicated Hamiltonian Eq.~\ref{eq:H-JM-many-copies} by a sign redefinition of $\gamma$ operators on a sublattice.)
By working separately on the case $\sigma=\pm$, one can then show that $\chi_{j,\mu}^{(\sigma a)} \chi_{j+1,\nu}^{(\sigma a)} = \gamma_{j,\mu}^{(\sigma a)} \gamma_{j+1,\nu}^{(\sigma a)}$, therefore obtaining Eq.~\ref{eq:H-JM-many-copies}.
Finally, integrating over $J$ and $M$, as explained in the previous Appendix, and using the cumulant expansion we arrive at Eqs.~\ref{eq:replicaevolutionbeforeaverage} and~\ref{eq:Heff-first-expr}.

\section{Boundary states and their stabilizers}
\label{app:sec:boundary-stabilizers}

In this section we introduce the boundary states relevant for the study of entanglement  entropies reported in the main text. Finally, we characterize them in terms of quadratic Majorana stabilizers. More of their properties, related to the ${\rm SO}(2N)$ irrep they belong to, are reported in Appendix~\ref{app:sec:boundary-state-irreps}.

We begin with the simplest one, corresponding to the initial density matrix, which we take to be
\be
	\rho(t=0) = 2^{- L N_F/2} \mathbb{I}.
\ee
After taking $N$ replicas $\rho^{\otimes N}(t=0)$ and mapping operators to states of a doubled Hilbert space (see Eq.~\ref{app:eq:operator-to-state}), this gives us as initial state  $\ket{\mathbb{I}}$, which can be defined in terms of its bosonic stabilizers.
The identity operator satisfies $V_k \mathbb{I} V_k = \mathbb{I}$ where $V=X,\,Y,\,Z$ and $k$ labels sites $j$ and flavors $\mu$. This then translates into
\be
    V_k^{(+a)} (V_k^{(-a)})^T \ket{\mathbb{I}} = \ket{\mathbb{I}}.
\ee
This implies that
\be
\label{eq:stabilizers-rho0}
    i \gamma_k^{(+a)} \gamma_k^{(-a)} \ket{\mathbb{I}} = \ket{\mathbb{I}},
\ee
as can be verified by rewriting the $\gamma$-operators in terms of Pauli matrices, and using that for Pauli matrices $V^T=V^*$.

The relevant boundary states at final time can be characterized in a similar fashion. For concreteness, we discuss the boundary state for the computation of
\be
	e^{-k(1-n)S_{n,A}} = \left(\Tr_A \rho_A^{\otimes n}(t)\right)^k,
\ee
where $A$ includes the sites from $j=\ell+1$ to $j=L$, and $\rho_A(t)=\Tr_{\bar A} \rho(t)$ is obtained by tracing the density matrix over $\bar A$ --- the complement of $A$. [Note that the computation of the entropy of the overall state is obtained by setting $\ell=0$.]
The computation above can be written as
\be
	e^{-k(1-n)S_{n,A}} = \Tr\left( \rho^{\otimes nk}(t) \mathcal{C}_{A,n}^{\otimes k}\right)
\ee
where $\mathcal{C}_{A,n}$ is defined in terms of its action on bosonic operators.
For any operator $V_{\bar A}$ supported in $\bar{A}$, $\mathcal{C}_{A,n}$ behaves like $\mathbb{I}$. Instead for operators $V_A$ supported in $A$, $\mathcal{C}_{A,n}$ acts as a cyclic permutation. Equivalently, after mapping operator to states through Eq.~\ref{app:eq:operator-to-state}, we have
\be
	e^{-k(1-n)S_{n,A}} = \left(\bra{\mathcal{C}_{A,n}}^{\otimes k}\right) \left(\ket{\rho(t)}^{\otimes{nk}}\right).
\ee
Here $\ket{\mathcal{C}_{A,n}}$ can be described by its stabilizers of Pauli matrices
\ba
    V_k^{(+a)} V_k^{(-a)} \ket{\mathcal{C}_{A,n}} &= \ket{\mathcal{C}_{A,n}},
    &
    k&\in \bar{A}
    \\
    V_k^{(+a)} V_k^{-(a-1\mod n)} \ket{\mathcal{C}_{A,n}} &= \ket{\mathcal{C}_{A,n}},
    &
    k&\in A.
\end{align}

Finally, by re-expressing the Majoranas in terms of Pauli operators, we have the following relations
\ba
    i \gamma^{(+a)} \gamma^{(-a)} \ket{\mathcal{C}_{A,n}} &= \ket{\mathcal{C}_{A,n}},
    &
    k&\in \bar{A}
\end{align}
and
\ba
    i \gamma^{+1} \gamma^{-n} \ket{\mathcal{C}_{A,n}} &= \ket{\mathcal{C}_{A,n}}\\
    i \gamma^{-a} \gamma^{+(a+1)} \ket{\mathcal{C}_{A,n}} &= \ket{\mathcal{C}_{A,n}}
\end{align}
for $a<n$ and $k\in A$.

We can equivalently characterize the boundary states by the expectation value of $S$ on each site. For all boundary states $\kket{B}$ discussed above we will have that $\langle L\rangle=\langle R \rangle=0$. Finally, the expectation value of $Q$ depends on the boundary state. For the initial state $\ket{\mathbb{I}}$, we have the expectation value of $Q$ is the $N\times N$ identity matrix. Finally, for $\ket{\mathcal{C}_{A,n}}$, and a site in $A$, the structure of $Q=q_n$, reported in Eq.~\ref{eq:qn-boundary}.

\section{Properties of the boundary states}
\label{app:sec:boundary-state-irreps}

We now proceed to show that the set of stabilizers above fix the $\mathrm{so}(2N)$ irrep on every site, and within this it holds that $S^T S = \mathds{1}+O(1/N_F)$.

We begin by observing that it is enough to characterize the state $\ket{\mathbb{I}}$, since the other boundary states can be obtained through local rotations from $\ket{\mathbb{I}}$, e.g. by rotating the $\gamma^+$ as specified by the orthogonal matrix $\langle Q \rangle$. 

To characterize the irrep of $\ket{\mathbb{I}}$, we begin by considering the $N_F=1$ case, where we will show that the state lies within the spin representation. For this purpose, we recall that the Dynkin diagram of $\mathrm{so}(2N)$ (viz. $D_N$) is given by 
\begin{center}
    \dynkin[labels={\alpha_1,\alpha_2,\alpha_{N-3},\alpha_{N-2},\alpha_{N-1},\alpha_{N}},%
        label directions={,,,right,,}, scale=2] D{}    
\end{center}
with each dot associated to a simple root $\alpha_n$. Given an orthonormal basis (w.r.t. the Killing form) $\{e_n\}$ the simple roots can be chosen to be $\alpha_j = e_{j} - e_{j+1}$  for $j<N$ and $\alpha_N = e_{N-1}+e_N$. For a single site and flavor, such an orthonormal basis is naturally represented by the set $e_n \mapsto X_n = i \gamma^{+n} \gamma^{-n}/2$.
We can then note that the state $\ket{\mathbb{I}}$ has weight $0$ in all $\alpha_n$ with $n<N$ and weight $1$ in $\alpha_N$.
Finally, we show that $\ket{\mathbb{I}}$ is an highest-weight state, i.e. it is annihilated by all raising operators $E_k$ defined by
\be
    [X_n, E_k] = (e_n \cdot \alpha_k) E_k.
\ee
For the purpose of verifying this, we write the $E_k$ explicitly in terms of Majorana bilinears
\ba
    E_{k<N} &= i \gamma^{+k} \gamma^{+(k+1)} + \gamma^{-k} \gamma^{+(k+1)}\nonumber\\
         &~~+i \gamma^{-k} \gamma^{-(k+1)} - \gamma^{+k} \gamma^{-(k+1)}\\
    E_{N} &= i \gamma^{+k} \gamma^{+(k+1)} + \gamma^{-k} \gamma^{+(k+1)}\nonumber\\
         &~~-i \gamma^{-k} \gamma^{-(k+1)} + \gamma^{+k} \gamma^{-(k+1)}.
\end{align}
Rewriting Eq.~\ref{eq:stabilizers-rho0} as
\be
\label{app:eq:minus-to-plus}
    \gamma_k^{-a} \ket{\mathbb{I}} = -i \gamma_k^{+a} \ket{\mathbb{I}},
\ee
it is immediate to see that $E_k \ket{\mathbb{I}}=0$ for all $k$. This therefore proves that $\ket{\mathbb{I}}$ lies in the spin representation.
Note that this irrep is $2^{N-1}$ dimensional and corresponds to the states with $\mathcal{R}=+1$. [The remaining $2^{N-1}$ states, which are not involved in the physics of the problem have $\mathcal{R}=-1$ lie in the conjugate representation.]

The case of larger $N_F$ follows immediately from the case $N_F=1$. Given the tensor-product structure of $\ket{\mathbb{I}}$ across different flavors, we have that in this case $\ket{\mathbb{I}}$ is a highest-weight state with weights $0$ w.r.t. all $\alpha_n$ ($n<N$) and weight $N_F$ w.r.t. $\alpha_N$.

Finally, we show that for states $\ket{\psi}$ within this irrep we have
\be
    \label{eq:S-orthogonal}
    (S^TS)^{\alpha\beta} \ket{\psi} = \left[ \delta^{ab} + O(1/N_F) \right] \ket{\psi},
\ee
To this end, we will show that Eq.~\ref{eq:S-orthogonal} holds when acting on the state $\ket{\psi}= \ket{\mathbb{I}}$. From this state, the property can be lifted to the whole irrep, by noting that an arbitrary state $\ket{\psi}$ within the irrep can be written as $W\kket{\mathbb{I}}$ for some $W$ representing a rotation $O\in \mathrm{SO}(2N)$, then
\be
    S^T S\ket{\psi} = W O^T \left(S^T S \ket{\mathbb{I}}\right) O = \mathds{1}\ket{\psi} + O(1/N_F).
\ee
To show that the property above holds for $\ket{\psi}= \ket{\mathbb{I}}$, we express  $S^T S$ as\begin{equation}
    S^T S = 
    \begin{pmatrix}
        -L^2 + Q Q^T & \rvline & -LQ-Q R\\
        \hline
        L Q^T + R Q^T & \rvline & -R^2 + Q^T Q
    \end{pmatrix}.
\end{equation}
Then, using Eq.~\ref{app:eq:minus-to-plus} we can transform any polynomials in $\gamma^+$ and $\gamma^-$ into ones involving only $\gamma^+$.
By doing this, we obtain
\ba
    LQ + QR &\equiv \f{N-1}{N_F} (L +  i \mathds{1})\\
    Q^T Q - L^2 &\equiv \mathds{1} \left( 1 + \f{2N-1}{N_F} \right) - 2i \f{N-1}{N_F} L\\
    Q Q^T - R^2 &\equiv \mathds{1} \left( 1 + \f{2N-1}{N_F} \right) + 2i \f{N-1}{N_F} L,
\end{align}
where the equivalence above means that the LHS and RHS are equal when applied on the state $\ket{\mathbb{I}}$. This therefore concludes the proof of Eq.~\ref{eq:S-orthogonal}.
The quadratic Casimir of the representation can be obtained as $\Tr S^T S$, which yields Eq.~\ref{eq:casimirvalue}.

In addition, we can further analyze higher-order Casimir. Most of these are already specified by the condition ~\eqref{eq:S-orthogonal}, however, the space of orthogonal antisymmetric matrices is split in two disconnected components distinguished by the Pfaffian of $\text{Pf}(S)$: a polynomial of degree $2N$ of its entries that is invariant under $S\mapsto O^T S O$ for $O\in \mathrm{SO}(2N)$. Due to its invariance properties, $\text{Pf}(S)$ commutes with all the generators of $\mathrm{so}(2N)$ and therefore with $\mathcal{H}$. At large $N_F$ we can see that
\be
    \braket{\mathbb{I} | \text{Pf}(S) | \mathbb{I}} = \text{Pf}\left(\braket{\mathbb{I} | S | \mathbb{I}}\right) + O(1/N_F)
\ee
as it is true for any polynomial in the entries of $S$ of degree $\leq 2N$, so that
\be
    \text{Pf}(S) = (-1)^{N(N-1)/2} + O(1/N_F)
\ee
in the irrep of interest.

\section{Path integral from coherent states}\label{app:coherentstates}

{ 
In Sec.~\ref{sec:mapping-to-NLSM} we used a semiclassical (large $N_F$) analysis to argue that there exists a path integral representation in terms of a field $S(x,t)$ that lives in the space of antisymmetric $\mathrm{SO}(2N)$ matrices.
Here we confirm, using generalized coherent states (see e.g. Ref.~\cite{stone2001note} and references therein) that the representation of the path integral in terms of the field $S(x,t)$ makes sense for any value of $N_F$, not necessarily large 
(with $x$ initially taking values on the spatial lattice).
This is consistent with our expectation that the same infra-red field theories and universality classes apply at both small $N_F$ and large $N_F$, even though for small $N_F$ we do not have quantitative control over nonuniversal constants when we take the continuum limit.}

We consider a single site $j$. Here our starting point is the state $\ket{\mathbb{I}}$. By definition, acting on $\ket{\mathbb{I}}$ with $O\in \mathrm{SO}(2N)$ rotations generates the whole irreducible representation the state belongs to. More precisely we act on $\ket{\mathbb{I}}$ with a unitary transformation $\hat W_O$ that are the image of $O\in \mathrm{SO}(2N)$ matrices under the representation map. The states $\hat W_O\ket{\mathbb{I}}$, commonly called coherent states, form an overcomplete basis of the irrep and can therefore be used to construct a path integral representation of the problem at hand.
It is a standard results that the identity (on the irrep) can be represented as an integral of the form
\be
    \mathds{1} \varpropto \int \dd \mu_H(O) \hat W_O \ket{\mathbb{I}} \bra{\mathbb{I}} W_O^\dag,
\ee
where $\dd \mu_H(O)$ is the Haar measure on the group $\mathrm{SO}(N)$, with some proportionality constant that is irrelevant for the subsequent treatment. To see this, we can note that the integral commutes with any unitary $\hat{U}_O$ in the representation; therefore, by Schur's lemma, the integral is proportional to the identity on the irrep.

We further note that the states $\hat W_O\ket{\mathbb{I}}$ can be identified, up to a phase, by 
\be
    S = \braket{\mathbb{I}|\hat W_O^\dag \hat S \hat W_O|\mathbb{I}} = O^T J_0 O,
\ee
viz. the expectation value of the spin operator $\hat S$ on them.
In this appendix only, we distinguish between operators and c-numbers by putting a hat on the former.
Here $J_0$ is the matrix
\be
	J_0 = 
	\begin{pmatrix}
		0 & \mathds{1}\\
		-\mathds{1} & 0
	\end{pmatrix}.
\ee
The value of $S$ uniquely defines the state up to a phase, given that the states in the same irrep of $\mathbb{I}$ are Gaussian by construction and are therefore uniquely specified by the expectation values of quadratic Majorana operators, which are contained in $\hat S$. 

Since the coherent states are defined by $S$, up to a phase, we can write the resolution of the identity as an integral over~$S$,
\be\label{eq:coherentstatesSintegral}
\mathds{1}\propto 
\int \dd \mu(S) \ket{S}\bra{S},
\ee
where $\ket{S}$ is defined by fixing some choice of phase convention, and $\mu(S)$ is the $\mathrm{SO}(2N)$-symmetric measure for $S$ induced by $\mu_H(O)$.

The path-integral representation for $\Hs$ can then be obtained through conventional means by trotterizing the time-evolution operator and inserting resolutions of the identity \eqref{eq:coherentstatesSintegral}.
The degrees of freedom in this path integral are orthogonal antisymmetric matrices, i.e. whose blocks satisfy Eq.~\ref{eq:constraints-on-S}, continuously connected to $J_0$, i.e. with $\text{Pf}(S)=(-1)^{N(N-1)/2}$.
This path integral has the schematic form 
\be\label{eq:coherentstatespathintegral}
\int \prod_j\mathcal{D}S_j 
\exp \lf
i N_F \sum_j \Omega[S_j]
- N_F \int \dd t' \mathcal{H}[\{S_j(t')\}]
\ri.
\ee
Here ${N_F\Omega[S_j]}$ is the Berry phase term for the $j$-th site, see e.g. Refs.~\cite{stone2001note, fradkin2013field} 
(it is proportional to  $N_F$
because the states $\ket{S}$ are product states over the flavors, and therefore the Berry phase is additive for the flavors).
The quantity  $\mathcal{H}[\{S_j(t')\}]$ appearing above is obtained by replacing the spin operators $\hat S_j$ with the c-number fields $S_j(t)$.
We will not need the explicit form of the Berry phase, but it is important  that the lattice action above is proportional to $N_F$.
Ultimately this is what allows a controlled derivation of a continuum theory at large $N_F$~(Sec.~\ref{subsec:equations-of-motion}).

The second important point is the precise  target space for $S$. We show that it is isomorphic to the quotient space $\mathrm{SO}(2N)/\mathrm{U}(N)$, where the isomorphism is given by the map $O\mapsto S=O^T J_0 O$ with $O\in \mathrm{SO}(2N)/\mathrm{U}(N)$. Here we think of $U(N)$ as a subgroup of $SO(2N)$. For this purpose we use a standard embedding of $U(N)$ in $\mathrm{SO}(2N)$ (see e.g. Ref.~\cite{Baker2002})
\be
    U = X+i Y \mapsto \begin{pmatrix}
		X & -Y\\
		Y & X
	\end{pmatrix},
\ee
where $X$ and $Y$ are the real and imaginary parts of the $\mathrm{U}(N)$ matrix $U$, respectively. One can explicitly verify that the image of a unitary matrix is orthogonal. By continuity of the determinant and connectedness of $\mathrm{U}(N)$, then the image lies within $\mathrm{SO}(2N)$.
From these observations, we can see that the image of $\mathrm{U}(N)$ is the set of $O\in\mathrm{SO}(2N)$ s.t. $O^T J_0 O = J_0$.

Finally, from this one can see that the map $O\mapsto S=O^T J_O O$ is bijective. In fact, any matrix $S$, which is antisymmetric admits a spectral decomposition of the form
\be
    O^T
    \begin{pmatrix}
		0 & \Lambda\\
		-\Lambda & 0
	\end{pmatrix}
    O,
\ee
with $\Lambda$ diagonal and $O\in SO(N)$.  $S$ being orthogonal and with Pfaffian $1$, we can further choose $\Lambda=\mathds{1}$. Finally, it is immediate to verify that the matrices $O$ are unique up to multiplication by a matrix in the image of $\mathrm{U}(N)$, therefore showing that the manifold os $S$ is isomorphic to $\mathrm{SO}(2N)/\mathrm{U}(N)$.

One route to the continuum theory would be via a derivative expansion of the lattice action
Eq.~\ref{eq:coherentstatespathintegral} --- see e.g. Ref.~\cite{fradkin2013field} for similar derivations for spin chains.  In the main text, we instead obtained the continuum theory by analyzing the equations of motion. The ultimate result should be the same.

\section{Details of the numerical simulations}
\label{app:sec:numerics}

In this Appendix we provide details about the numerical simulations of the monitored dynamics. Our approach follows that of Ref.~\cite{turkeshi2021measurementinduced}, which in turn is a generalization of the method introduced in Ref.~\cite{cao2019entanglement} for free-fermionic dynamics with conserved particle numbers. 

Throughout this section we will consider a single fermionic species and set $N_F=1$. Given that there will be no flavor indices, we will use greek and latin indices interchangeably to label physical sites. We further assume that the number of sites is even and define $M=L/2$. We begin by recalling that the monitored dynamics described in Appendix~\ref{app:sec:discrete-time} can be formulated within the so-called quantum-state-diffusion formalism~\cite{caves1987quantum,diosi1998non,gisin1992quantum}, yielding the stochastic differential equation
\begin{align}\label{eq:QSD_equation}
\!d\ket{\psi_t}=-&iH(t)dt\ket{\psi_t}\!\nonumber\\
+&\!\sum_{j}d \xi_j\left(i\gamma_{j}\gamma_{j+1}\!
-\!\braket{i\gamma_j\gamma_{j+1}}_t\right)\ket{\psi_t}\nonumber\\
-&\frac{ d t}{2}\sum_{j}\Gamma_j(i\gamma_{j}\gamma_{j+1}
-\braket{i\gamma_j\gamma_{j+1}}_t)^2\ket{\psi_t}\,,
\end{align}
where $H(t)=\sum_{j=1}^{L-1}J_j(t)i\gamma_j\gamma_{j+1}$ (for our numerical simulation we chose open boundary conditions). Here $d \xi_j$ are real random variables satisfying 
\begin{equation}
d\xi_j^2=\Gamma_j dt\,,    
\end{equation}
where we employed It\^{o} notation~\cite{oksendal2003stochastic}, while
\begin{equation}
\braket{\gamma_j\gamma_{j+1}}_t=\braket{\psi_t|\gamma_j\gamma_{j+1}|\psi_t}\,.
\end{equation}
In the following, we will also set 
\begin{equation}
    \Gamma_{o}=\left[1-\Delta\right]\Gamma\,,\quad 
    \Gamma_{e}=\left[1+\Delta\right]\Gamma\,,
\end{equation}
for the staggered value of $\Gamma_j$ at odd/even position. 
Using a standard derivation~\cite{kurt2006straightforward}, one can show that Eq.~\ref{eq:QSD_equation} describes the continuous-time limit of the discrete dynamics discussed in Appendix~\ref{app:sec:discrete-time}. This formulation is convenient from the numerical point of view, because one does not need to sample measurement outcomes according to the Born rule. Note that at the first relevant order in It\^{o} calculus, we may split the infinitesimal time evolution into three parts: a unitary step
\begin{equation}\label{eq:unitary}
    d|\psi_t\rangle=-iH(t)dt\ket{\psi_t}\,,
\end{equation}
and the measurements at even/odd pairs of Majoranas
\begin{align}\label{eq:measurement_step}
d|\psi_t\rangle=+&\sum_{j\in e/o}d \xi_j\left(i\gamma_{j}\gamma_{j+1}
-\braket{i\gamma_j\gamma_{j+1}}_t\right)\ket{\psi_t}\nonumber\\
-&\frac{ d t}{2}\sum_{j\in e/o}\Gamma_j(i\gamma_{j}\gamma_{j+1}-\braket{i\gamma_j\gamma_{j+1}}_t)^2\ket{\psi_t}\,,
\end{align}
where $j\in e$ ($j\in o$) means that the sum is restricted to $j$ even ($j$ odd).

For an initial Gaussian state~\cite{bravyi2004lagrangian}, the state of the system remains Gaussian at all times, and the dynamics can be simulated efficiently. Indeed, Gaussian states satisfy Wick theorem, and can be described entirely in terms of the covariance matrix
\begin{equation}\label{eq:majorana_covariance}
\Omega_{a b}[\phi]=\frac{i}{2} \langle \phi|\left[\gamma_{a}, \gamma_{b}\right]| \phi\rangle\,.
\end{equation}
In our numerics, we initialize the system in the vacuum state $\ket{\boldsymbol{0}}$ associated with the fermions $c_j=(\gamma_{2j-1}-i\gamma_{2j})/2$. Its covariance matrix reads
\begin{equation}
	\Omega_0=\begin{pmatrix}
		0 & \openone\\
		-\openone & 0
	\end{pmatrix}\,.
\end{equation}
Here and throughout this Appendix, we order the Majoranas as $\boldsymbol{\gamma}=(\gamma_1,\gamma_3,\ldots, \gamma_{2M-1}, \gamma_2,\gamma_{4},\ldots \gamma_{2M})$. Given a Gaussian state $\ket{\phi}$, its covariance matrix can be computed if we know a Gaussian unitary operator $\mathcal{O}$ s.t.
\begin{equation}\label{eq:phi_state}
    \ket{\phi}= \mathcal{O}\ket{\boldsymbol{0}}\,.
\end{equation}
Indeed, denoting by $O$ the single-particle orthogonal matrix corresponding to $\mathcal{O}$, viz. $\mathcal{O}^\dagger \gamma_j \mathcal{O}= O_{ij} \gamma_i$, we have
\begin{equation}
    \Omega[\phi]=O^T\Omega_0O\,.
\end{equation}
For what follows, it is important to note that the state~\eqref{eq:phi_state} is annihilated by the operators
\begin{equation}\label{eq:u_v_matrices}
	a_\mu=(\mathcal{O} c_\mu \mathcal{O}^{\dagger})= \sum_{j} U_{j,\mu}c_j+V_{j,\mu} c^{\dagger}_j\,,
\end{equation}
where the $M\times M$ matrices $U$, $V$ are defined by
\begin{equation}\label{eq:o_uv_relation}
 \begin{pmatrix}
		U^{T} & V^T \\
		V^{\dagger} & U^{\dagger}
\end{pmatrix}
 =\frac{1}{2}
	\begin{pmatrix}
		\openone & -i \openone \\
		\openone & i \openone
	\end{pmatrix}
O
 	\begin{pmatrix}
		\openone & \openone \\
		i \openone & -i \openone
	\end{pmatrix}
\,.
\end{equation}
Note that the matrices $U$ and $V$ in the LHS are well-defined since $O$ is a real matrix. In addition, because of unitarity, it must be 
\begin{equation}\label{eq:relation_1}
	V^\dagger U^\ast+U^\dagger V^\ast=0\,,\qquad U^{\dagger}U+V^{\dagger}V=\openone\,.
\end{equation}

The algorithm works as follows. We first discretize time and, for all time steps, we compute the orthogonal matrix $O(t)$ corresponding to an operator $\mathcal{O}(t)$ s.t. $\ket{\psi_t}=\mathcal{O}(t)\ket{\boldsymbol{0}}$ (clearly, $O(0)=\openone$). The infinitesimal time interval $dt$ is replaced by a finite Trotter time $\Delta t$, which is taken to be sufficiently small (we have always checked robustness of our data with respect to decreasing $\Delta t$).

Let us first consider the unitary step~\eqref{eq:unitary}. Denoting by $\ket{\psi_t}$ the state of the system at time $t$ and by $O(t)$ the corresponding orthogonal matrix, 
the state of the system is updated as
\begin{equation}
    \ket{\psi_t}\mapsto \mathcal{W}\ket{\psi_t}\,,
\end{equation}
where $\mathcal{W}=\exp(-iH \Delta t)$, with $J_{j,j+1}(t)$ drawn from the random Gaussian distribution with average $0$ and variance $J^{2}/(\Delta t)$. Therefore, the matrix $O(t)$ is updated as
\begin{equation}\label{eq:multiplication}
O(t)\mapsto O(t) W\,,
\end{equation}
where 
\begin{equation}
W=e^{2\sum_{j,j+1}J_{j,j+1}(t)(E_{j,j+1}-E_{j+1,j})\Delta t}\,,
\end{equation}
and we introduced the matrices $E_{j,k}$ defined by $(E_{j,k})_{\alpha,\beta}=\delta_{j,\alpha}\delta_{\beta,k}$. Note that in Eq.~\ref{eq:multiplication} we have right matrix multiplication.

Next, we consider the non-unitary step~\eqref{eq:measurement_step} corresponding to weak measurements of the odd Majorana pairs. Again, we denote by $\ket{\psi_t}$ the state of the system right before this step, and by $O(t)$ the corresponding orthogonal matrix. After this step, the state is updated as
\begin{equation}\label{eq:non_unitary_update}
    \ket{\psi_t}\mapsto 
    \ket{\psi^\prime_t}=
    \frac{\mathcal{D}\ket{\psi_t}}{||\mathcal{D}\ket{\psi_t}||}\,,
\end{equation}
where
\begin{equation}
	\mathcal{D}=\exp\left\{\sum_{j}n_j\left[2d\xi_j+4\Gamma_o \Delta t\left(2\braket{n_j}-1\right)\right]\right\}\,,
\end{equation}
where $n_j=c^\dagger_{j}c_{j}$ and $c_{j}=(\gamma_{2j-1}-i\gamma_{2j})/2$. Note that here $d\xi_j$ are random numbers with average zero and variance equal to $\Gamma_o\Delta t$. It is immediate to show that this state is annihilated by the operators
\begin{equation}
	b_\mu=(\mathcal{D}\mathcal{O} c_\mu \mathcal{O}^{\dagger}\mathcal{D}^{-1})=\sum_{j} \tilde{U}_{j,\mu}c_j+\tilde{V}_{j,\mu} c^{\dagger}_j\,,
\end{equation}
where
\begin{equation}\label{eq:tildeU_tildeV}
	\tilde{U}=D^{-1}U\,,\qquad \tilde{V}=DV\,,
\end{equation}
and 
\begin{eqnarray}
	D={\rm diag}(e^{-\varepsilon_1},e^{-\varepsilon_2}, \ldots, e^{-\varepsilon_M})\,,
\end{eqnarray}
where
\begin{equation}
\varepsilon_j=\left[2 d\xi_j+4\Gamma_o \Delta t\left(2\braket{n_j}-1\right)\right]\,.
\end{equation}
In Eq.~\ref{eq:tildeU_tildeV}, $U$ and $V$ are defined as in~\eqref{eq:o_uv_relation}, where $O$ is the orthogonal matrix $O(t)$. 

Since $D^\dagger=D^\ast=D$, we have
\begin{equation}\label{eq:tilde_relation_1}
	\tilde{V}^\dagger \tilde{U}^\ast+\tilde{U}^\dagger \tilde{V}^\ast=V^\dagger U^\ast+U^\dagger V^\ast=0\,.
\end{equation}
However, because $\mathcal{D}$ is not a unitary operator, $\tilde{U}$ and $\tilde{V}$ in Eq.~\ref{eq:tildeU_tildeV} do not satisfy the second relation in Eq.~\ref{eq:relation_1}, and therefore do not define an orthogonal matrix. However, since $\ket{\psi^\prime_t}$ is annihilated by $b_\mu$, it is also annihilated by different operators that can be written as $\tilde{b}_\mu = T_{\mu \nu} b_\nu$ as long as $T$ is a non-singular matrix. This condition can be most conveniently formulated as
\begin{equation}
	\tilde{b}_\mu =\sum_{j=1}^M X_{\mu, j} c_j+ \sum_{j=M+1}^{2M} X_{\mu, j} c^\dagger_j\,,
\end{equation}
where $X$ is any $M\times 2M$ matrix, whose rows form a basis for the row-space of the $M\times 2M$ matrix
\begin{equation}\label{eq:to_be_QRed}
	\begin{pmatrix}
		\tilde{U}^T , \tilde{V}^T
	\end{pmatrix}\,.
\end{equation}
By definition, we can find such a matrix by performing a QR decomposition 
\begin{equation}\label{eq:QR}
	\begin{pmatrix}
		\tilde{U}\\
		\tilde{V}
	\end{pmatrix}
	=
	\begin{pmatrix}
		Q_{11} R_{11}\\
		Q_{21} R_{11}
	\end{pmatrix}\,.
\end{equation}
Indeed, the rows of the matrix
\begin{equation}
	\begin{pmatrix}
		Q_{11}^T, Q^{T}_{21}
	\end{pmatrix}
\end{equation}
form a basis for the row-space of the $M\times 2M$ matrix~\eqref{eq:to_be_QRed}. It follows that the state $\ket{\psi^\prime_t}$ is also annihilated by
\begin{align}
	\tilde{b}_\mu &=\sum_{j=1}^M [Q^{T}_{11}]_{\mu, j} c_j+ [Q^{T}_{21}]_{\mu, j} c^\dagger_j\,,\nonumber\\
	&=\sum_{j=1}^M [Q_{11}]_{j,\mu} c_j+ [Q_{21}]_{j,\mu} c^\dagger_j\,.
\end{align}
Next, we show that $Q_{11}$, and $Q_{21}$ satisfy both the relations~\eqref{eq:relation_1}. First, by the property of the QR decomposition we immediately have 
\begin{align}
	Q_{11}^\dagger Q_{11}+Q_{21}^\dagger Q_{21}=\openone\,.
\end{align}
Next, plugging Eq.~\ref{eq:QR} into Eq.~\ref{eq:tilde_relation_1} we have
\begin{equation}\label{eq:to_test_1}
	R_{11}^\dagger (Q_{21}^\dagger Q_{11}^\ast+Q_{11}^\dagger Q_{21}^\ast )R_{11}=0\,.
\end{equation}
Because the matrix~\eqref{eq:QR} is full-ranked, $R_{11}$ is invertible and so necessarily
\begin{equation}\label{eq:to_test_2}
	Q_{21}^\dagger Q_{11}^\ast+Q_{11}^\dagger Q_{21}^\ast =0\,.
\end{equation}
It follows that we can update the orthogonal matrix
\begin{equation}
    O(t)\mapsto O^\prime(t)
\end{equation}
where
\begin{equation}
O^\prime(t)=
\frac{1}{2}
	\begin{pmatrix}
		\openone & \openone \\
		i \openone & -i \openone
	\end{pmatrix}
	\begin{pmatrix}
		Q_{11} ^{T} & Q_{12}^T \\
		 Q_{12}^{\dagger} & Q_{11} ^{\dagger}
	\end{pmatrix}
	\begin{pmatrix}
		\openone & -i \openone \\
		\openone & i \openone
	\end{pmatrix}\,.
\end{equation}
By construction, $O'(t)$ is orthogonal and $\ket{\psi^\prime_t}=\mathcal{O}^\prime(t)\ket{\boldsymbol{0}}$. 

Finally, let us consider the non-unitary step~\eqref{eq:measurement_step} corresponding to weak measurements of the even Majorana pairs. This can be implemented by simply shifting the covariance matrix by one, and repeating the step described above for the non-unitary step~\eqref{eq:measurement_step} corresponding to weak measurements of the odd Majorana pairs. The shift corresponds to the matrix multiplication
\begin{equation}
	\tilde{O}_t=O_tS^T\,,
\end{equation}
where
\begin{equation}
	S=
	\begin{pmatrix}
		0 & 1 & 0 &\ldots &0\\
		0 & 0 & 1 & \ldots &0 \\
		\vdots & \vdots & \vdots & \ddots & 0\\
		0 & 0 & 0 & \ldots & 1\\
		1 & 0 & 0 & \ldots & 0\\
	\end{pmatrix}\,.
\end{equation}
After the QR-decomposition step described above is applied, yielding an orthogonal matrix, we finally shift back performing a right multiplication by the matrix $S$.

For completeness, we also briefly recall  how the R\'enyi entropies can be obtained from the covariance matrix~\cite{vidal2003entanglement}. Let us denote by $\Gamma_\ell$ the matrix obtained by selecting the rows and columns $j,k=1,\ldots 2\ell$ from Eq.~\ref{eq:majorana_covariance}. $\Gamma_\ell$ is real and antisymmetric and its $2\ell$ eigenvalues form $\ell$ pairs of imaginary numbers $\pm i\nu_j$, with $\nu_j\in \mathbb{R}$. The R\'enyi and von Neumann entanglement entropies of the subsystem associated with the first $2\ell$ Majoranas read
\begin{align}
S_n&=\frac{1}{1-n}\sum_{j=1}^\ell \log \left[\left(\frac{1-\nu_j}{2}\right)^n+\left(\frac{1+\nu_j}{2}\right)^n\right]\,,\\
S_1=&-\sum_{j=1}^\ell \left[\left(\frac{1-\nu_j}{2}\right)\log \left(\frac{1-\nu_j}{2}\right)\right.\nonumber\\
+&\left.\left(\frac{1+\nu_j}{2}\right)\log \left(\frac{1+\nu_j}{2}\right)\right]\,.
\end{align}

\section{Further numerical data}
\label{app:sec:further-data}

\begin{figure*}
    \centering
    \includegraphics[width=0.48\linewidth]{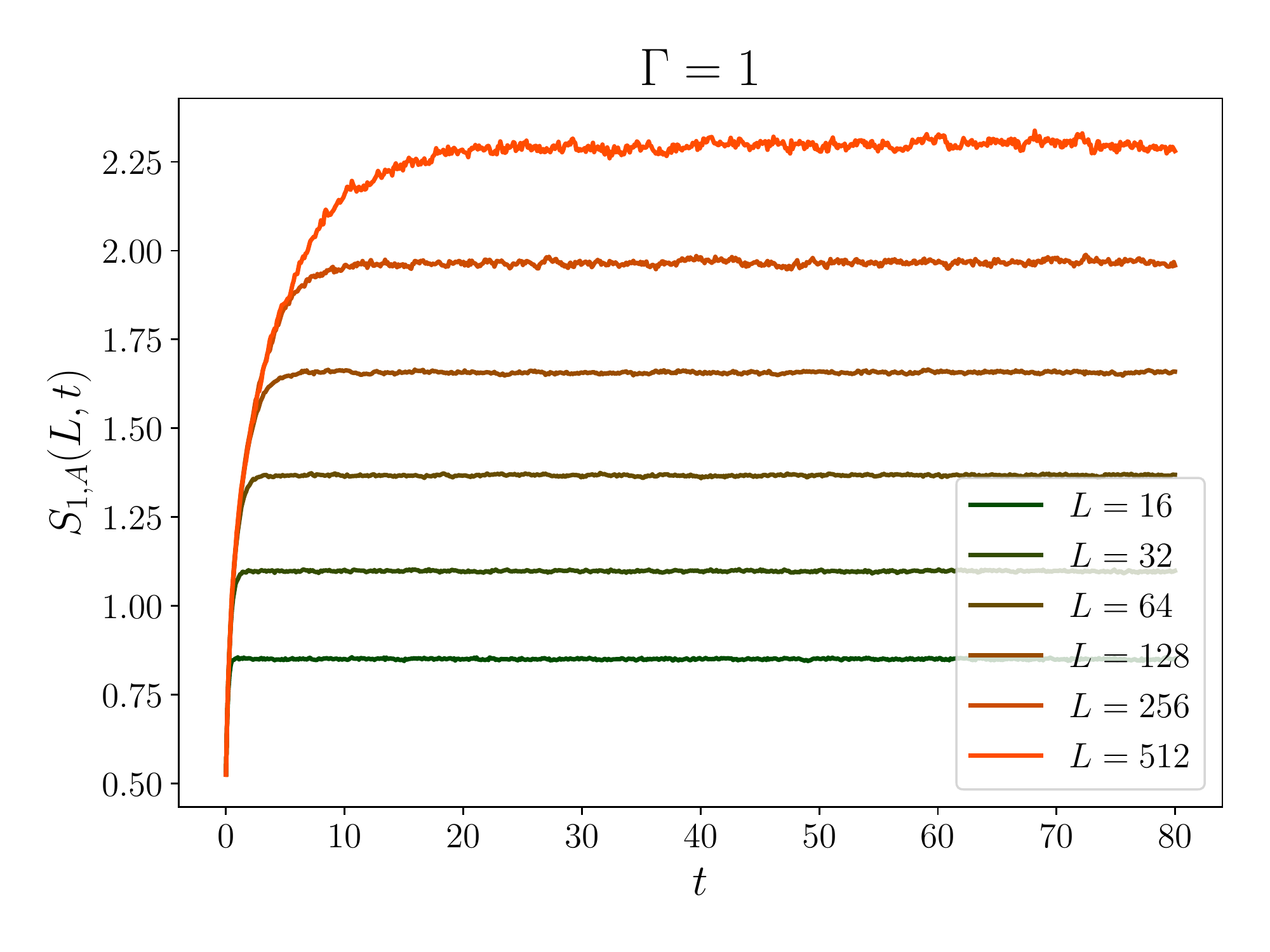}
    \includegraphics[width=0.48\linewidth]{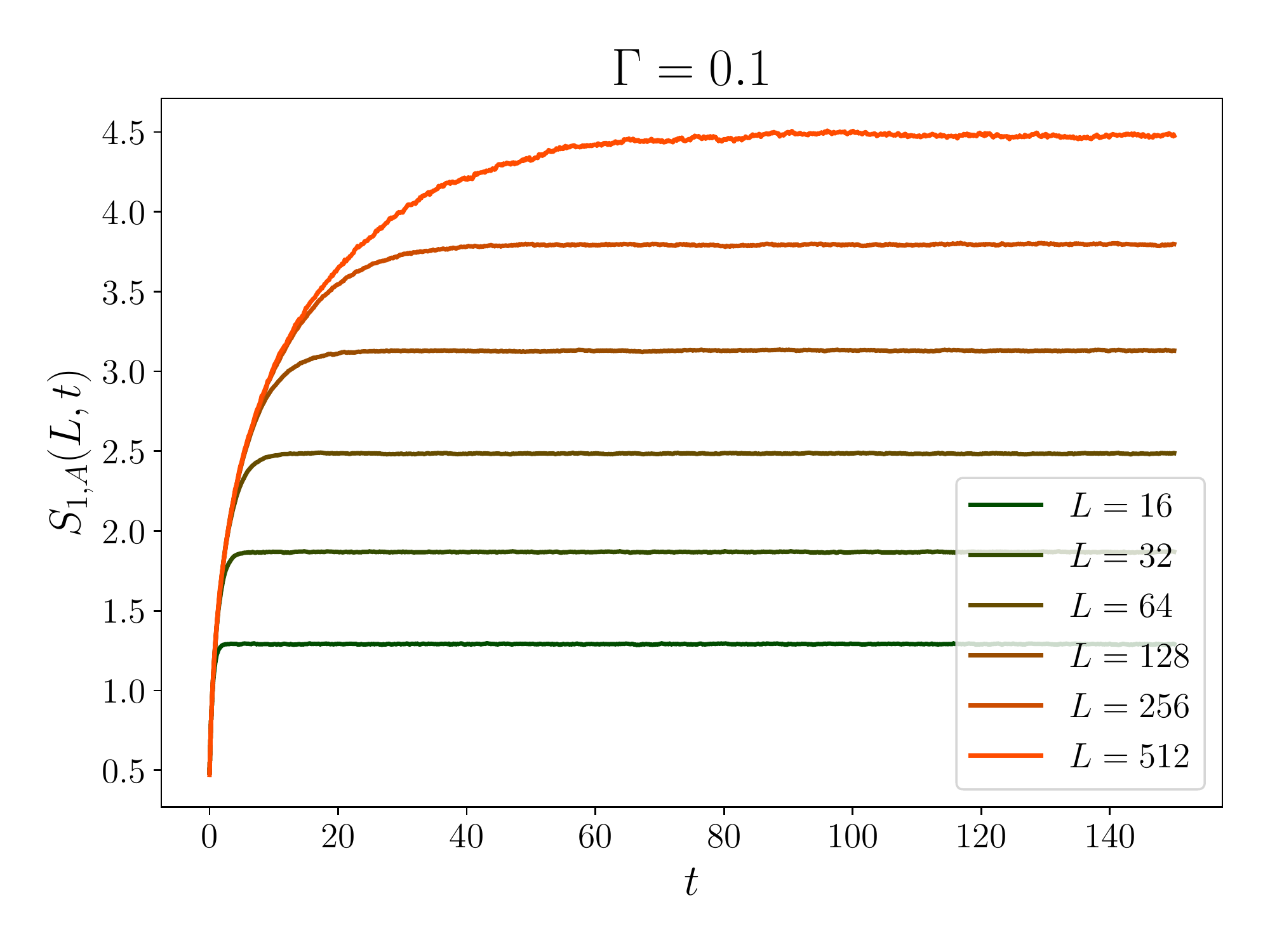}
    \caption{Bipartite Von Neumaunn entanglement entropy $S_{1,A}$ as a function of time for different system sizes $L$. The data in this plot has been obtained with $\dd t=2.5 \times 10^{-3}$}
    \label{fig:time-traces}
\end{figure*}

In this appendix we report further plots related to the data presented in Sec.~\ref{sec:numerics}. First we show $S_{n,A}(t)$ as a function of $t$ averaged over quantum trajectories in Fig.~\ref{fig:time-traces}. We see that for $\Gamma=1$ and $\Gamma=0.1$, the entropy has already reached a plateau at times $t=40$ and $t=100$ respectively, therefore justifying the time interval we used to compute the average.

\begin{figure*}
    \centering
    \includegraphics[width=0.48\linewidth]{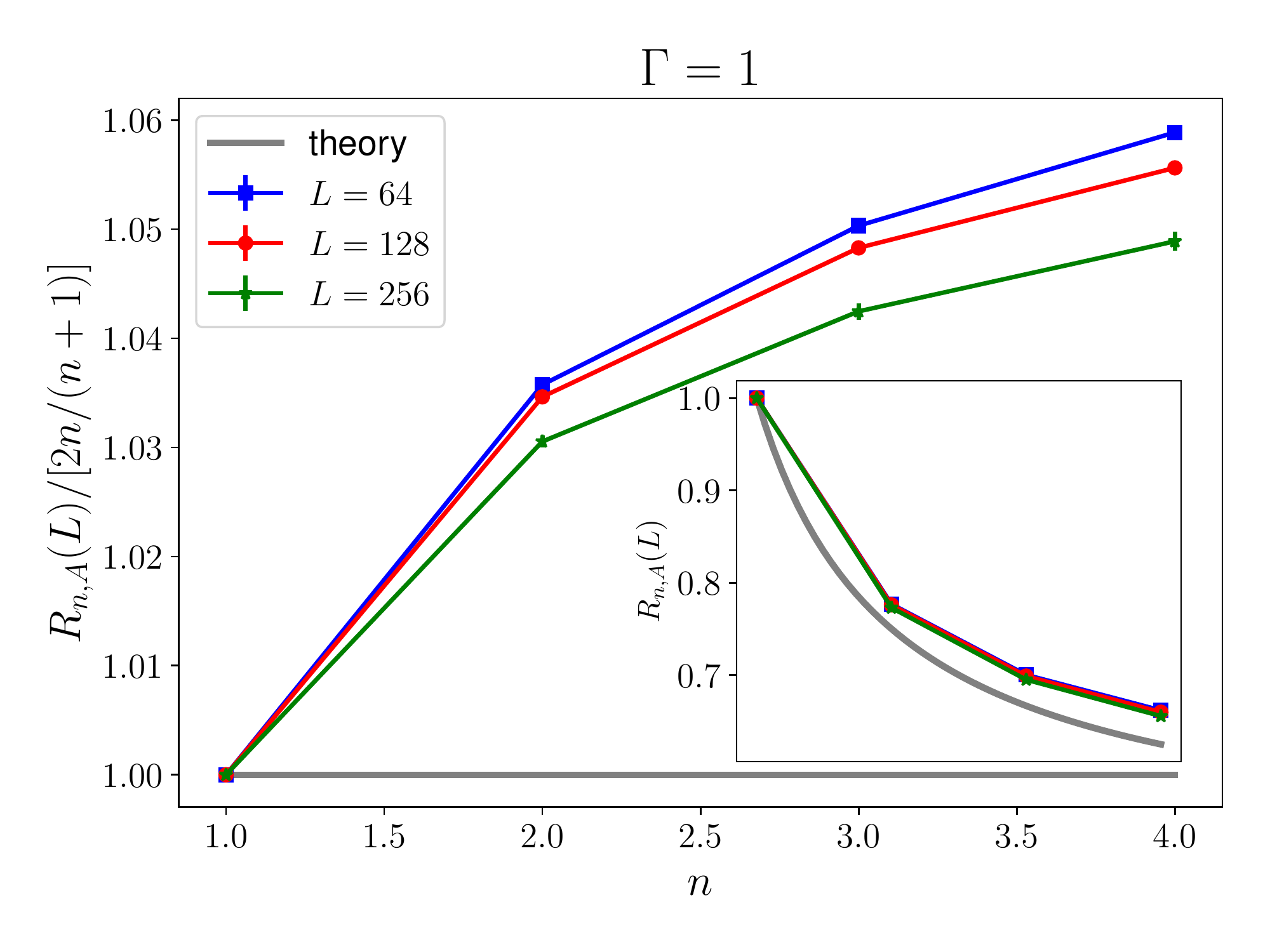}
    \includegraphics[width=0.48\linewidth]{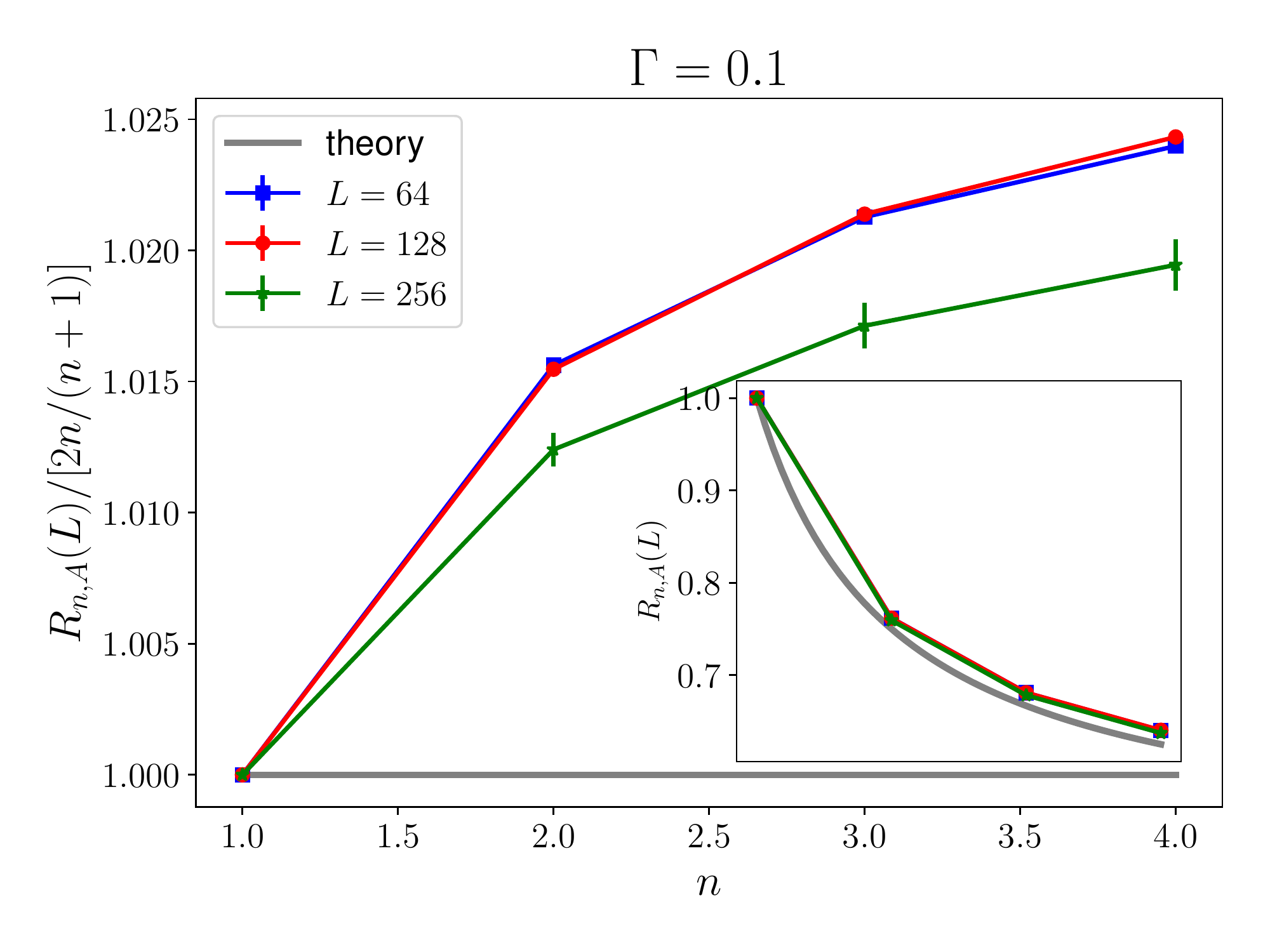}
    \caption{The ratio $R_{n,A}(L)$ (Eq.~\ref{eq:ratio-R}) for $\Delta = 0$ and respectively $\Gamma=1$ (Left) and $\Gamma=0.1$ (Right).
    The main plot shows the ratio of $R_{n,A}(L)$ with the theoretical prediction of $(n+1)/(2n)$, highlighting that, as $L$ increases the ratio becomes closer to $1$ (also note the scale of the $y$-axis). For comparison an inset shows the value of $R_{n,A}$ and the theoretical curve.
    The plot is obtained from the same data used in Figs.~\ref{fig:entanglement-scaling-Gamma-1} and \ref{fig:entanglement-scaling-Gamma-01}. For graphical convenience only data obtained with a time step $\dd t=2.5\times 10^{-3}$ is reported. Errors are estimated using jackknife resampling after blocking quantum trajectories into $20$ subsets.}
    \label{fig:numerics-n-dependence}
\end{figure*}

Finally, as a further test of the theory developed in Sec.~\ref{sec:entanglement}, we verify that $S_{n,A}$ is proportional to $(n+1)/n$. For the purpose of verifying this explicitly we define the ratio
\be
\label{eq:ratio-R}
    R_{n,A}(L) := \f{\Delta S_{n,A}(L)}{\Delta S_{1,A}(L)},
\ee
which, according to our theory should converge to $(n+1)/(2n)$ in the limit of large $\log L$. Indeed, Fig.~\ref{fig:numerics-n-dependence} shows that there is a { reasonable} agreement between the theory and the numerics and the numerical data appear { consistent with convergence} to the theoretical curve as $\log L$ increases.

\section{Vortices and Berry phase}
\label{app:sec:vortices}

In this appendix, we complete the argument in Sec.~\ref{sec:dimerization-and-vortices} by showing that the fugacity of a vortex depends on  position as
\be
    y_j \propto (-1)^{N_F j}.
\ee
The argument is a standard one in the context of magnets, see e.g. Ref~\cite{haldane19883}
(indeed the $N=2$ case of the model is simply an SU(2)  chain), but we spell it out for completeness.

For simplicity we imagine expressing the amplitude
\be\label{eq:1to1}
    Z = \braket{\mathbb{I}| e^{- \Hs t} |\mathbb{I}}
\ee
using the microscopic coherent states path integral: we wish to isolate sign factors associated with vortex configurations in this integral.
Eq.~\ref{eq:1to1} gives  boundary conditions $Q(x,t)=Q(x,0)=\mathds{1}$ for all $x$:
these boundary conditions are convenient below for defining winding numbers, but since we are ultimately interested in a bulk property of the field theory,
this choice of boundary conditions is not crucial.
Similarly it is convenient to set $Q(0,t')=Q(L,t')=\mathds{1}$ for all $t'\in[0,t]$,  which can be done by imposing boundary fields.
Finally, we will assume that in order to fix the phase factor of interest
it is sufficient to consider configurations of a simple form where vortices are well-separated in space and, far away from the vortex cores, the field $S$ is slowly varying and lives in the ``$Q$'' subspace, i.e. $L$ and $R$ are zero.
However, the structure of the field configuration close to the vortex core can be arbitrary.

Recall that, given a continuum field configuration
$Q(x,t')$ 
we can define a winding number
${w\in \pi_1(\mathrm{SO}(N))=\mathbb{Z}_2}$
for a generic closed path $\gamma$ in spacetime,
with $w=1$ if  $Q\circ \gamma$ is a non-contractible path in $\mathrm{SO}(N)$.
Equivalently, $w=1$ if $\gamma$ encloses an odd number of vortices. 

On the lattice, we may define an integer  winding number $w_j$ for each site $j$,
describing the winding of the field along the path with fixed position $j$ and ${t'\in[0,t]}$. 
For each bond of the lattice between sites $j$, $j+1$, we define a vortex number $n_{j+1/2}$ by
\be
    n_{j+1/2} = w_{j+1} + w_j  
    \quad (\operatorname{mod} 2).
\ee
This definition is natural if 
we think of the vortices as located at spacetime positions 
$(j+1/2, t')$
associated with bonds of the 1D chain
sites \cite{haldane19883}. 
However, the microscopic centering is not crucial for the following since we will need only the winding numbers far from the vortex cores.

On sites far from the vortex cores, we take the field configuration to be of the form
\be
    Q(t) = \begin{pmatrix}
        \cos(\theta_t) & -\sin(\theta_t) &\rvline& 0\\
        \sin(\theta_t) & \cos(\theta_t) &\rvline& 0\\
        \hline
        0 & 0 &\rvline& \mathds{1}
    \end{pmatrix}
\ee
where $\theta_t$ is a smoothly-evolving function with $\theta(t=0)=0$ and $\theta(t)=2\pi m$ for some $m\in \mathbb{Z}$.
Since $L=R=0$ for these sites, 
the configuration of $Q$ fully specifies the coherent state in terms of fermions (see Appendix~\ref{app:coherentstates}).
The winding number associated with such a site is then ${w=m \, (\operatorname{mod }2)}$.

The imaginary part of the action comes from a Berry phase for each site, cf. Eq.~\ref{eq:coherentstatespathintegral}. Below we show that for a configuration of the above form the Berry phase is of the form ${\exp({i N_F \Omega[w]})=(-1)^{N_F w}}$.

We can then ask how the total action 
for the entire configuration changes when 
a vortex is translated by one lattice spacing, say from $j+1/2$ to $j+3/2$.
Let the winding number of the sites to the left of this vortex 
(sufficiently far from the core) be $w$, and let the winding number of the sites to the right (sufficiently far from the core) be ${{w+1}\,(\operatorname{mod}\,2)}$.
The shift is done in such a way that the configuration in a region of some large spatial width  $R$ around the core is simply translated.
As a result, the contribution to the action from  this core region is unaltered. 
However, outside this core region, 
the effect of the translation is that there is one additional site with winding number $w$, and one fewer site with winding number $w+1$.
Therefore, translating a vortex changes the overall weight $e^{-S}$ in the path integral by $(-1)^{N_F}$. 
Translated into the effective field theory picture (\ref{eq:partition-fun-with-vortices}) 
this implies that the vortex fugacity $y_j$ alternates in sign with position if $N_F$ is odd.

Note that we have only fixed the position-dependence of the sign of the fugacity: the prefactor may have either sign. 

After coarse-graining we obtain the position-independent fugacity $y$ as described in Sec.~\ref{sec:dimerization-and-vortices}.
Using standard ideas from the spin chain context one can argue that reversing the sign of $y$ exchanges a disordered phase with gapless projective modes  at the boundary for a disordered phase without such modes. 
In terms of Majoranas this is the presence or absence of undimerized boundary Majoranas. 
E.g. at $N_F=1$ we have the two oppositely dimerized gapped phases (discussed in Sec.~\ref{sec:firstlookatphasediagram}) which are related by translation (since for $N_F$ odd translation changes the sign of $y$).
For $N_F=2$ 
we expect two gapped phases, one at $y>0$ that is trivial and one at $y<0$ that is analogous to the Haldane or AKLT phase.

Computing  the Berry phase for  a given winding reduces to the spin-1/2 problem. First, we note that the relevant coherent states factorize between different flavors, so $N_F$ just appears as an exponent in the form $e^{iN_F\Omega[w]}$.
[From now on we consider a single site, so suppress the site index.] 
{ Next we note that}, for the specified $Q(t)$, the coherent state spans a $2$-dimensional subspace. We can see this by explicitly by choosing the orthogonal matrix $O$ and its unitary representation $W_O$ appearing in the definition of the coherent states $\ket{S}=\hat{W}_O\ket{\mathbb{I}}$. We choose
\be
    O =
    \begin{pmatrix}
        \cos(\theta) & +\sin(\theta) &\rvline& 0\\
        -\sin(\theta) & \cos(\theta) &\rvline& 0\\
        \hline
        0 & 0 &\rvline& \mathds{1}_{2N-2}
    \end{pmatrix}.
\ee
[We have denoted the dimension of the identity matrix as a subscript.] This can be represented by the unitary transformation
\be
    W_O = \exp\left(-\f{\theta}{2} \gamma^{+1}\gamma^{+2} \right).
\ee
We can then map the space of the $2N$ Majorana modes into a collection of $N-1$ spin-$1/2$, e.g. we can identify $i\gamma^{+(a+1)}\gamma^{-(a+1)}$ with $X_a$, the $X$ Pauli matrix on the $a$-th site. The fact that we can encode the $2N$ Majorana modes into $N-1$ spins rather than $N$ is due to the conservation of $\mathcal{R}= \prod_a (i\gamma^{+a}\gamma^{-a})$, which halves the dimension of the Hilbert space.
Through this mapping $W_O\ket{\mathbb{I}}$ corresponds to a $2$-dimensional space where $X_a=1$ for all $a>1$, whereas the state of the first spin depends on $O$.
For the first spin we have $X_1 = i \gamma^{+1} \gamma^{-1} = i \gamma^{+2} \gamma^{-2}$, with the second equality originating from $\mathcal{R}=1$.
Given that $i\gamma^{+1}\gamma^{-2}=-i\gamma^{+2}\gamma^{-1}$ anti-commutes with $X_1$, but commutes with all other $X_a$, we can identify it with $Y_1$.

From the mapping above we now recognize that the coherent states $W_O\ket{\mathbb{I}}$ correspond to states of a spin-$1/2$. 
{ Computing} the expectation values of $X_1$, $Y_1$ and $Z_1=-i X_1 Y_1$ as a function of $\theta$, we have that the Berry phase associated with $Q(t)$ is equal to that of a spin-$1/2$ coherent state trajectory with
\be
    \begin{pmatrix}
        X\\ Y \\ Z
    \end{pmatrix}
    =
    \begin{pmatrix}
        \cos \theta_t \\ \sin\theta_t\\ 0
    \end{pmatrix}.
\ee
Using the explicit form of the Berry phase for spins~\cite{altland2010condensed} we see that the Berry phase only depends on the winding number $m$ of the angle modulo $4\pi$, and that $e^{i\Omega(1)}=-1$, thus completing our argument.

\section{Frobenius norm of $a_{n}$}
\label{app:sec:bipartite-entanglement}

In this section we report a detail related to Sec.~\ref{sec:entanglement}: the computation of the Frobenius norm of the matrices $a_{n} = \log\left(q_{n}\right)$, defined in Eq.~\ref{eq:qn-boundary}. We recall that the branch of the $\log$ should be chosen in such a way that $\left\lVert a_n\right\rVert$ is minimum.
We start by writing the characteristic polynomial for $q_{n}$, by noting that
\be
    \left(q_n\right)^n = (-1)^{n+1} \mathds{1}.
\ee
This fixes the eigenvalues of $a_n$ to be
\ba
    \lambda_j &= \frac{2\pi}{n}j,
    &
    j &= -\f{n-1}{2},\dots,\f{n-1}{2},
    &
    \text{for }n\text{ odd}\\
    \lambda_j &= \frac{2\pi j - \pi}{n},
    &
    j &= -\f{n}{2}+1,\dots,\f{n}{2},
    &
    \text{for }n\text{ even}.
\end{align}
Then the Frobenius norm is given by
\be
    \left\lVert a_n \right\rVert^2 = \f{\pi^2 (n^2-1)}{3n}
\ee
independently of the parity of $n$.

%

\end{document}